\documentclass[traditabstract]{aa}
\usepackage{longtable,lscape,amssymb}
\usepackage{float}
\usepackage{graphicx}
\usepackage{epsfig}
\usepackage[authoryear]{natbib}

\usepackage{lscape}
\usepackage{txfonts}
\usepackage{graphicx}
\usepackage{natbib}

\bibpunct{(}{)}{;}{a}{}{,} 

\newcommand{\Ha}{{H$\alpha$}}
\newcommand{\Hb}{{H$\beta$}}
\newcommand{\Hg}{{H$\gamma$}}
\newcommand{\Hd}{{H$\delta$}}

\newcommand{\pab}{{Pa$\beta$}}
\newcommand{\pag}{{Pa$\gamma$}}
\newcommand{\pad}{{Pa$\delta$}}
\newcommand{\brg}{{Br$\gamma$}}

\newcommand{\Ll}{{$L_{\rm line}$}}

\newcommand{\Lacc}{{$L_{\rm acc}$}}
\newcommand{\Macc}{{$\dot{M}_{\rm acc}$}}

\newcommand{\Msun}{{$M_{\odot}$}}
\newcommand{\Lsun}{{$L_{\odot}$}}
\newcommand{\Rsun}{{$R_{\odot}$}}

\newcommand{\Mstar}{{$M_{\star}$}}
\newcommand{\Lstar}{{$L_{\star}$}}
\newcommand{\Rstar}{{$R_{\star}$}}
\newcommand{\Teff}{{$T_{\rm eff}$}}
\newcommand{\Av}{{$A_{\rm V}$}}

\newcommand{\Mdisk}{{$M_{\rm disk}$}}

\newcommand{\papI}{{Paper I}}
\newcommand{\papII}{{Paper II}}

\newcommand{\vsini}{$v\sin i$}

\begin{document}

\title{GIARPS High-resolution Observations of T Tauri stars (GHOsT). III. A pilot study of stellar and accretion properties \thanks{
Based on observations made with the GIARPS instrument at the Telescopio Nazionale Galileo under program A36TAC\_22 (PI: Antoniucci).}
}

\author{
      J.M.~Alcal\'a\inst{1} 
   \and M.~Gangi\inst{2} 
   \and K.~Biazzo\inst{2}  
   \and S.~Antoniucci\inst{2} 
   \and A.~Frasca\inst{3} 
   \and T.~Giannini\inst{2} 
   \and U.~Munari\inst{4} 
   \and B.~Nisini\inst{2} 
   \and A.~Harutyunyan\inst{5} 
   \and C.~F.~Manara\inst{6} 
   \and F.~Vitali\inst{2} 
}

\offprints{J.M. Alcal\'a}
\mail{juan.alcala@inaf.it}

\institute{ 
      INAF-Osservatorio Astronomico di Capodimonte, via Moiariello 16, 80131 Napoli, Italy
 \and INAF-Osservatorio Astronomico di Roma, Via di Frascati 33, 00078 Monte Porzio Catone, Italy
 \and INAF-Osservatorio Astrofisico di Catania, via S. Sofia 78, 95123 Catania, Italy
 \and INAF-Osservatorio Astronomico di Padova -- Via dell'Osservatorio 8, 36012, Asiago (VI), Italy
 \and Fundaci\'on Galileo Galilei -- INAF -- Telescopio Nazionale Galileo, 38700, Bre\~na Baja, Santa Cruz de Tenerife, Spain
 \and European Southern Observatory, Karl Schwarzschild Str. 2, 85748 Garching, Germany
}

\date{Received ; accepted  }

\abstract{ 
The mass-accretion rate, \Macc, is a crucial parameter for the study of the evolution of accretion disks around 
young low-mass stellar objects (YSOs) and for planet formation studies. The Taurus star forming region (SFR) 
is rich in pre-main sequence (PMS) stars, most of them of the T Tauri class. A variety of methodologies have been 
used in the past to measure mass accretion in samples of YSOs in Taurus, but despite being a general benchmark for 
star formation studies, a comprehensive and systematic analysis of the Taurus T Tauri population, where the stellar and accretion properties are derived homogeneously and simultaneously, is still missing. 
As part of the GIARPS High-resolution Observations of T Tauri stars (GHOsT) project, here we present 
a pilot study of the stellar and accretion properties of seven YSOs in Taurus using the spectrograph GIARPS
at the Telescopio Nazionale Galileo (TNG).
Contemporaneous low-resolution spectroscopic and photometric ancillary observations allow us to perform an accurate 
flux calibration of the high-resolution spectra. The simultaneity of the high-resolution, wide-band spectroscopic 
observations, from the optical to the near-infrared (NIR), the veiling measurements in such wide 
spectral range, and many well-calibrated emission line diagnostics allows us to derive the stellar and accretion 
properties of the seven YSOs in a homogeneous and self-consistent way. The procedures and methodologies presented 
here will be adopted in future works for the analysis of the complete GHOsT data set. We discuss the accretion 
properties of the seven YSOs in comparison with the 90\% complete sample of YSOs in the Lupus SFR and investigate 
possibilities for the origin of the continuum excess emission in the NIR.
}


\titlerunning{A Pilot study of mass accretion in T Tauri stars with GIARPS}
\authorrunning{Alcal\'a et al.}
\maketitle

\section{Introduction} 
\label{intro}

The way in which circumstellar disks evolve and form proto-planets is deeply influenced by the processes of 
mass accretion onto the star, ejection of outflows, and photo-evaporation of disk material through winds  
\citep[][and references  therein]{hartmann16,ercolanopascucci17}.
In order to understand planet formation it is necessary to explain how optically thick accretion disks surrounding 
the youngest low-mass (\Mstar$\lesssim$2.0\Msun) stars evolve into optically thin debris disks
\citep[][]{morbidelliraymond16}. 
In this framework, the mass accretion rate, \Macc, is a fundamental parameter for the evolution of 
accretion disks around young low-mass stellar objects (YSOs). \Macc ~measurements set important constraints for 
disk evolution models \citep{hartmann98, hartmann16} and disk clearing mechanisms 
\citep[][and references therein]{alexander14,ercolanopascucci17}. 

In the current magnetospheric accretion paradigm for classical T Tauri (CTT) stars , the strong stellar 
magnetic fields truncate the inner disk at a few stellar radii \citep[][]{DonatiLandstreet09, johns-krull13}. 
Gas flows from this truncation radius onto the star along the stellar magnetic field lines, crashing onto 
the star and forming an accretion shock. The $\sim$10$^4$\,K optically thick post-shock gas and optically thin 
pre-shock gas produce emission in the Balmer and Paschen continua and in many lines, including the Balmer 
and Paschen series and the Ca II IR triplet \citep{hartmann98, hartmann16} observed in the optical spectra of  CTTs.  

The mass accretion rate can be derived from the energy released in the accretion shock 
\citep[accretion luminosity \Lacc; see][]{gullbring98, hart98} given the stellar properties. Observationally, 
this requires measurements of excess flux in continuum and lines with respect to similar nonaccreting template 
stars. Such measurements are best performed at ultraviolet (UV) wavelengths ($\lambda$ $\lesssim$ 4000\,\AA) 
with the Balmer continuum excess emission and the Balmer jump 
\citep[see ][and references therein]{HH08, ingleby13, alcala14, alcala17, manara17a}. In the past, \Lacc\ has been 
calculated using veiling measurements in high-resolution optical spectra 
\citep[e.g.,][and references therein]{hartigan91, hartigan03, white04}. Also, it is well known that \Lacc, and therefore \Macc, 
is correlated with the line luminosity, \Ll, of $\ion{H}{i}$, $\ion{He}{i,}$ and $\ion{Ca}{ii}$ lines  
\citep[e.g.,][and references therein]{muzerolle98, calvet04, HH08, rigliaco12, alcala14, alcala17}. These latter works 
 provide \Lacc--\Ll\ correlations simultaneously and homogeneously derived from the UV to the near-infrared (NIR), 
underlying the importance of these emission features as accretion diagnostics. These accretion tracers are 
key diagnostics with which to estimate \Lacc via the correlations mentioned above when flux-calibrated spectra below $\lambda$$\sim$3700\,\AA
~are missing.

On the other hand, accretion is a highly variable process \citep[][]{basribatalha90, jayawardhana06, cody10, venuti14}, 
which leads to a range of \Macc ~values for a given object when measured at different epochs \citep[see][]{costigan12, costigan14, biazzo12}. 
Variability in YSOs  induces dispersion in the observed \Macc--\Mstar\ and \Macc--\Mdisk\ scaling relationships, but cannot 
explain the large scatter of more than 2\,dex in $\log{}$\Macc\ at a given YSO mass. Such scaling relationships 
are predicted by the theory of viscous disk evolution \citep[][and references therein]{lynden-bell74,hartmann16,rosotti17},
but the \Macc--\Mdisk\ relationship has been confirmed observationally only recently by spectroscopic surveys in 
strong synergy with ALMA surveys of disks in star forming regions \citep[][]{ansdell16,manara16,pascucci16,mulders17}. 

As concluded in previous works \citep[e.g.,][]{rigliaco12,alcala14}, the average \Lacc\ and \Macc\ derived from 
several diagnostics, {measured simultaneously}, has a significantly reduced error. This suggests the 
need to use spectroscopic observations simultaneously performed from visible to NIR wavelengths and with 
the highest possible resolution to overcome problems due to line blending. These requirements can be achieved 
with the GIARPS \citep[GIAno and haRPS,][]{claudi17} high-resolution spectrograph at the Telescopio Nazionale Galileo (TNG)
in the Canary Islands, Spain. The spectral coverage and resolution of GIARPS allows one  
to probe the properties of the accretion columns, hot spots, the inner gaseous disk, the stellar and 
disk winds, and the collimated jets \citep[see][]{gangi20, giannini19}, making TNG/GIARPS a powerful 
instrument with which to investigate accretion in YSOs of the northern hemisphere, such as the Taurus-Auriga star 
forming region (SFR). 

The Taurus SFR contains a rich population of pre-main sequence (PMS) stars, most of 
them T Tauri stars \citep[see][and references therein]{kenyon08}. Several works have addressed the problem of 
measuring \Macc\  in the Taurus population using the different methodologies mentioned above 
\citep[see][and references therein]{HH08,HH14}. Also, a large number of CTTs have already been observed with ALMA 
\citep[][]{andrews18,long19}, probing the outer regions of their disk, and allowing 
the scaling relationships predicted by the viscous disk evolution theory to be investigated  further. However, despite being 
a general benchmark for star formation studies, a comprehensive and systematic analysis of the Taurus  population, 
where the relevant stellar and accretion parameters are derived simultaneously with sufficient accuracy, 
is still missing. As a first step in filling this gap,  here we present a pilot study of the accretion of seven 
YSOs in Taurus observed with TNG/GIARPS as part of the GHOsT (GIARPS High-resolution Observations of T Tauri stars) 
project. GHOsT is a survey of a flux-limited complete sample of T Tauri stars in the Taurus star forming 
region that is to be used to derive, in a homogeneous fashion (thus avoiding systematic errors due to the use of different sets of 
nonsimultaneous observations), the stellar, accretion, and outflow parameters, and to constrain the 
properties of both the inner disk and the associated winds and jets. The ultimate goal of GHOsT is to provide 
reliable measures of the mass-accretion and mass-loss rates of the Taurus population and to put them in 
relation with the properties of the central star and its disk, in synergy with the complementary ALMA 
observations \citep[][]{andrews18,long19}.
 
A comprehensive study of the jet line emission of the targets investigated in this paper was published 
by \citet[][henceforth \papI]{giannini19} and an investigation of the link between atomic and molecular winds 
was published in \citet[][henceforth \papII]{gangi20}, while the results of the complete GHOsT survey will 
be presented in forthcoming papers. One of the main goals of the present paper 
is the definition and assessment of the methodologies for the determination of the stellar and accretion properties, 
in a  homogeneous and self-consistent way, to be adopted for the analysis of the complete set of GHOsT data.  

The paper is organized as follows. In Sect.~\ref{obs} we present the target selection, the observations, and the data 
processing. In Section~\ref{datanalysis} we show the procedures used for the determination of the stellar parameters,
and veiling estimates both in the optical and in the NIR. In Sect.~\ref{acc_diag} the methodologies to derive the 
accretion parameters are presented, while in Sect.~\ref{results} we discuss results on the stellar and accretion properties of 
the studied sample, comparing with those of the 90\% complete sample of Lupus YSOs. The results of the veiling measurements
are also reported and discussed in the same section. Conclusions are outlined in Sect.~\ref{conclusions}.

\section{Target selection, observations, and data processing} 
\label{obs}
The targets, observations, and data reduction are extensively described in \papI\ and \papII, 
and are briefly summarized in this section.

\subsection{The targets}
For the pilot study, we selected six well-known CTTs in Taurus-Auriga with clear signatures of accretion.
The sources are listed in Table~\ref{ctts_prop} together with relevant properties from the 
literature (see also \papI). 
Noteworthy is the range of values for both the stellar physical parameters and mass accretion rate, 
which is likely due to variability, but also to the distinct methodologies used to derive them in previous works.
The extreme CTT RW\,Aur\,A \citep[][]{alencar05} was included in the sample as a case of highly veiled object 
spectrum \citep[][]{HH14} to test the limits of our procedures. 
In addition, we also study the well-known intermediate-mass CTT CQ\,Tau. This object was 
recently investigated in detail with ALMA \citep[][]{ubeira19} and has a low level of accretion \citep[e.g.,][]{mendigutia11}; 
therefore, it has been included to test the ability of our methods to trace very low accretion rates. 
However, this object displays a remarkable UX\,Ori-type variability \citep[][]{shakhovskoj05}, which complicates
the analysis of the data \citep[][]{dodin21}.
Several of the selected CTTs have been included in the recent ALMA survey by \citet[][]{long19}. 
Hereafter, we use the terms CTT and YSO interchangeably to refer to the objects in our sample.

\setlength{\tabcolsep}{5pt}
\begin{table*}[!ht]
\caption[ ]{\label{ctts_prop} Selected CTTs for this study with their physical and accretion properties from the literature.} 
\begin{tabular}{l|llccccl|cll|l}
\hline \hline

Source  & SpT   & \Teff &   \Av   & \Lstar   &   \Rstar  & \Mstar    &  Ref. & $\log$\Macc  &  Method for \Macc &  Ref.  & d  \\     
        &       &  (K)  &   (mag) &  (\Lsun) &  (\Rsun)  &  (\Msun)  &      &  (\Msun\ / yr)  &  determination   &  \Macc & (pc) \\
\hline                                                                                                                      
\hline                                                                                                                      
RY\,Tau & G1    & 5945  &   2.20   &  9.60  &   2.9    & 2.00   &  1     & $-$7.04/$-$7.19 &  UV-excess  &  1  & 138 \\
        &       & 5750  & 0.6-1.7  &  6.30  &          & 1.90   &  2     & $-$7.30          &            & 15  &     \\
        & F7    & 6220  &  1.95    &  12.30 &          & 2.04   &  6     &                  &            &     &     \\
        &       & 5920  &  2.35    &        &          &        &  18    & $-$7.21         & \Ha, He\,{\sc i}6678 &  18 &     \\
        & K1    & 5080  &          &  7.60  &  2.92    & 2.00   &  19    & $-$7.11         & \Ha          &  19 &     \\
\hline          

DG\,Tau & K7.0  &       &   1.60   &  0.51  &          & 0.76   &  3     &    $-$8.20      & \Ha &  3  & 125 \\
        & K6    &       &          &        &          &        & 17     &    $-$7.00      & \brg        &  4  &     \\
        &       &       &          &        &          &        &        &    $-$6.30      & UV-excess &  5  &     \\  
        &       & 4350  &  2.20    &        &          &        & 18     &    $-$7.30      & \Ha, He\,{\sc i}6678 &  18 &     \\  
        & M0    & 3890  &          &  1.70  &   2.87   &  0.30  & 19     &    $-$6.39      & \Ha        &  19 &     \\  
        & K3    & 4775  &          &  3.62  &   2.80   &  2.20  & 20     &    $-$6.13      & Excess emiss. &  21 &  \\  
\hline  

DL\,Tau & K5.5  & 4277  &   1.80   &  0.65  &          & 0.98   &  6     &    $-$8.6       &  \Ha        &  3  & 160 \\
        & K5.5  &       &   1.80   &  0.50  &          & 0.92   &  3     &    $-$6.3       &  UV-excess &  5  &    \\  
        &       &       &          &        &          & 0.75   &  16    &                 &             &     &    \\  

\hline  

HN\,Tau\,A & K3 &       &   1.15   &  0.17  &          & 0.70   &  3     &    $-$8.69      &  \Ha        &  3  & 134  \\
        &  K3  & 4730   &   1.15   &  0.16  &          & 1.53   &  6     &    $-$8.89      &  UV-excess &  7  &    \\  
        &  K5  &        &   0.65   &  0.19  &   0.76   & 0.81   &  7     &    $-$8.37      & Several line lum. &  8  &  \\  
        &      &        &          &        &          & 0.78   &  16    &                 &             &     &      \\  
\hline                                                                                                          

DO\,Tau & M0   &        &   2.27   &  1.01  &   2.25   & 0.37   &  7     &    $-$6.84      &  UV-excess &  7  & 139  \\
        & M0.3 & 3806   &   0.75   &  0.23  &          & 0.59   &  6     &    $-$8.21      &  \Ha        &  3  &     \\
        &      &        &          &        &          & 0.56   &  16    &                 &             &     &     \\

\hline  

RW\,Aur\,A & K3 &       &   0.5    &  0.50  &   1.10    & 0.90  &  9     &    $-$7.70      &  UV-excess &  9  & 183  \\  
        & K0-K3 & 5082  &   0.4    &  1.70  &          & 1.34   &  10    &    $-$7.50      &  UV-excess &  10 &     \\
        & K0    & 5250  &   0.0    &  0.99  &          & 1.20   &  6     &    $-$7.39      & Several line lum. &  11 &   \\ 
        & K2    & 4955  &          &  1.70  &  1.70    & 1.34   &  20    &    $-$7.51      &  Excess emiss.    &  21 &   \\ 
\hline  

CQ\,Tau & F2    &       &   1.90   &  10.0  &          & 1.67   &  12    &   $<-$8.3       &  UV-excess &  13 & 149 \\
        &       &       &          &        &          &        &        &    $-$7.0       &  \brg       &  14 &     \\
        & F3    & 6900  &   1.40   &   3.4  &          &        &  22   &                 &              &      &     \\
                                                                                                        
\hline
\end{tabular}
\tablefoot{~\\
-- Parameters of Long et al. are revisited values of \citet[][]{HH14}. \\
-- \Av, SpT, \Lstar\ in \citet[][]{simon16} are those in \citet[][]{HH14}.\\
-- Distances in the last column are from Gaia EDR3 \citep[][]{gaia16, gaiaedr3}.\\
~~\\
{\bf References:} \\
(1)\,\citet[][]{calvet04}; (2)\,\citet[][]{garufi19}; (3)\,\citet[][]{simon16}; 
(4)\,\citet[][]{agra11}; (5)\,\citet[][]{gullbring00}; (6)\,\citet[][]{long19}; (7)\,\citet[][]{gullbring98}; 
(8)\,\citet[][]{fang18}; (9)\,\citet[][]{ingleby13}; (10)\,\citet[][]{white11}; (11)\,\citet[][]{facchini16}; 
(12)\,\citet[][]{ubeira19}; (13)\,\citet[][]{mendigutia11}; (14)\,\citet[][]{donehew11}; 
(15)\,\citet[][]{skinner18};(16)\,\citet[][]{rigliaco15}; (17)\,\citet[][]{hessman97}; (18)\,\citet[][]{frasca18}; 
(19)\,\citet[][ and references therein]{isella09}; (20)\,\citet[][and references therein]{akeson05}; 
(21)\,\citet[][]{white04}; (22)\,\citet[][]{meeus12}.}\\
\end{table*}

\subsection{Observations and data reduction}
\label{obs_datared}
The GIARPS observing mode combines the HARPS-N \citep[][]{pepe02,cosentino12} and GIANO-B \citep[][]{oliva12,origlia14} 
high-resolution (resolving power of 115,000 and 50,000, respectively) spectrographs, simultaneously covering 
a wide spectral range of 390--690\,nm for HARPS-N, and 940--2420\,nm for GIANO-B. 
In order to perform a flux calibration as accurately as possible, avoiding additional uncertainties due 
to variability, the GHOsT survey is complemented with simultaneous/contemporaneous  (within less than two days 
of the GIARPS observations) low-resolution spectroscopy in the optical, as well as with optical and NIR 
photometry (see \papI).

The TNG/GIARPS observations were performed in 2017 during two nights, one on October 29 and the other 
on November 13. The journal of the GIARPS observations is reported in \papI. 
The reduction steps of the GIARPS spectra are thoroughly described in \papI\ and \papII, but a summary is 
provided here. 

The HARPS-N spectra were reduced using the latest version (Nov. 2013) of the HARPS-N instrument Data 
Reduction Software and applying the appropriate mask depending on the spectral type of the object 
\citep[][]{pepe02}. The basic processing steps for the data reduction consist in bias and dark subtraction, 
flat fielding, wavelength calibration, spectrum extraction, and cross-correlation computation. For the removal 
of the spurious features caused by the telluric lines we first used the {\em molecfit} tool \citep[][]{smette15, kausch15}
to produce a synthetic telluric spectrum and then the package {\em telluric} in IRAF\footnote{IRAF is distributed by 
the National Optical Astronomy Observatory, which is operated by the Association of the Universities for 
Research in Astronomy, inc. (AURA) under cooperative agreement with the National Science Foundation.} 
to remove the telluric lines.

In order to flux-calibrate the HARPS-N spectra, the contemporaneous absolute flux-calibrated spectroscopy was acquired 
within 2 nights of our GHOsT runs with the 1.22m telescope at the Asiago observatory, Italy. The spectra cover the 
wavelength range 330--790\,nm and were fully reduced and flux-calibrated against a spectrophotometric standard observed 
during the same night. Their flux zero-point was also checked with $BVR_CI_C$ photometric measurements collected during 
the same nights with the ANS Collaboration telescopes \citep[see][]{munari12}. The photometry is reported in \papI.
Given the short temporal distance between the two datasets we assume that the continuum shape did not change significantly 
between the Asiago and TNG observations. Thus, for each source we first fitted the continuum of the Asiago spectrum and 
then multiplied it for the continuum-normalized HARPS-N spectrum. 

The GIANO-B spectra were extracted following the data-reduction prescriptions described
in \citet[][]{carleo18}. Halogen lamp exposures were employed to map the order geometry and for flat-field 
correction, while wavelength calibration was based on lines from a U-Ne lamp acquired at the end of each 
observing night. The {\em molecfit} tool \citep[][]{smette15} was used for removing the telluric contribution 
in the near infrared. 

To flux-calibrate the GIANO-B spectra, NIR photometry in the $JHK_S$ bands was acquired with the 
REMIR instrument on the Rapid-Eye-Mount (REM) telescope \citep[][]{vitali03}, at the La Silla Observatory, Chile, 
during the night of  November 11, 2017. We assume that the NIR magnitudes did not change significantly between TNG,  
Asiago, and REM observations. We then performed an interpolation between the considered flux measurements using 
a spline function to derive a smooth continuum function in the wavelength range 940--2420\,nm, that is a 
response function, which was then applied to flux-calibrate the various (continuum-normalized) segments of 
the GIANO-B spectrum. Finally, the flux-calibrated spectral segments most relevant for this pilot study of 
accretion were selected. 

We estimate that the flux calibration of the GIARPS spectra is precise within 20\%. The final products
do not contain the region of the Balmer Jump, but include 17 well-resolved accretion diagnostics 
(12 in the optical and 5 in the NIR) that form the basis for our measurements of accretion in 
the sample (see Sect.~\ref{acc_diag}).

\section{Stellar parameters}
\label{datanalysis}

 A first step in any study of accretion is the determination of the YSO physical properties. Estimates of the physical
 parameters, derived using a number of methodologies, exist in the literature for all the YSOs in our sample.  
 In order to minimize uncertainties due to the different procedures used in the literature, and in view of the 
 forthcoming GHOsT data, we need to adopt a methodology that allows us to determine the YSO properties in a self-consistent 
 and homogeneous way.

\setlength{\tabcolsep}{5pt}
\begin{table*}[!ht]
\caption[ ]{\label{rotfit_stel_prop} Properties of the CTT sample derived in this work} 
\begin{tabular}{lllcclll}
\hline \hline

Name &\Teff\ ($\pm$err) &   SpT   & \Av\ $^a$  & \Av\ $^b$ & \Lstar  & \Rstar & \Mstar   \\
     & (K)            &         & (mag)    & (mag)   & (\Lsun)         &  (\Rsun)   & (\Msun)   \\

 (1)       &     (2)  &  (3)     & (4)    &   (5)   &  (6)  & (7) &  (8)   \\
\hline     

RY\,Tau    & 5856 (151)  &   G1   & 2.25 & 2.32 & 8.87 & 2.89 & 1.80  \\
DG\,Tau    & 4004 (153)  &   K7   & 1.50 & 1.79 & 0.44 & 1.38 & 0.70  \\
DL\,Tau    & 4188 (100)  &   K5   & 1.50 & 1.45 & 0.40 & 1.20 & 0.90  \\
HN\,Tau\,A$^\dagger$ & 4617 (97) & K4 & 1.25 & 1.53 & 0.15 & 0.60 & 0.80$^\star$  \\
DO\,Tau    & 3694 (104)  &   M1   & 1.40 & 1.69 & 0.42 & 1.58 & 0.50  \\
RW\,Aur\,A &  4870        &   K0   & 1.00 & 0.57  & 1.64 & 1.80 & 1.50 \\
CQ\,Tau$^\ddagger$  & 6823 (136) & F4 & 0.50 & 0.60 & 2.71 & 1.18 & 1.50$^\star$ \\

\hline
\end{tabular}
\tablefoot{~\\
$^a$: extinction derived using spectral templates only. \\
$^b$: extinction derived using spectral templates and veiling. \\
$\dagger$ : subluminous YSO (Sect.~\ref{luminosity}). The \Lstar, \Rstar, and \Mstar\ values are underestimated. \\ 
$\ddagger$ : UX\,Ori type variable (Sect.~\ref{luminosity}). Parameters affected by variable circumstellar extinction. \\
$\star$: corrected values of \Lstar, \Rstar, and \Mstar\ for the two subluminous objects are estimated in 
Sect.~\ref{HN_Lacc}. The mass reported here is that of the closest PMS track on the HR diagram, with the
uncorrected values of \Lstar.}
\end{table*}

\subsection{Effective temperature and optical veiling}
\label{rotfit}
 We used the ROTFIT code to determine effective temperature,  \Teff, surface gravity, $\log{g}$, radial velocity, $RV$, 
 projected rotational velocity, $v \sin{i}$, and  veiling as a function of wavelength, $r_\lambda$. 
 The code has been successfully applied on YSOs optical spectra at different resolutions \citep[][]{frasca06, frasca15}, 
 and is particularly well suited for the HARPS-N spectra.
 In short, the code finds the best photospheric rotationally broadened template spectrum that reproduces the target spectrum 
 by minimizing the $\chi^2$ of the difference between the observed and template spectra in specific spectral segments. 
 The spectral segments ($\sim 100$\,\AA\ each) are normalized to unity, and therefore a measure of extinction is not provided by the code. 
 We adopted as templates a grid of high-resolution spectra of slowly rotating, low-activity stars with well-known 
 atmospheric parameters, which were retrieved from the ELODIE Archive \citep[][]{Moultaka2004}. 
 The HARPS-N spectra were degraded to the ELODIE resolution ($R=42,000$)  before running the analysis code. 
 We have chosen this template grid because real spectra are best suited for the determination of veiling and \vsini\ 
 \citep[][]{frasca15,frasca19}.
 Moreover, they allow us to perform a careful subtraction of the photospheric spectrum (see Section~\ref{line_fluxes}). 
 The spectral intervals analyzed with ROTFIT contain features that are sensitive to the effective 
 temperature and/or $\log{g}$.

 The whole set of physical parameters derived with ROTFIT (\Teff, $\log{g}$, $RV$,  $v \sin{i}$, and  veiling) 
 for the complete GHOsT sample will be presented in a forthcoming  paper, while the effective temperature, 
 essential for this work, is provided in Table~\ref{rotfit_stel_prop}. 
 Optical veiling values, also crucial for this pilot study, are given in Table~\ref{veilings}. 
 The spectral types, determined using the \Teff\ from ROTFIT and the relation between spectral type 
 versus \Teff\ by \citet[][]{HH14}, are also reported in column 3 of Table~\ref{rotfit_stel_prop}.
  
 For RW\,Aur\,A it was not possible to derive the parameters using the ROTFIT method, as the spectrum is totally crossed
 by emission lines that hide the photospheric lines at optical wavelengths. As discussed in Sect.~\ref{nir_veiling}, 
 the templates best matching the NIR 
 spectrum have a spectral type between K1 and K0, the latter being consistent with the determination by \citet[][]{HH14}
 for this star.  We therefore adopted the \Teff\ value as determined from the K0 spectral type and the 
 SpT versus \Teff\ relationship by these authors. For this star, \citet[][]{HH14} report a veiling of 0.5 at $\lambda=$750\,nm.
 Based on a linear relationship between the veiling at 580\,nm and 710\,nm drawn from our previous studies of the
 Lupus population \citep[][]{alcala14,alcala17},  we estimate that the veiling at $\lambda=$600\,nm  
 is $r600 \approx 2.3 \times r710$, where r710 is the veiling at $\lambda=$710\,nm. Within the errors,
 this is consistent with the measured veiling in \citet[][]{frasca17} where $r620 \approx 2.0 \times r710$.
 For RW\,Aur\,A, we calculate a  value of $r600 \sim 1.2$. We find a similar value of veiling considering 
 the equivalent width of the lithium line at 6708\,\AA\,and applying the method described in \citet[][]{biazzo11}.
 However, we consider this a tentative value only, as the veiling of this CTT may 
 be strongly variable \citep[0.3$ < r_\lambda < $6.1, for $\lambda$ in the range 500--650\,nm,][]{stout-batalha00}. 
  
 The corresponding \Teff\ values derived from ROTFIT are in good agreement, within the errors,  with those drawn from the  
 \citet[][]{HH14} spectral types. As a logical consequence,  the spectral types are also in good agreement, within a spectral 
 subclass. We note that the values for CQ\,Tau, which have not been included in the \citet[][]{HH14} study, are in 
 agreement with those published in \citet[][]{mendigutia11} and \citet[][]{meeus12}. Also, the negligible veiling at 
 optical wavelengths for this star is in agreement with the recent results by \citet[][]{dodin21}.

\setlength{\tabcolsep}{2pt}
\begin{table*}[!ht]
\caption[ ]{\label{veilings} Veiling in the optical from ROTFIT and in the NIR with the procedure described in text. The projected
 rotational velocity determined with this procedure is also given in the last column.} 
\begin{tabular}{l|ccccc|cccccccccccc|c}
\hline \hline

Name &  $r450$ & $r500$ & $r550$  & $r600$ & $r650$    & $r968$ & $r983$  & $r1178$ & $r1256$ & $r1298$ & $r1565$ & $r1597$ & $r1666$ & $r1741$ & $r2130$ & $r2255$ & $r2322$ & $v \sin{i}$ ($\pm$err) \\
     &         &       &         &        &          &      &       &       &       &       &       &       &       &       &       &       &     & (km/s) \\
\hline      
     &      &      &      &      &      &      &      &      &       &       &       &       &       &       &       &     &      \\
RY\,Tau     & 0.0 & 0.0 & 0.0 & 0.0 & 0.0 & 0.2 & 0.6 & 0.4  & 0.7  & 0.3 & 0.5 & 0.4 & 0.5 & 0.9 & 1.4  & 1.2 & ... & 49.8 (1.6) \\
DG\,Tau     & ... & ... & 2.0 & 1.5 & 1.0 & 1.6 & 1.2 & 1.8  & 1.4  & 1.7 & 1.8 & 1.6 & 1.7 & 2.4 & 3--4 & 3--4 & ... & 26.3 (3.3) \\
DL\,Tau     & 3.0 & 2.5 & 2.0 & 1.2 & 1.5 & 1.1 & 1.1 & 1.1  & 1.0  & 1.0 & 0.9 & 0.6 & 0.8 & 1.7 & 2.5 & 2.1 & 2.5 & 12.0 (3.0)   \\
HN\,Tau\,A   & ... & 0.8 & 0.8 & 0.8 & 0.5 & ... & ... &        1.1  & 1.2  & 1.5 & 1.8 & 1.6 & 1.3 & 2.3 & 3.3 & 5.0 & ... & 50.7 (2.7) \\
DO\,Tau     & 1.8 & 1.5 & 1.5 & 1.0 & 1.5 & 0.5 & 0.6 & 0.7  & 1.4  & 1.4 & 1.9 & 1.7 & 1.8 & 2.1 & 3.4 & 3.7 & 2.7 &  12.0 (2.2)\\
RW\,Aur\,A  & ... & ... & ... & 1.2 & ... & ... & ... & ...  & 1.5  & ... & ... & 2.5  & 2.7 & 3.5 & 3.6 & 5.7 & 5.4 & 20.0 (3.4) \\
CQ\,Tau     & 0.0 & 0.0 & 0.0 & 0.0 & 0.0 & ... & 0.5 & 1.0 &  0.8 & ... & $<$3.0 &  2.0 & ... & ... & ... & ... & ... & 79.0 (5.0) \\
\hline
\end{tabular}
\end{table*}

\subsection{Veiling and \vsini\ estimates in the NIR}
\label{nir_veiling}
The veiling of IR photospheric lines is usually attributed to the emission of the inner edge of the dusty disk, 
where the dust is heated by stellar and accretion radiation. As such, it can be used to infer properties of 
the emission in excess of the stellar photosphere originating in the inner disk and complementary to the 
UV and optical excess. 
In addition, veiling may have an important impact on extinction \citep[see][and references therein]{fischer11, HH14}.

\begin{figure}[!ht]

\resizebox{0.95\hsize}{!}{
{\includegraphics[bb=30 0 460 490]{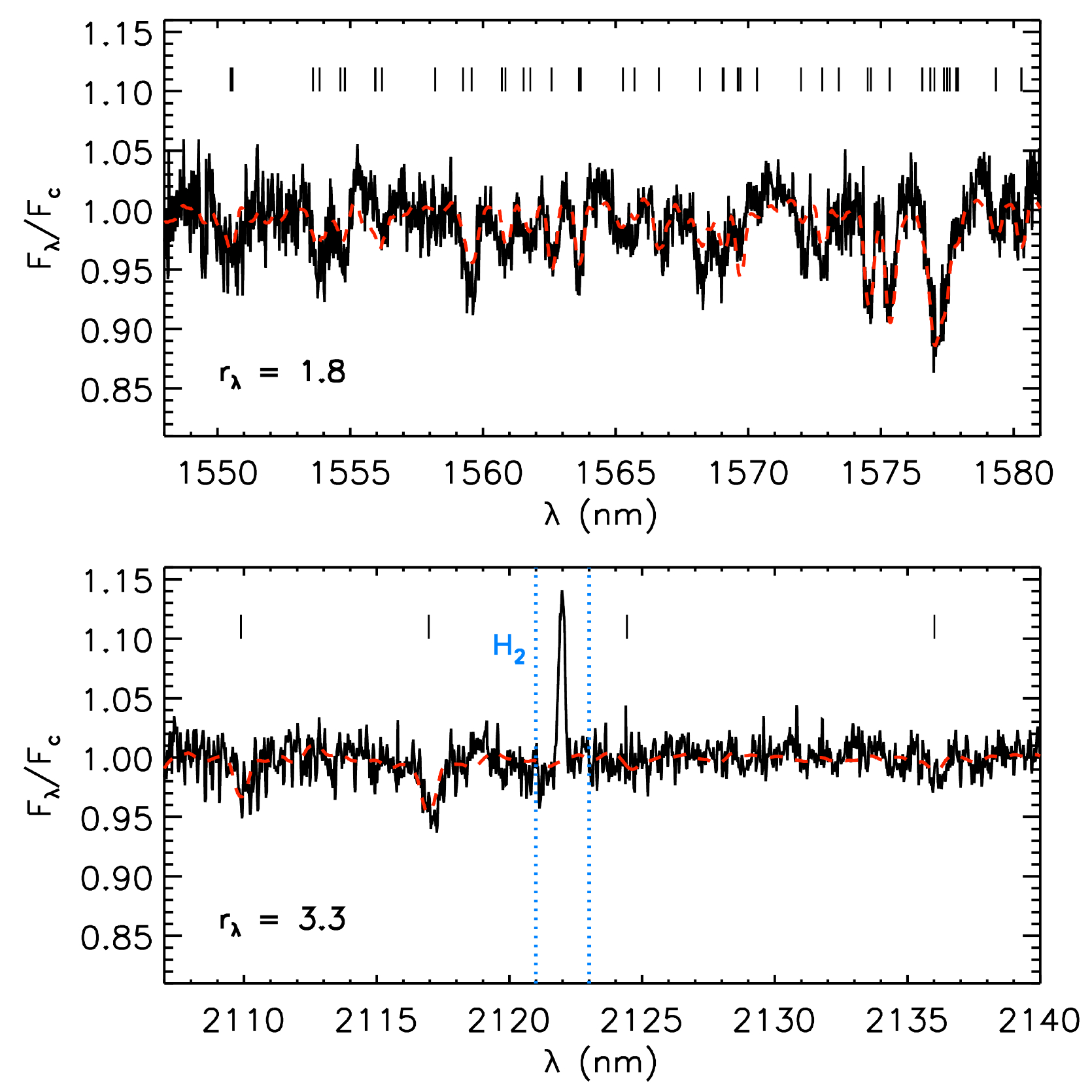}}
}
\caption{Two portions of the continuum-normalized spectrum of HN\,Tau\,A (black solid line) in the $HK$ bands, 
with the rotationally broadened and veiled spectrum of the stellar template overplotted (red dashed line). 
Absorption features, mainly of the iron-peak group, are indicated with small vertical lines. We refer
to \citet[][]{rayner09} for the details on the absorption features. Emission lines such as the H$_2$ line,
indicated with the blue dotted lines, are excluded from the analysis. The value of veiling, which minimizes 
residuals with respect of the rotationally broadened and veiled template, is indicated in the lower left.
   \label{hntau_veiling}}
\end{figure}

The procedure we used to measure veiling of the photospheric lines in the NIR follows the prescription thoroughly described 
in \citet[][]{antoniucci17}. Briefly, we use template spectra acquired with GIANO-B and with similar spectral types to the targets. 
The templates are chosen to have very low $v \sin{i}$ values.
Each template spectrum is broadened by convolution with a rotational profile \citep[][]{gray05} by increasing $v \sin{i}$ and 
it is artificially veiled by adding a continuum excess, with both \vsini\ and $r_\lambda$ as free parameters 
\citep[see Eq.~1 in][]{antoniucci17}, until the photospheric features match those of the target and minimum residuals between 
the target spectrum and the rotationally broadened and veiled template are obtained. Figure~\ref{hntau_veiling} shows an example of the result
of the fitting procedure for HN\,Tau\,A in two spectral intervals.
The $r_\lambda$ values are assumed to be constant within each spectral order. Values of $r_\lambda$ derived from this procedure are given in Table~\ref{veilings}. The procedure also yields an estimate of $v \sin{i}$, which is also given in Table~\ref{veilings}.  The errors in veiling are on the order of 20\%.

As mentioned in the previous section, the optical HARPS-N spectrum of RW\,Aur\,A is heavily veiled and crossed 
by numerous emission lines. Yet its NIR 
GIANO-B spectrum shows some photospheric lines making it possible to estimate veiling and \vsini\ . During this process, we were also able to determine that the spectral type of the template best matching the photospheric lines corresponds to a type 
between K1 and K0. We stress that in the case of DG\,Tau, the  $r2130$ and $r2255$ estimates are rather uncertain, 
and so we were only able to determine a range of values (see Table~\ref{veilings}). 

\subsection{Extinction}
\label{extinction}
As a first step to derive the visual extinction, \Av, we adopt the methods described in our previous works \citep[][]{alcala14,alcala17,manara17a}. 
As input data we use our primary flux calibrated spectra, that is the Asiago spectra, which cover a sufficient wavelength 
range (330--790\,nm) for the purpose. To derive \Av\ for a given YSO, its spectrum was compared with the spectra of 
nonaccreting YSOs best matching in spectral type. All our nonaccreting templates were taken 
from \citet[][]{manara13,manara17} and have very low or negligible extinction \citep[$A_{\rm V} <$ 0.5\,mag; see][]{manara13,manara17}. 
The templates were rebinned to the Asiago spectra and then artificially reddened by $A_{\rm V}$ in the range 0.0--4.0\,mag 
in steps of 0.25\,mag until the best match to the continuum slope of the YSO spectrum was found. To redden the spectra 
we used the extinction law by \citet[][]{WD01} for R$_{\rm V}=5.5$, which has been found to be particularly well suited
for YSOs in general \citep[][and references therein]{evans09}. The \Av\ values derived in this way are listed in column~4 
of Table~\ref{rotfit_stel_prop}. 

The main sources of uncertainty on \Av\ are the errors in spectral type when associating a template to a given YSO,
the error in the extinction of the templates, and the errors in flux calibration. The combined effect leads to an 
error of $\sim$0.35\,mag, except for the case of RW\,Aur\,A for which we estimate a larger error of about 0.5\,mag. 
Adopting extinction laws by different authors yields results consistent, within the errors of about 0.35\,mag, with 
those of \citet[][]{HH14}. For instance, the latter authors find that for a star with their measured \Av$=$1.0\,mag, 
adopting R$_{\rm V}=$5 from \citet[][]{cardelli89} or R$_{\rm V}=$5.1 from \citet[][]{WD01} would yield \Av$=$ 1.2 and 
1.15 mag, respectively.

 \begin{figure*}[!ht]

 \resizebox{0.95\hsize}{!}{
 {\includegraphics[bb=0 0 850 550]{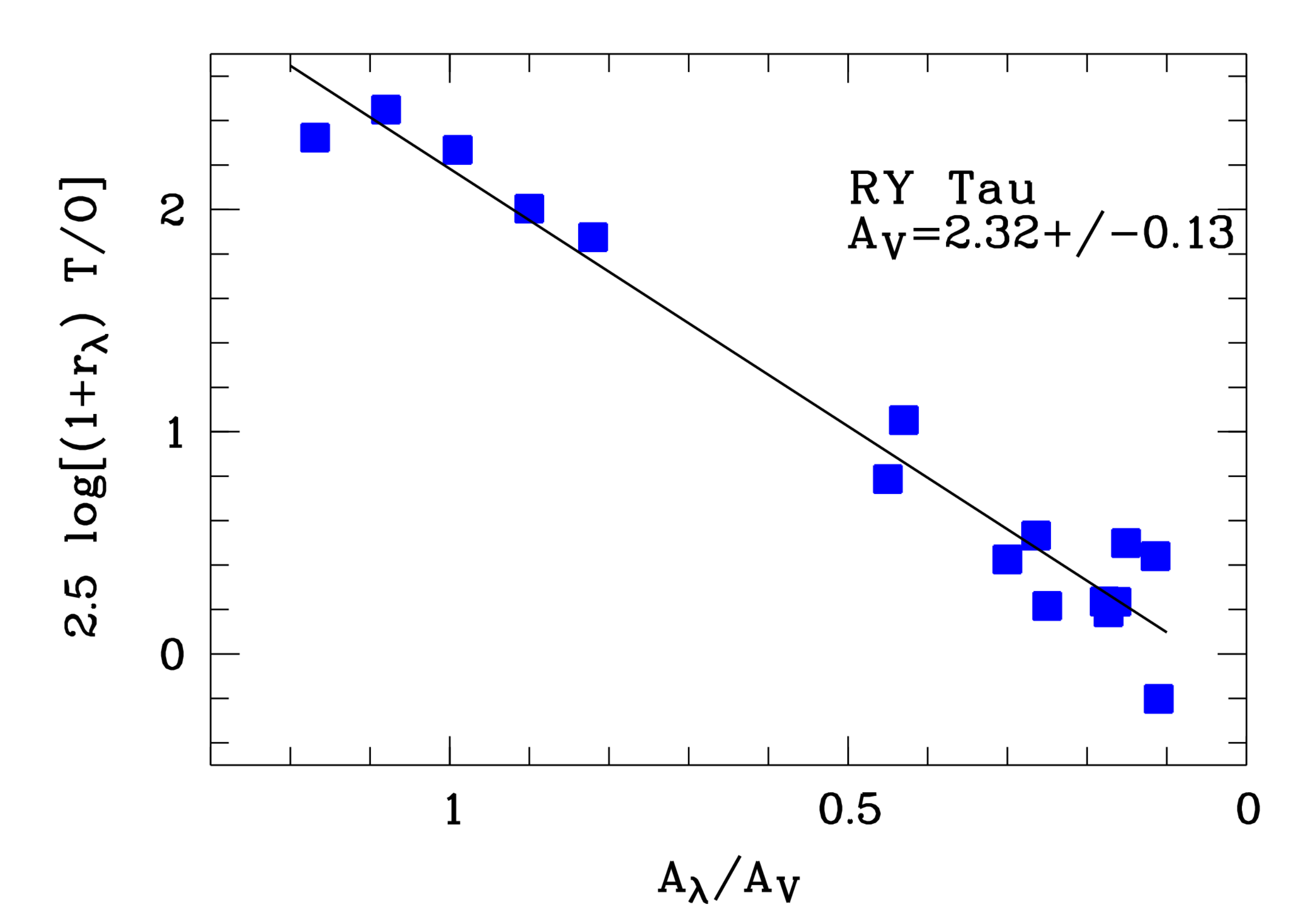}}
 {\includegraphics[bb=0 0 850 550]{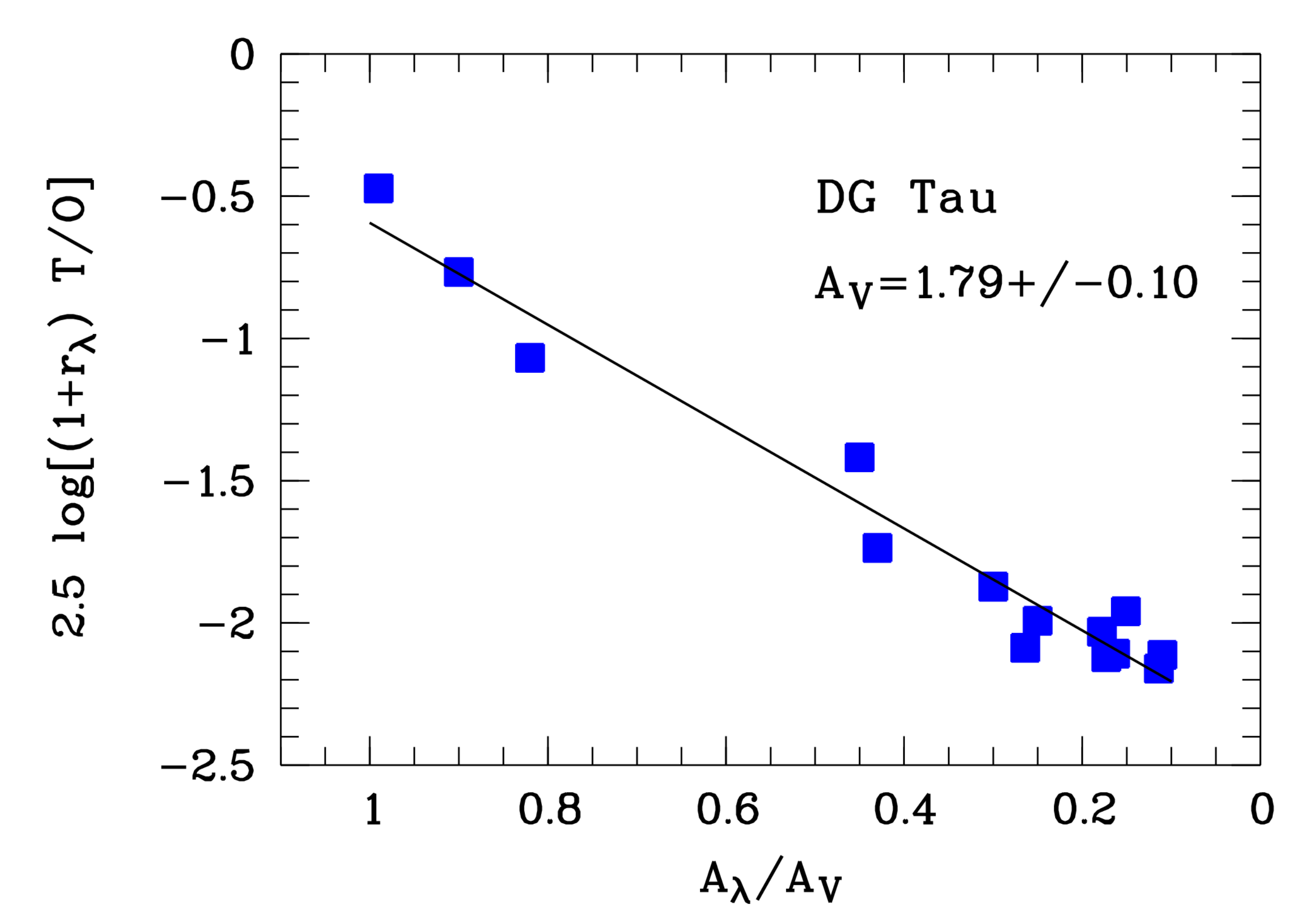}}
 }
 \resizebox{0.95\hsize}{!}{
 {\includegraphics[bb=0 0 850 550]{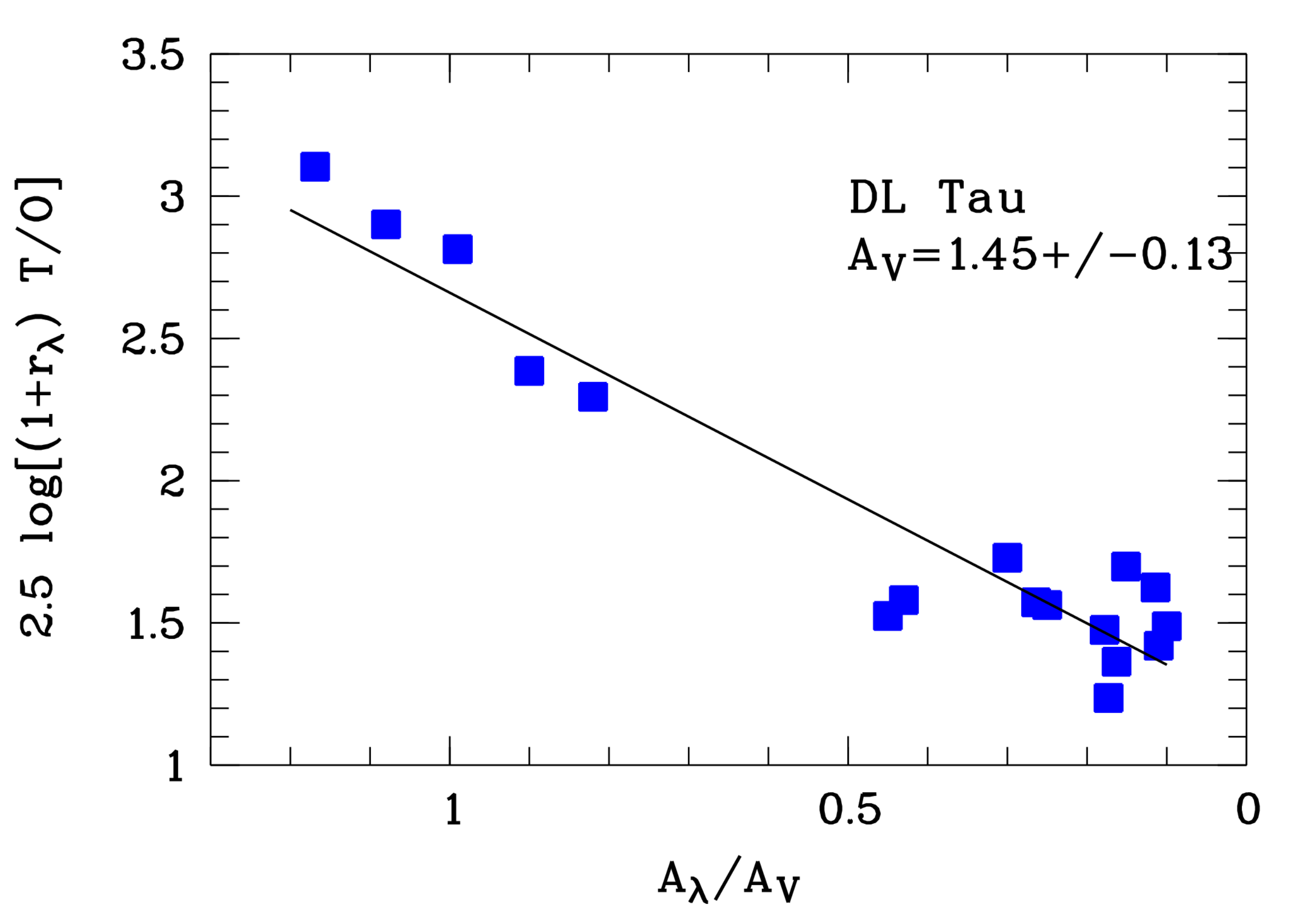}}
 {\includegraphics[bb=0 0 850 550]{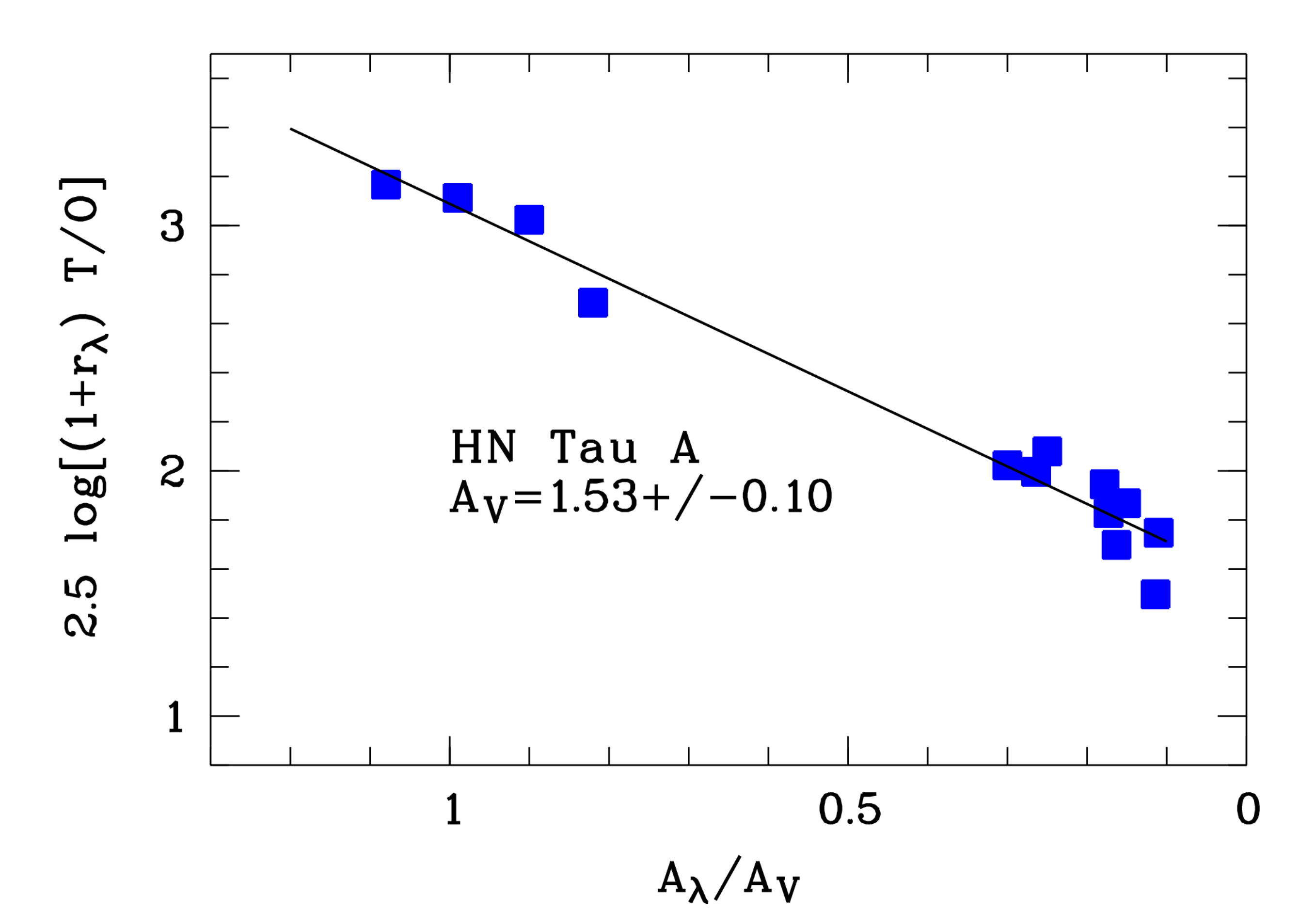}}
 }
 \resizebox{0.95\hsize}{!}{
 {\includegraphics[bb=0 0 850 550]{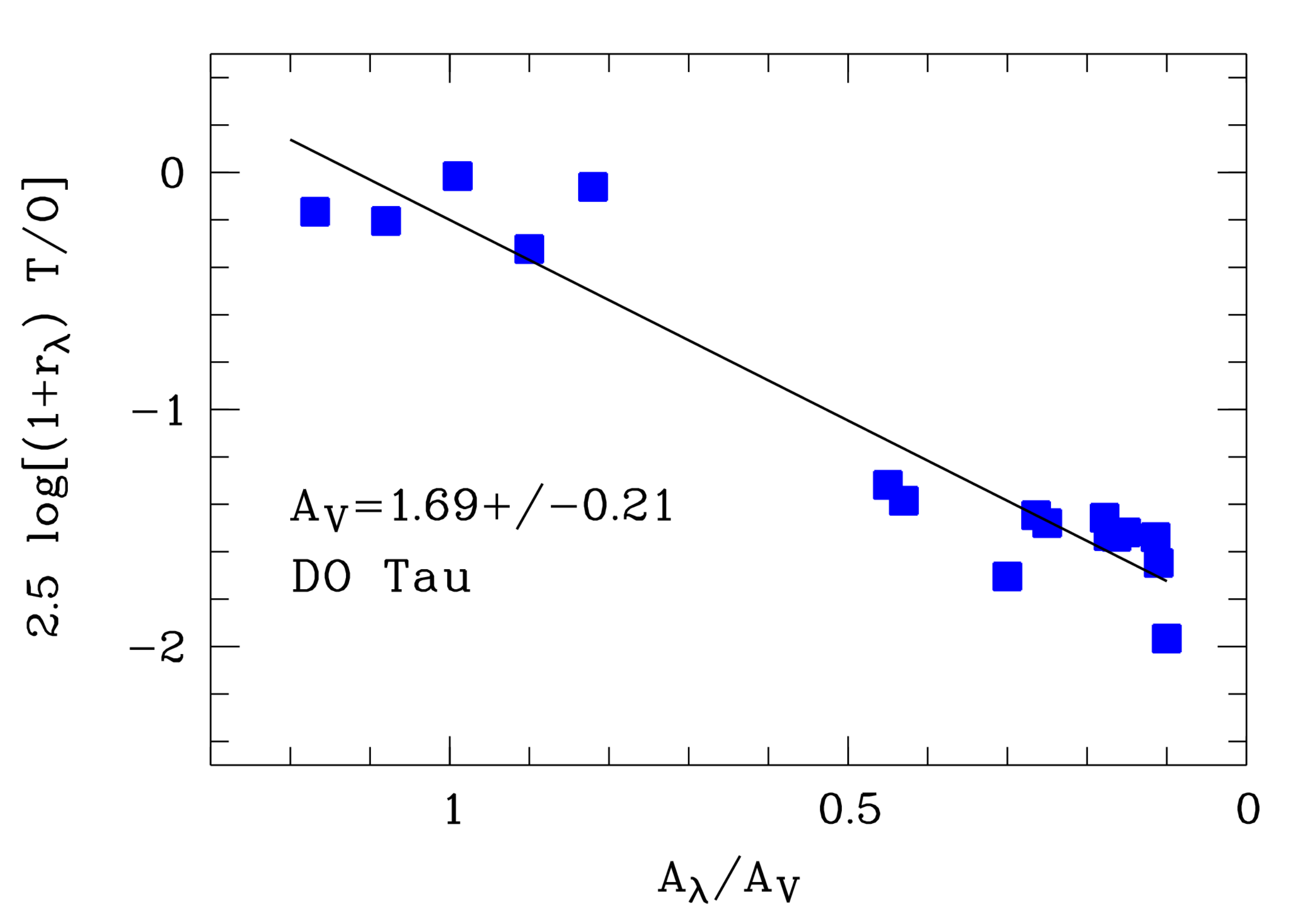}}
 {\includegraphics[bb=0 0 850 550]{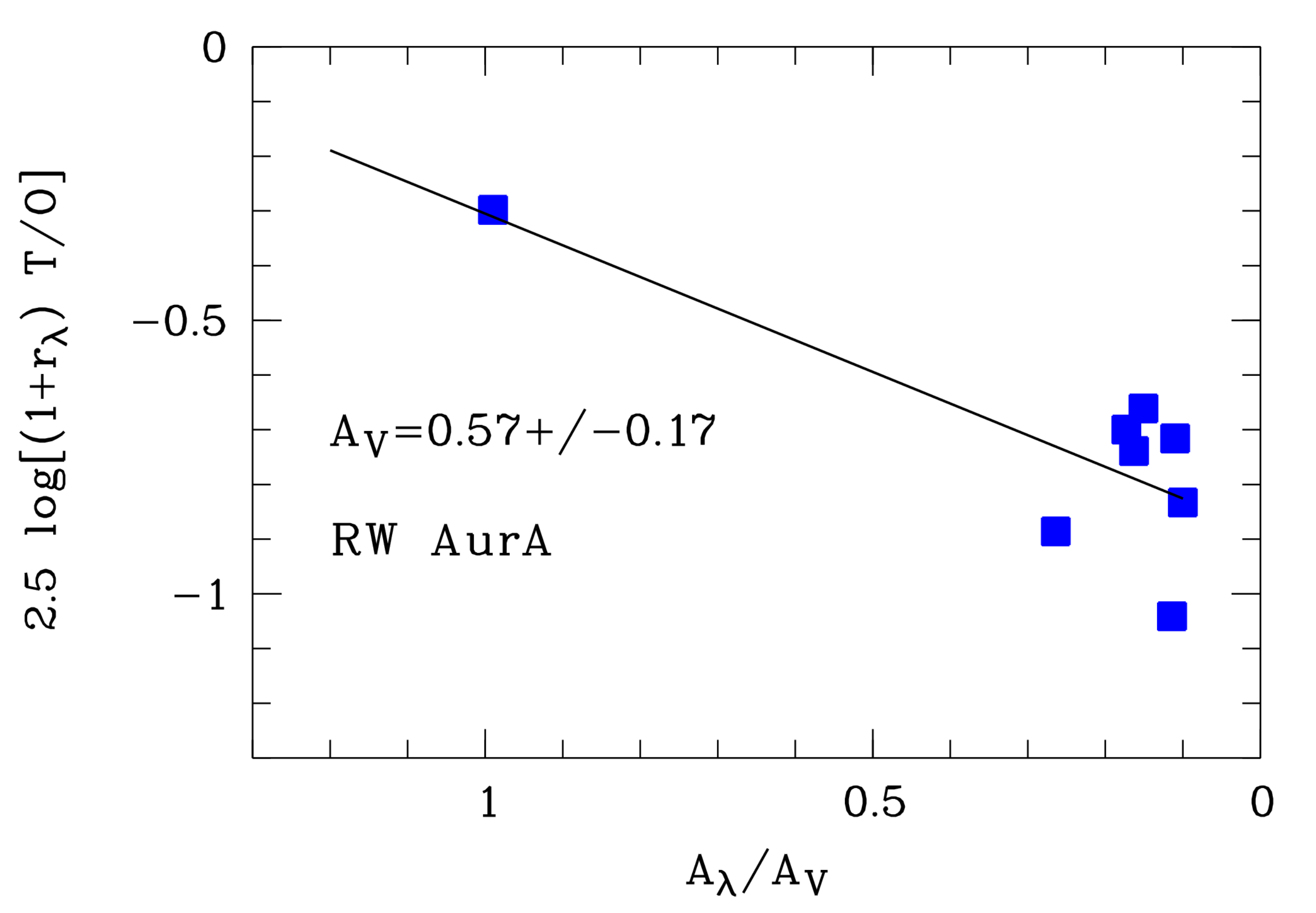}}
 }
 \resizebox{0.73\hsize}{!}{
 {\includegraphics[bb=-400 0 850 550]{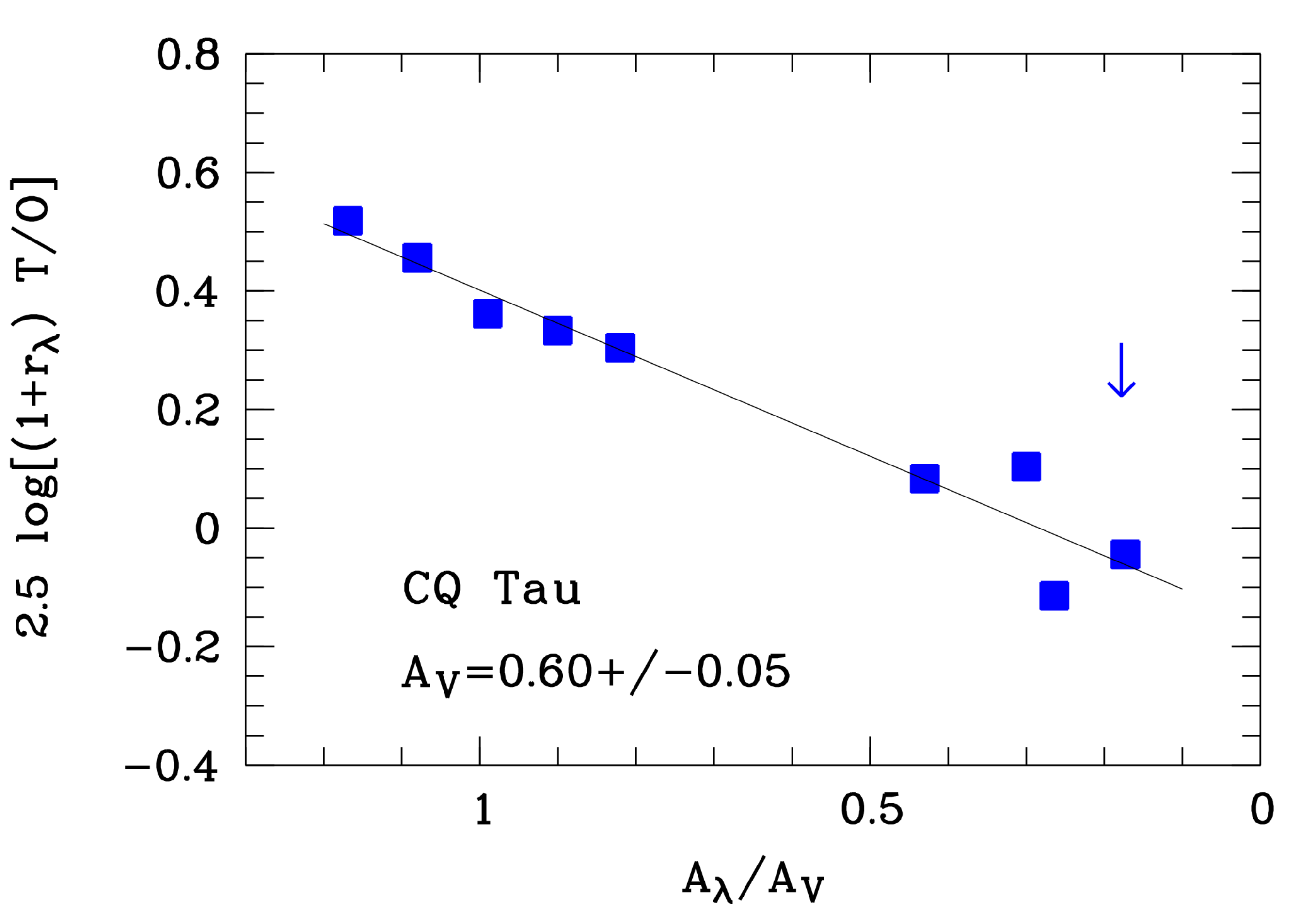}}
 }
\caption{Plots of  $2.5\cdot\log{[(1 + r_\lambda) \cdot F^T_\lambda / F^O_\lambda ]}$ vs. $A_\lambda/$\Av\ for the YSOs in the sample
(blue squares). The best linear fit for each object is shown as a black line. The value of the slope \Av\ is shown in each panel.
In the case of CQ\,Tau the $r1565$ value is shown as an upper limit.
   \label{lin_fits_Av}}
\end{figure*}

As mentioned in Section~\ref{nir_veiling}, another source of uncertainty is strong veiling, 
which makes the YSO spectra intrinsically bluer than the templates. The extinction as derived above is not 
severely affected by the veiling, as soon as the analysis 
is restricted to wavelengths longer than about 450\,nm \citep[see analysis in Appendix~B of][]{alcala14}, but also shorter
than about 800\,nm to avoid the effects of veiling in the NIR \citep[][]{fischer11}. 
Thus, to minimize the impact of veiling, the extinction was derived only from the red portion of the 
Asiago spectra, starting at 500\,nm. In this case, the effects of NIR veiling are automatically excluded as the Asiago 
spectra do not cover wavelengths longer than 790\,nm.

We also applied an alternative method to derive \Av\  using the prescription explained in \citet[][see their Eq.~4]{fischer11}, originally proposed 
in \citet[][their Eq.~5]{gullbring98}. This requires measurements of the veiling, $r_\lambda$, of the observed flux of 
the spectral template, $F^T_\lambda$, and of the observed flux of the object, $F^O_\lambda$, as a function of wavelength.
The method is based on the fact that the quantity  
$\Gamma_\lambda=2.5\cdot\log{[(1 + r_\lambda) \cdot F^T_\lambda / F^O_\lambda ]}$ is a linear function of the extinction 
curve, $A_\lambda/$\Av, and is equal to $A_\lambda/$\Av $\cdot$(\Av$^O-$\Av$^T$)$-2.5\cdot\log{C}$, where \Av$^O$ and \Av$^T$ 
are the visual extinction of the object and template, respectively, and $C$ is a constant. The slope of a linear fit 
to the  $\Gamma_\lambda$ versus ($A_\lambda/$\Av) relationship yields the difference between the extinction of the object and 
the template, assuming the same extinction law for both. In our case, \Av$^T$ is always very close to zero and therefore the slope 
of the linear fit yields \Av$^O$. 

Figure~\ref{lin_fits_Av} shows the plots of $\Gamma_\lambda$ versus $A_\lambda/$\Av\  for the YSOs in the sample. 
For every YSO, we used the $r_\lambda$ values of Table~\ref{veilings}, and the corresponding continuum $F^T_\lambda$ and 
$F^O_\lambda$ values reported in Tables~\ref{veil_anal_ry}--\ref{veil_anal_cq} in Appendix~\ref{Ext_Av_veil}. In these tables,
we also report the $r_\lambda$ values for convenience. The plots show the best linear fit and the corresponding 
slope \Av$^O$. The error on \Av\ indicated in each panel of Figure~\ref{lin_fits_Av} is only the error on the slope
of the fit and does not represent the full error on \Av, which we estimate to be also on the order of 0.35\,mag 
based mainly on flux calibration errors and mismatch in spectral type of the templates and the CTTs. 
The \Av\ results using this method are reported in column~5 of Table~\ref{rotfit_stel_prop}. 

In the case of RW\,Aur\,A, where the fit is not well constrained because only one rough estimate of veiling in the 
optical was possible, the error may be as high as 0.5\,mag. Nevertheless, the resulting extinction values as derived 
from both methods are consistent, within the errors.
We also stress that the UX\,Ori-type variability of CQ\,Tau may lead to large uncertainties on the \Av\ value,
as part of the extinction may be gray \citep[][]{dodin21}.

The \Av \  derived from the spectral templates alone and those including the veiling measurements 
(respectively columns~4 and 5 in Table~\ref{rotfit_stel_prop}) are in very good agreement within the error 
of about 0.35\,mag, although the veiling method provides systematically higher values by up to $\sim$30\%. 
As discussed in \citet[][]{fischer11}, the adoption of different values of the total-to-selective 
extinction, R$_{\rm V}$, has little impact on the results. We adopt the \Av\ values derived from the 
"veiling" method for the subsequent analysis.

The veiling methodologies used in the NIR by \citet[][]{fischer11} yielded \Av\  values systematically 
higher than those derived from other previous measurements. Yet, applying the same methods  
here, but extended to the optical veilings, provides \Av\ values that are significantly lower than those in 
\citet[][]{fischer11} for the same stars. \citet[][]{HH14} also calculate much lower \Av\ values than \citet[][]{fischer11} 
and interpret the higher values by the latter authors as possibly due to spectral mismatch between 
the templates and the CTTs and to the lower sensitivity of NIR spectra to extinction.
Our extinction values are slightly higher than those of \citet[][]{HH14}, but still consistent 
within the small amount of about 0.5\,mag, and in some cases the agreement is very good. 
This is consistent with the fact that the veiling method gives \Av\ values that are systematically higher 
by $\sim$30\% with respect to the templates method alone.

\subsection{Luminosity and mass}
\label{luminosity}

\begin{figure}[!ht]



{\includegraphics[trim= 30 30 110 20,width=0.9\columnwidth]{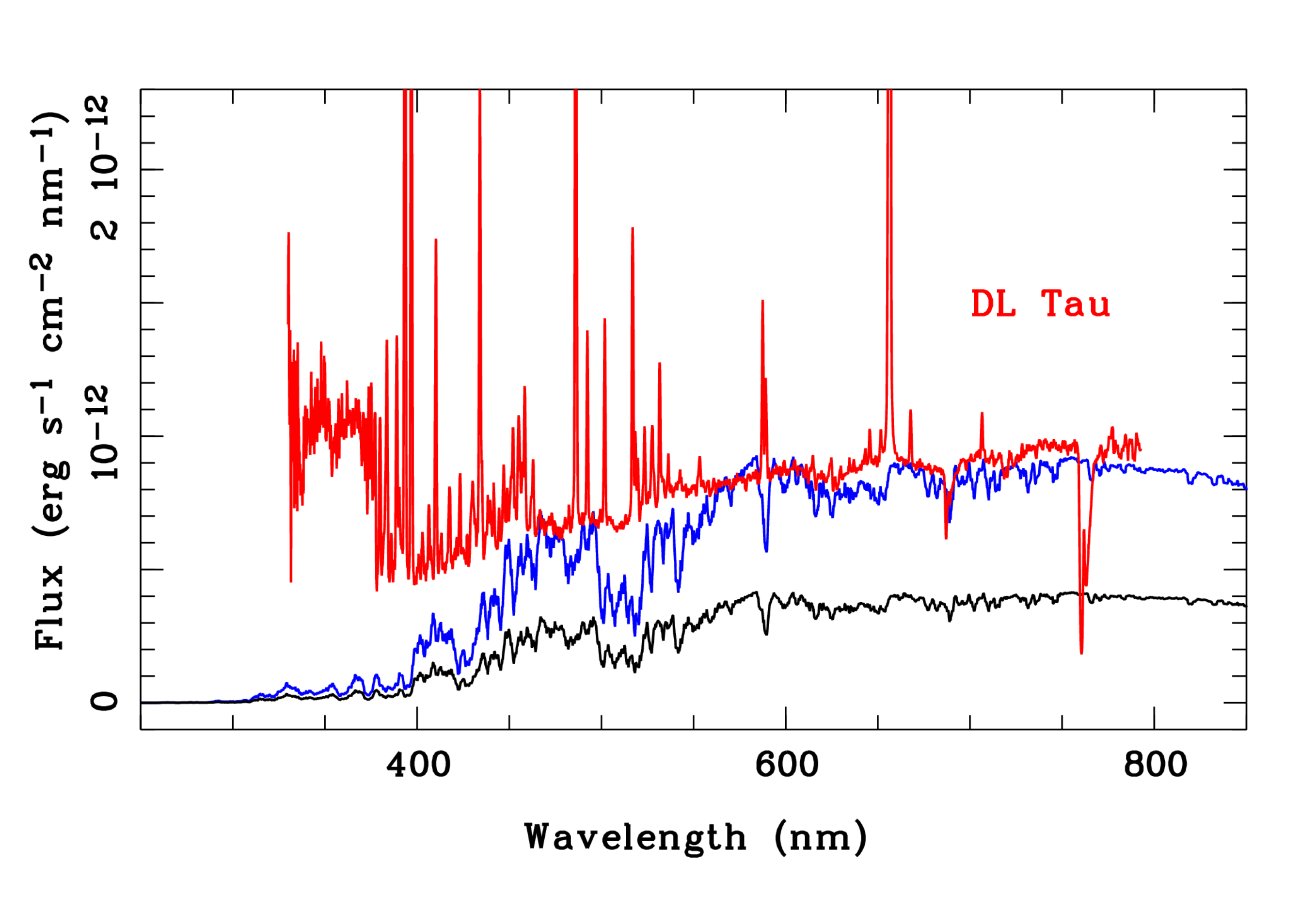}}
\caption{Example of BT-settl model normalization. The extinction-corrected Asiago spectrum of DL\,Tau is shown with
       the red line, while the BT-settl model normalized to DL\,Tau at $\lambda =$600\,nm is plotted with the blue 
       line. We note the increase of excess emission of DL\,Tau at wavelengths shorter than about 450\,nm. The veiling-corrected BT-settl model is plotted with the black line. See text for details
   \label{synth_model_norma}}
\end{figure}

The stellar luminosity, \Lstar, was derived as follows. For a given object, we extracted first the BTsettl model \citep[][]{allard12} 
best matching the \Teff\  values derived with ROTFIT above (see Sect.~\ref{rotfit}). The model was then normalized
to the extinction corrected flux, $F^{corr}_{600}$, of the Asiago spectrum at $\lambda=$600\,nm, where 
the veiling estimates are expected to be more reliable, and corrections subject to less uncertainty. 
To take into account the contribution of veiling, the extinction-corrected flux at 600\,nm was multiplied by the 
factor $1 / (1 + r600)$, where $r600$ is the veiling at $\lambda=$600\,nm derived from ROTFIT and listed in 
Table~\ref{veilings} (see example for DL\,Tau in Figure~\ref{synth_model_norma}). This factor, which is always
$\le$1, effectively reduces the observed flux to that emitted by the stellar photosphere only, excluding the 
emission due to accretion.

We therefore assume that the BT-settl model normalized in this way best represents the spectral energy distribution (SED) 
of the object's photosphere at the star distance. Integration of the normalized model at all wavelengths yields the bolometric flux corresponding to the object's photosphere. The stellar luminosity was then calculated using this 
flux and adopting the distance reported in Table~\ref{ctts_prop}.
The main sources of uncertainty on \Lstar\ are the errors in flux calibration of the Asiago spectra and the
error on the veiling correction. On this basis, we estimate an average uncertainty of about 0.2\,dex in $\log$\Lstar.
 We verified that using veiling values at other wavelengths yields consistent results of the veiling-corrected bolometric flux. For instance, in the case of DL\,Tau, which has the strongest 
veiling at 450\,nm, we derive a corrected bolometric flux of 5$\times$10$^{-10}$\,erg\,s$^{-1}$\,cm$^{-2}$ and
4.55$\times$10$^{-10}$\,erg\,s$^{-1}$\,cm$^{-2}$ when using the $r600$ and $r450$ values, respectively.
The stellar radius, \Rstar, was calculated from the effective temperature and stellar luminosity.

In some cases (e.g., DG\,Tau) the \Lstar\ values from the oldest literature may be overestimated most likely
because such determinations did not consider the contribution of veiling and/or overestimate extinction. 
Our stellar luminosities are well consistent with those derived in \citet[][]{HH14} within the 
error of about 0.2\,dex in $\log$\Lstar. It is worth noting that in the case of RY\,Tau our \Teff, \Av, 
and \Lstar\ values are in good agreement, within the errors, with those reported in the recent and thorough 
study by \citet[][]{garufi19}. These values are also fairly consistent with those reported in the
paper by \citet[][]{calvet04}. 


We note that HN\,Tau\,A is subluminous on the HR diagram with both the  value of \citet[][]{HH14} for \Lstar\ and the one 
derived here.
This is likely due to obscuration of the stellar photosphere by the highly inclined disk \citep[$i\approx70^\circ$;][]{long19}.
The \Lstar, \Rstar, and \Mstar\ values for this star may be underestimated by a factor of $\sim$17, $\sim$4.3, and $\sim$2, 
respectively (see Sect.~\ref{HN_Lacc}).
Likewise, our derived luminosity for CQ\,Tau places the star slightly below the main sequence, most likely as a consequence 
of the fact that during the periods of dimming, the extinction may be gray, and therefore the luminosity is underestimated.
The GHOsT observations of this star were performed when it was in a faint stage, that is, $B=$10.89\,mag as compared with 
$B=$9\,mag in its bright phase  \citep[][]{grinin08}, and therefore its luminosity may be underestimated by a factor of about six 
(see also Sect.~\ref{HN_Lacc}).

Finally, the mass, \Mstar, of the seven CTTs was estimated by comparison of the position of the objects on the HR diagram with the 
theoretical PMS evolutionary tracks by \citet[][]{siess00}. The uncertainties on \Lstar\ and
\Teff\ lead to a typical uncertainty of $\sim$0.15\,dex in \Mstar. The derived \Lstar, \Rstar,\ and \Mstar\ values 
for the sample are given in Table~\ref{rotfit_stel_prop}.
The mass reported in this table for the subluminous objects HN\,Tau\,A and CQ\,Tau corresponds 
to that of the evolutionary track closest to  these stars on the HR diagram, 
but corrected values are estimated in Section~\ref{HN_Lacc}.


\section{Accretion diagnostics}
\label{acc_diag}
The GIARPS spectra include several emission lines that are 
well correlated with \Lacc\ \citep[][]{alcala14,alcala17} and will be the basis of our measurements of accretion 
in the sample. In particular, we extracted 17 spectral portions with well-resolved and flux-calibrated accretion 
diagnostics, namely, eight hydrogen recombination lines (\Ha, \Hb, \Hg, \Hd, \pab, \pag, \pad, Pa8), eight helium lines 
(He\,{\sc i}4026, He\,{\sc i}4471, He\,{\sc i}4713, He\,{\sc i}4922, He\,{\sc i}5016, He\,{\sc i}5876, He\,{\sc i}6678, 
He\,{\sc i}10830), and the Ca\,{\sc ii}3934 line.

\begin{figure*}[!ht]


\resizebox{1\hsize}{!}{
{\includegraphics[trim= 500 340 90 20,width=0.36\columnwidth, angle=180]{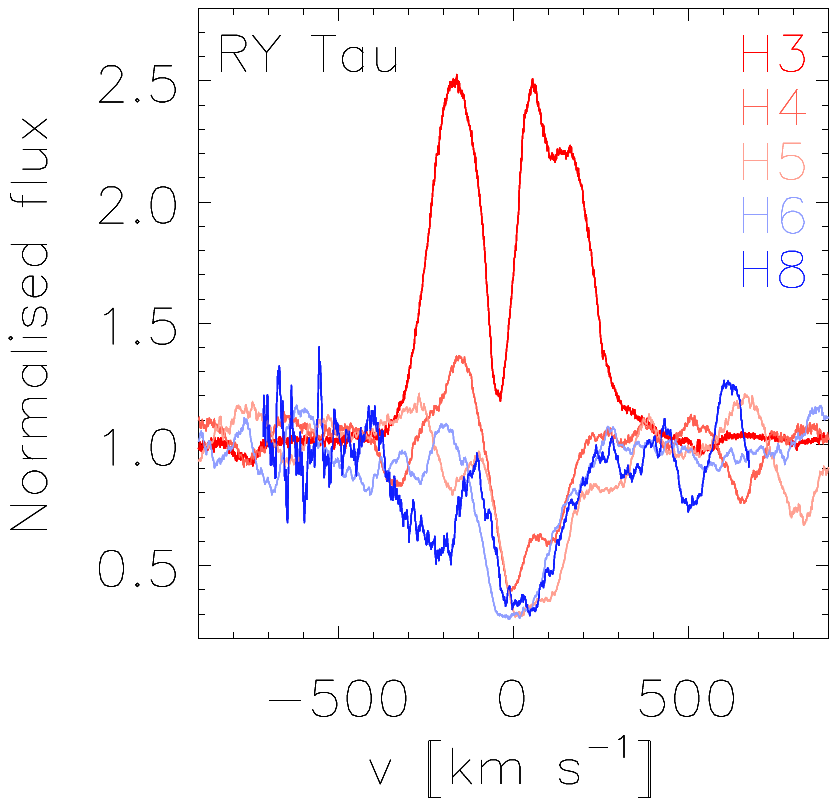}}\qquad
{\includegraphics[trim= 500 340 90 20,width=0.36\columnwidth, angle=180]{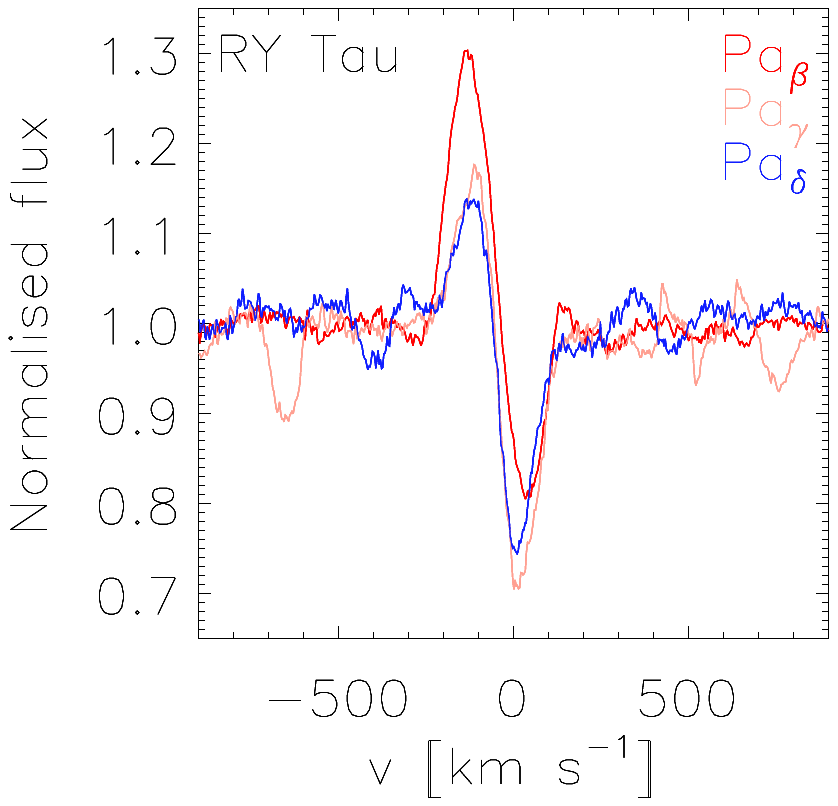}}\qquad
{\includegraphics[trim= 500 340 90 20,width=0.36\columnwidth, angle=180]{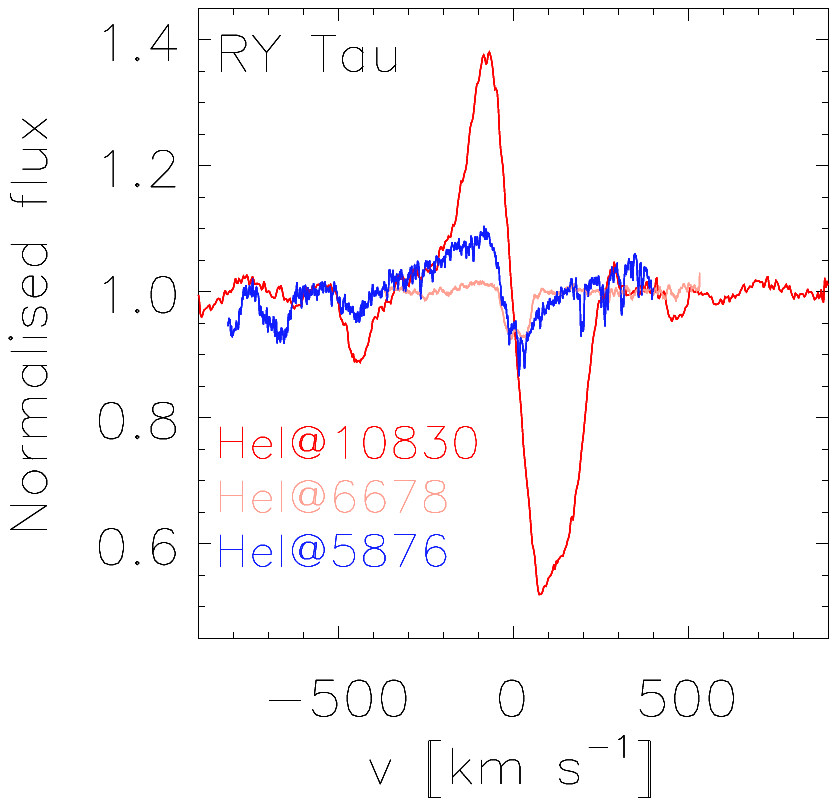}}\qquad

{\includegraphics[trim= 500 340 90 20,width=0.36\columnwidth, angle=180]{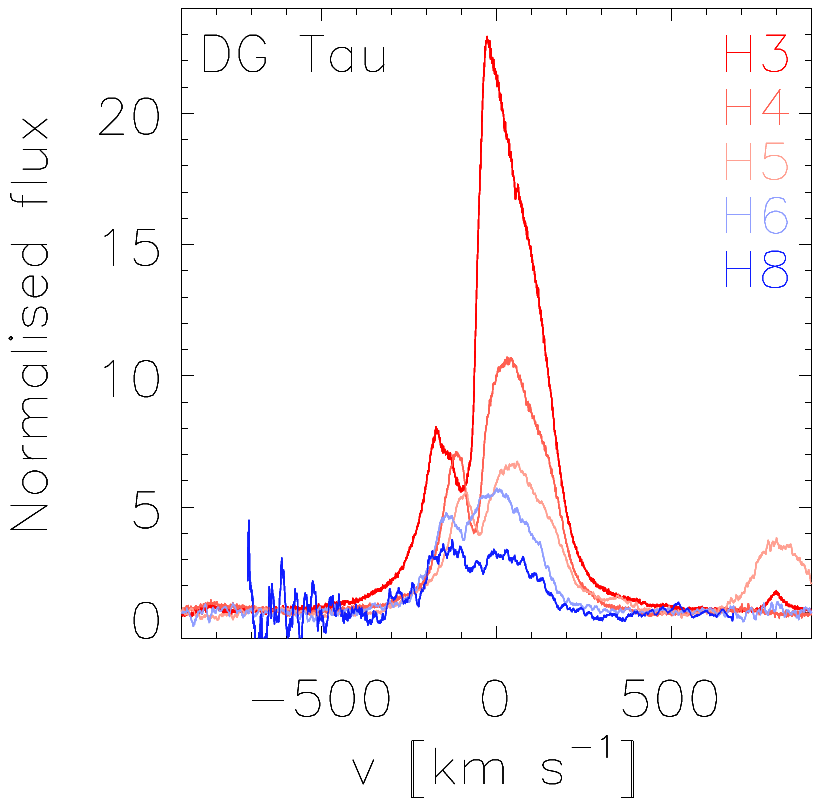}}\qquad
{\includegraphics[trim= 500 340 90 20,width=0.36\columnwidth, angle=180]{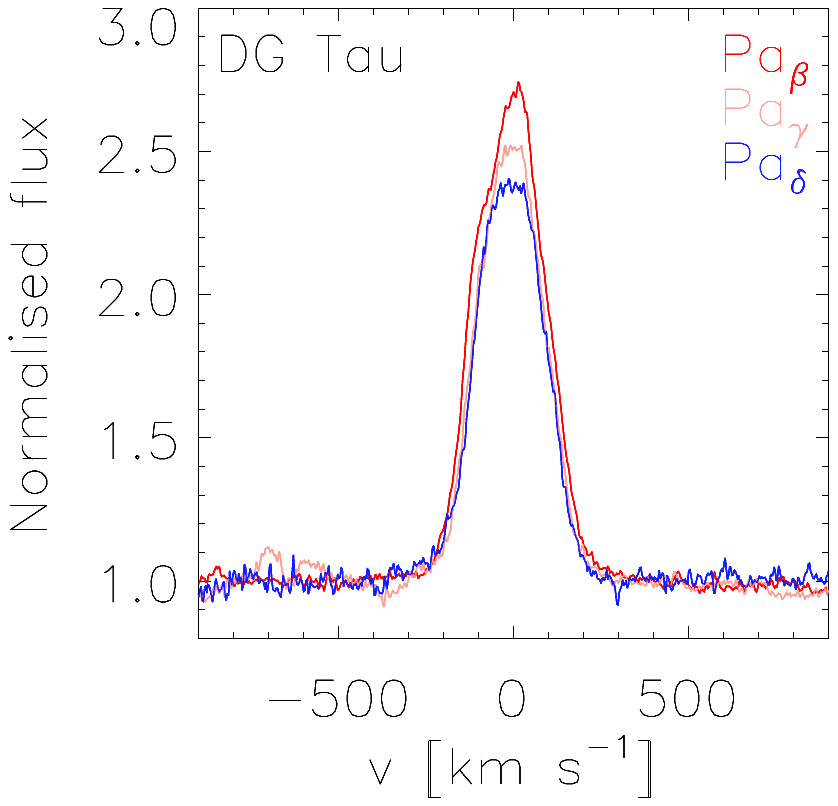}}\qquad
{\includegraphics[trim= 500 340 90 20,width=0.36\columnwidth, angle=180]{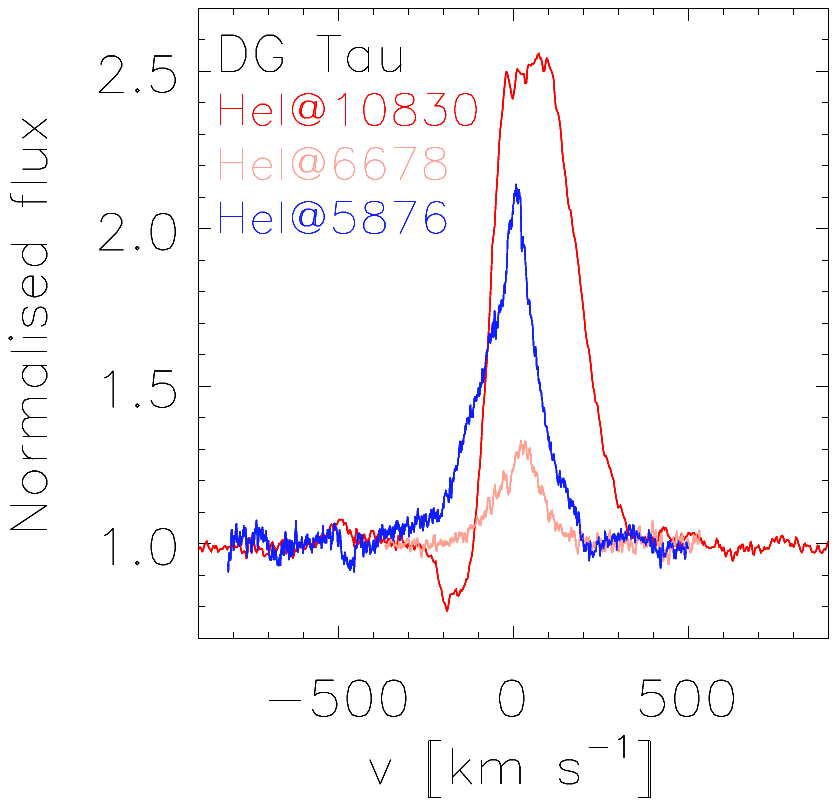}}\qquad
}

\resizebox{1\hsize}{!}{
{\includegraphics[trim= 500 340 90 20,width=0.36\columnwidth, angle=180]{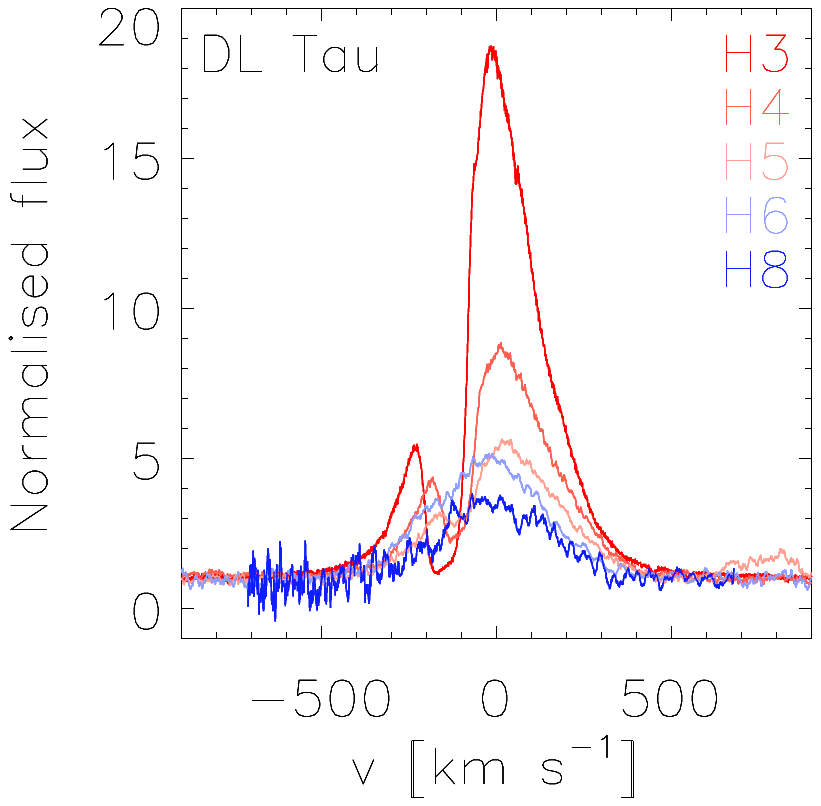}}\qquad
{\includegraphics[trim= 500 340 90 20,width=0.36\columnwidth, angle=180]{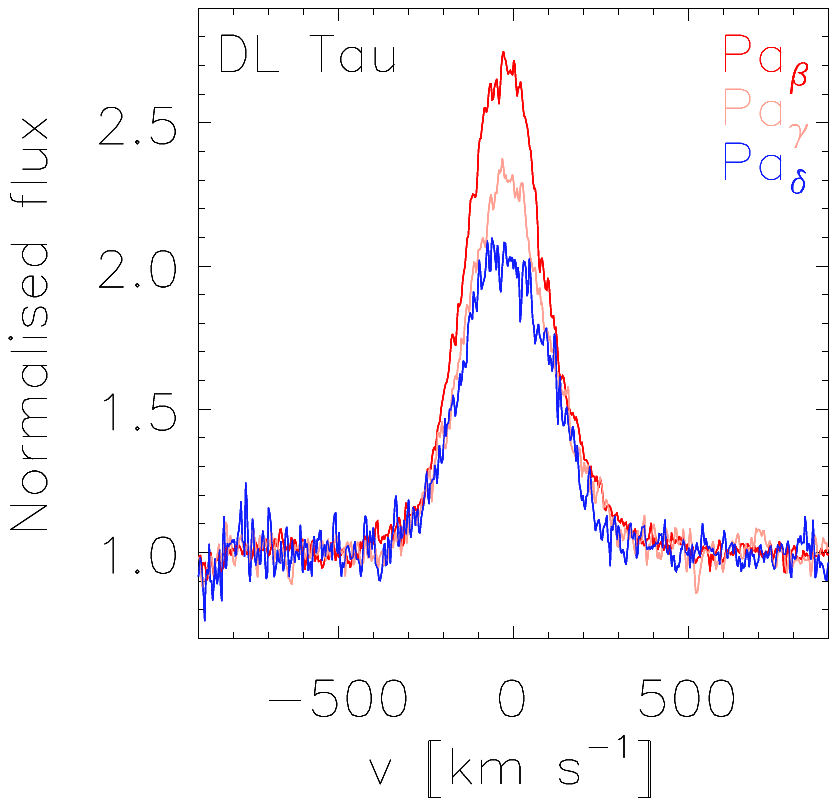}}\qquad
{\includegraphics[trim= 500 340 90 20,width=0.36\columnwidth, angle=180]{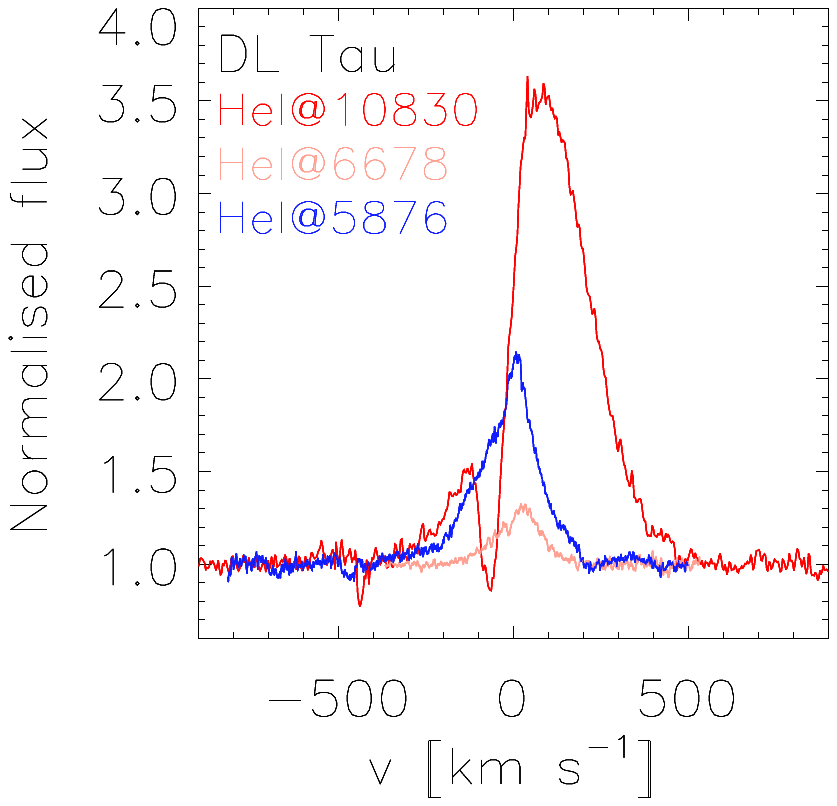}}\qquad

{\includegraphics[trim= 500 340 90 20,width=0.36\columnwidth, angle=180]{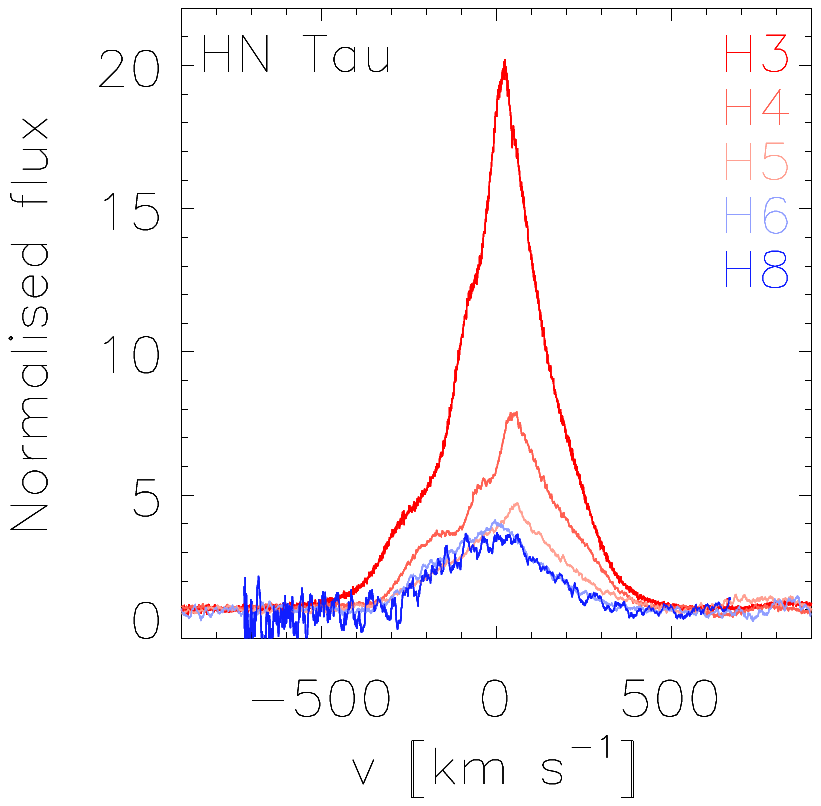}}\qquad
{\includegraphics[trim= 500 340 90 20,width=0.36\columnwidth, angle=180]{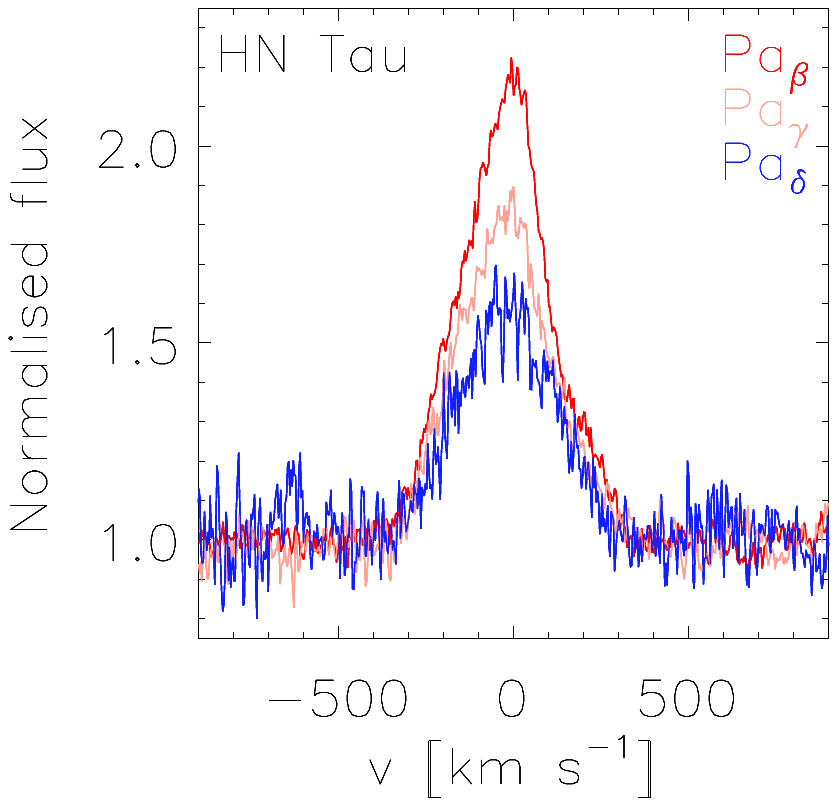}}\qquad
{\includegraphics[trim= 500 340 90 20,width=0.36\columnwidth, angle=180]{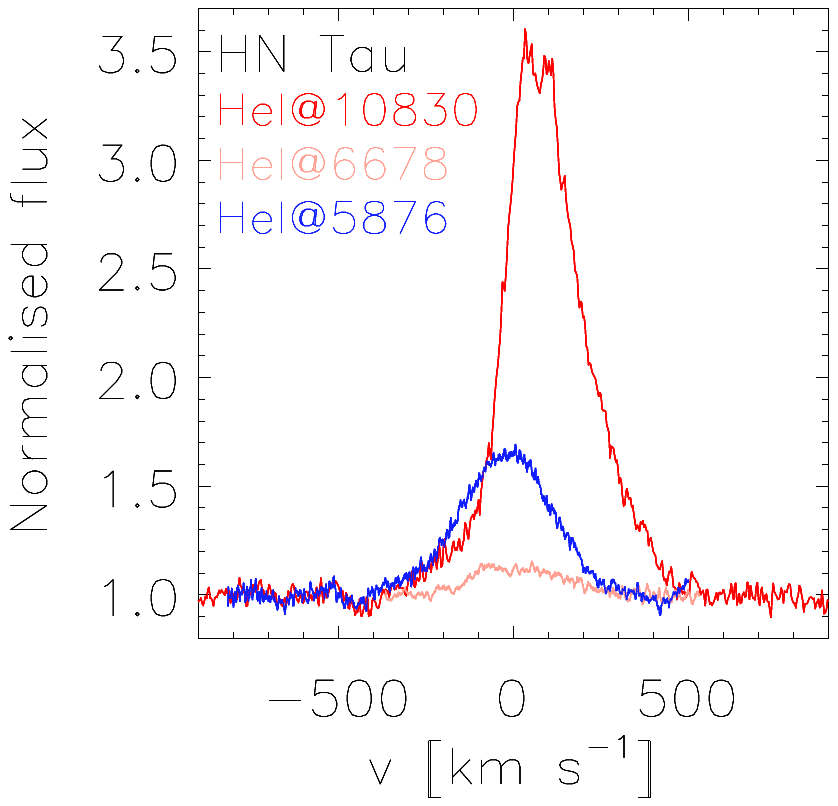}}\qquad
}

\resizebox{1\hsize}{!}{
{\includegraphics[trim= 500 340 90 20,width=0.36\columnwidth, angle=180]{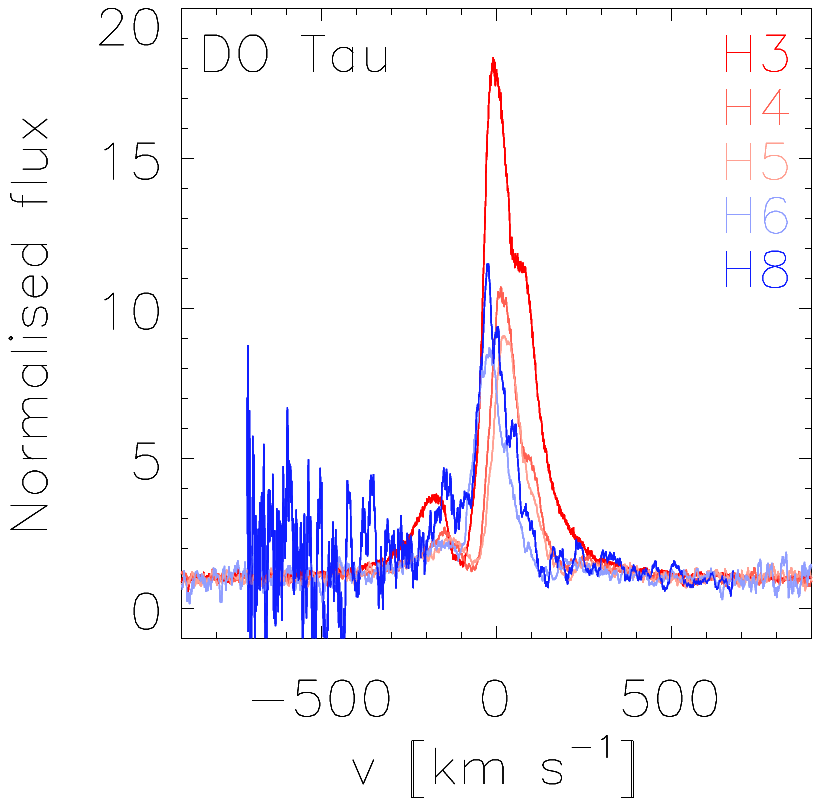}}\qquad
{\includegraphics[trim= 500 340 90 20,width=0.36\columnwidth, angle=180]{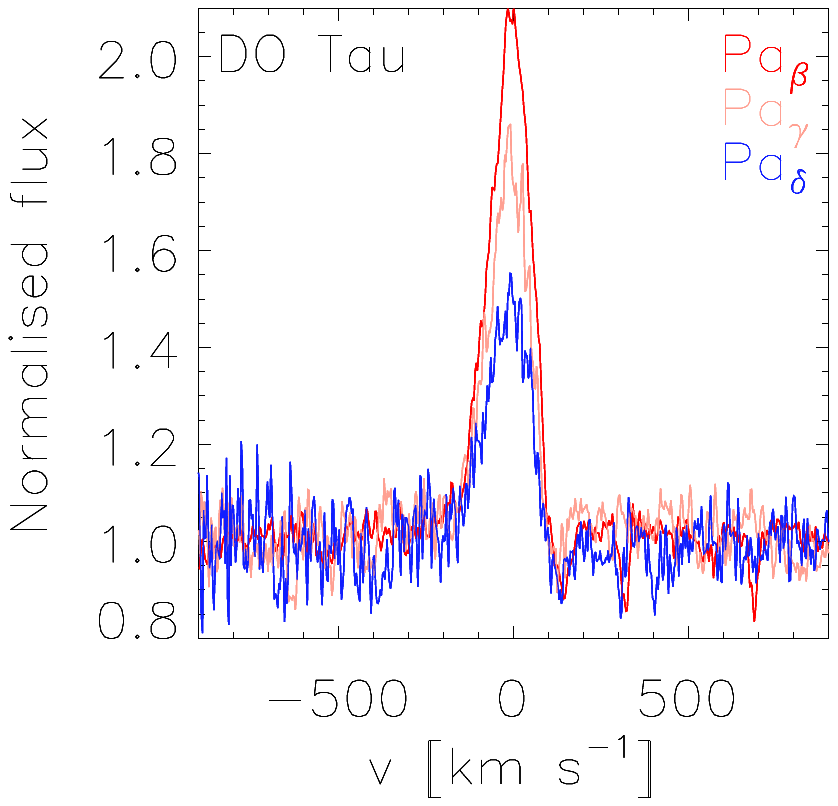}}\qquad
{\includegraphics[trim= 500 340 90 20,width=0.36\columnwidth, angle=180]{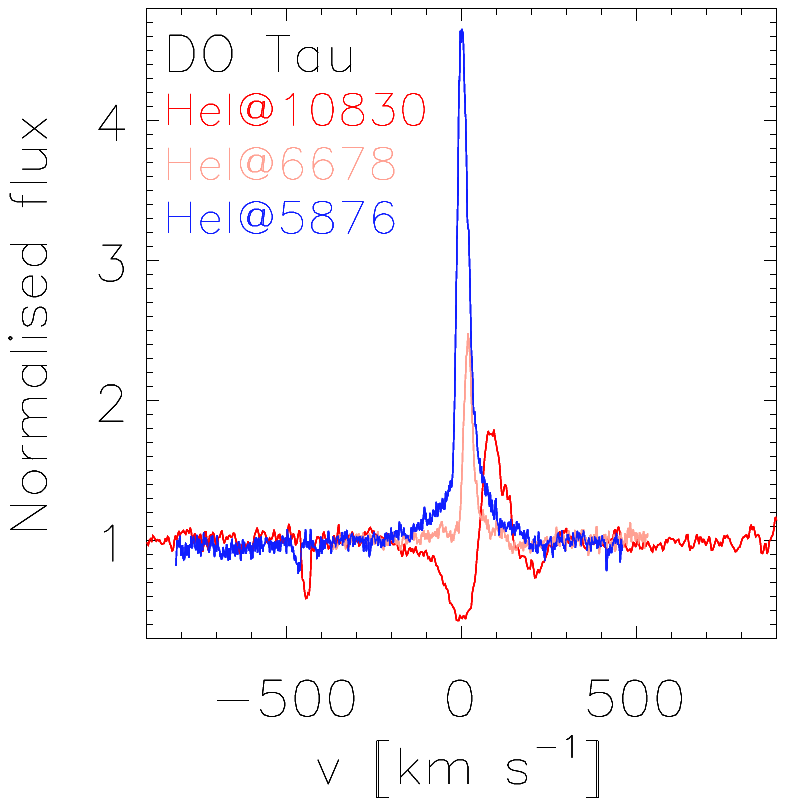}}\qquad

{\includegraphics[trim= 500 340 90 20,width=0.36\columnwidth, angle=180]{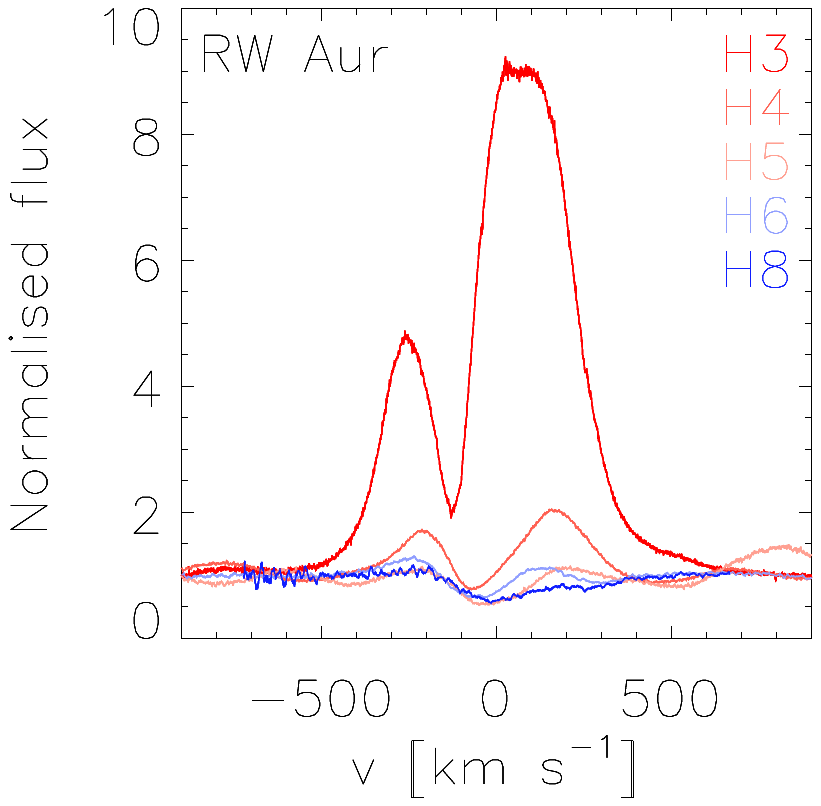}}\qquad
{\includegraphics[trim= 500 340 90 20,width=0.36\columnwidth, angle=180]{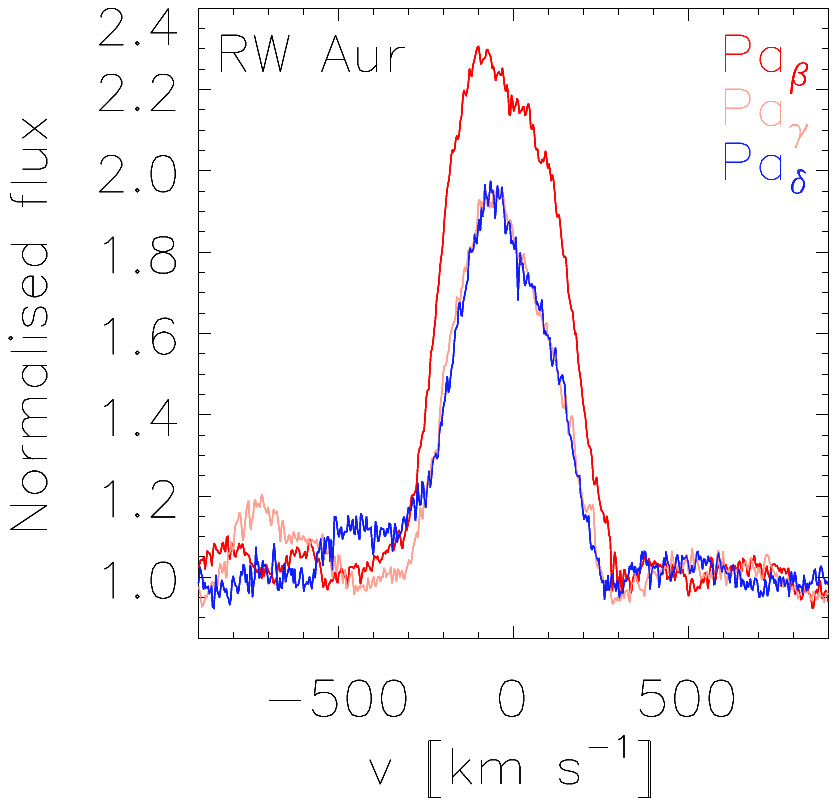}}\qquad
{\includegraphics[trim= 500 340 90 20,width=0.36\columnwidth, angle=180]{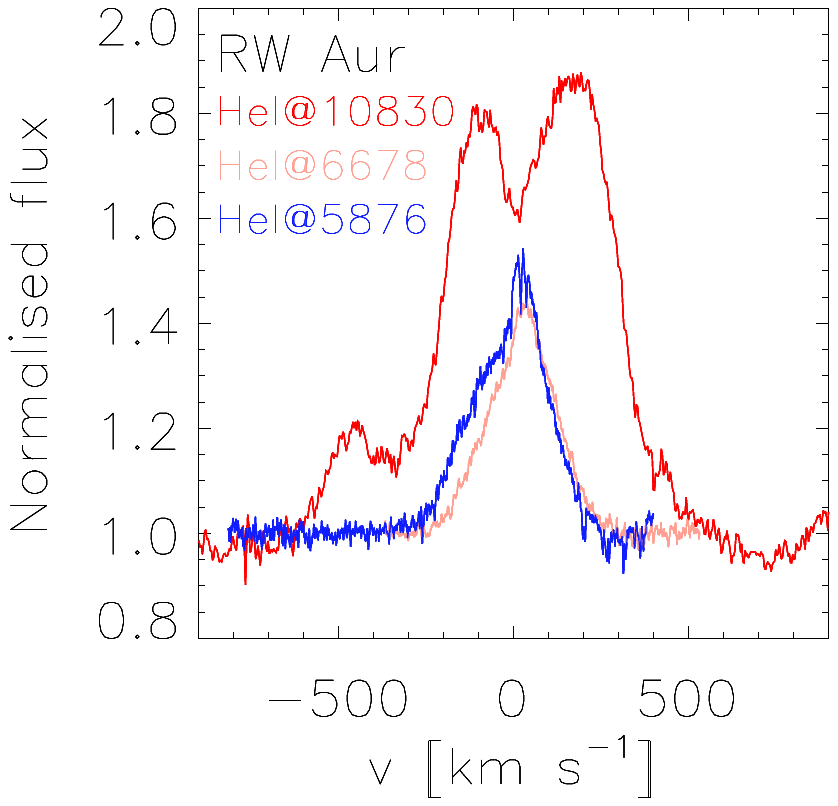}}\qquad

}

\resizebox{0.5\hsize}{!}{
{\includegraphics[trim= 500 340 90 20,width=0.36\columnwidth, angle=180]{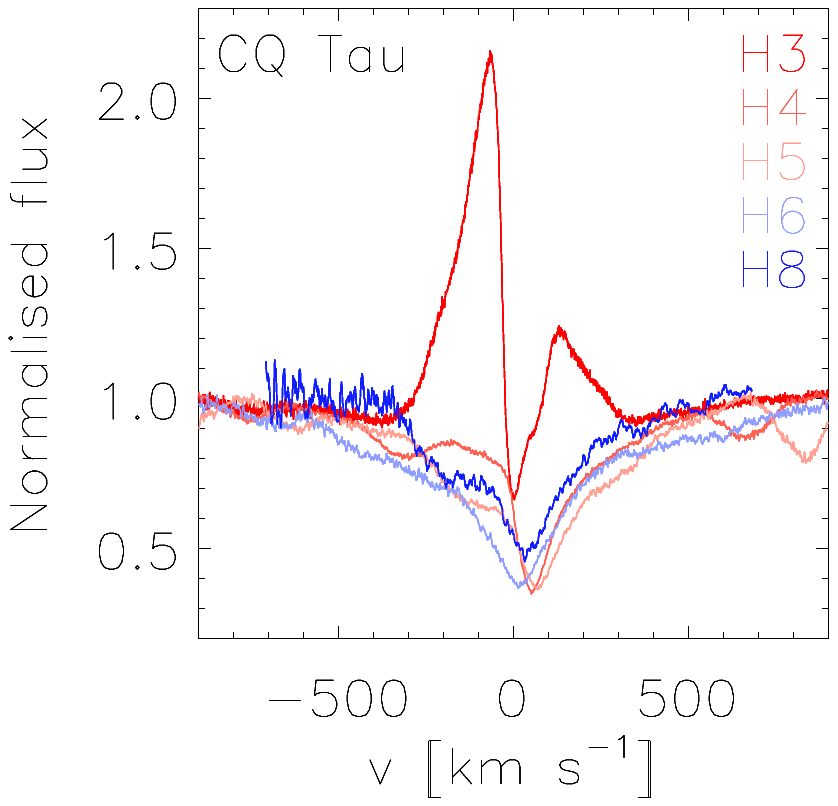}}\qquad
{\includegraphics[trim= 500 340 90 20,width=0.36\columnwidth, angle=180]{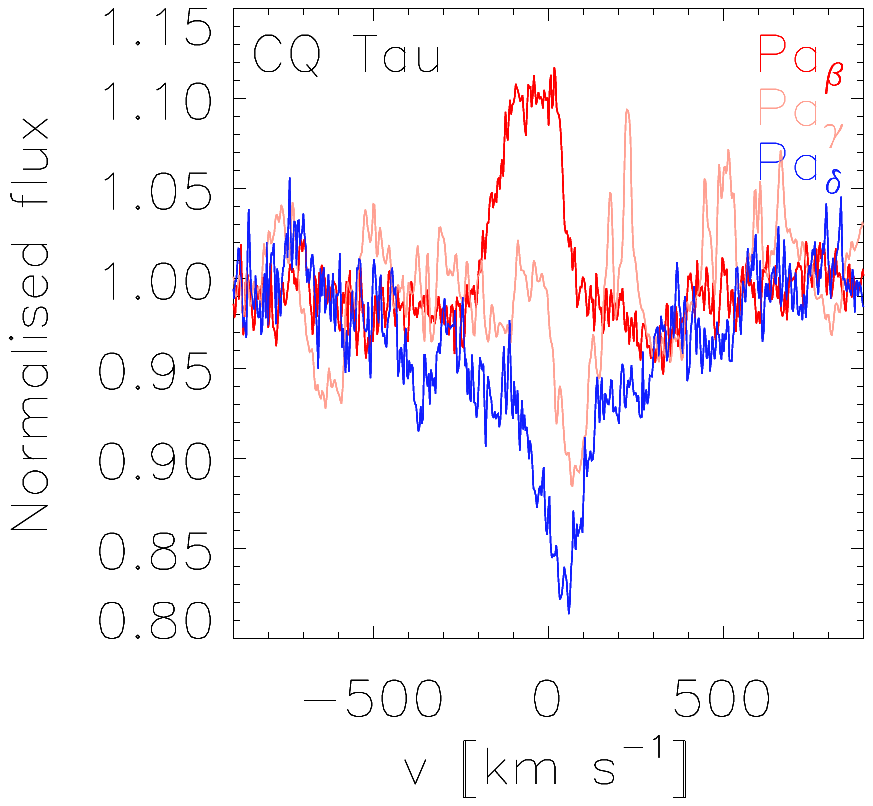}}\qquad
{\includegraphics[trim= 500 340 90 20,width=0.36\columnwidth, angle=180]{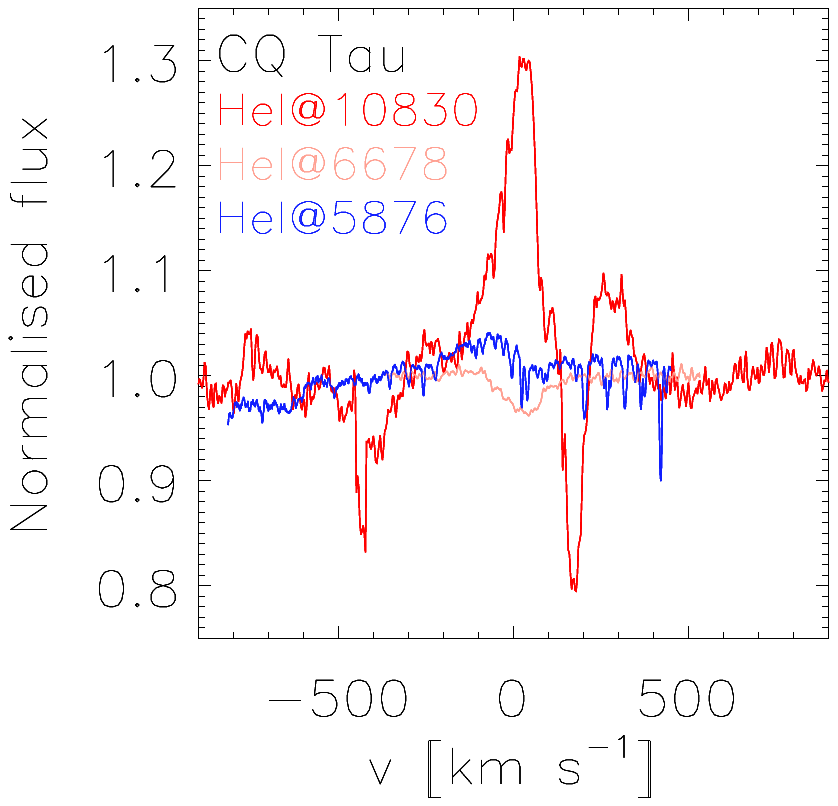}}\qquad
}

\caption{Examples of GIARPS spectra showing the line profiles of CTTs in our sample
   \label{profiles1}}
\end{figure*}






Figure~\ref{profiles1} shows examples of GIARPS spectra for several emission lines of 
every CTT star in our sample.
For comparison, the spectra are normalized to the local continuum and shifted in velocity to 
the rest wavelength. The latter was determined from the profiles of the  \ion{Li}{I} $\lambda6708$\,\AA\
 photospheric line, assuming weighted $\rm \lambda = 670.7876 nm$, and in the NIR from the \ion{Al}{I} lines 
at $\lambda = 2019.884, 2116.958, 2121.396$\,nm. All the spectra are shown in the same velocity range
for comparison.

There is a variety of line profiles with a range of widths and intensities, typical of accreting YSOs.
The widest lines are from RY\,Tau, RW\,Aur\,A, and HN\,Tau\,A, while DG\,Tau and DL\,Tau show lines with 
intermediate width. The narrowest lines are from DO\,Tau, yet the width of the H$\alpha$ line at 10\% of the line 
peak that we measure for the latter is more than 400\,km/s, pointing to significant accretion activity
\citep[][]{white03,natta06}.
Within the qualitative classification scheme of the Balmer lines proposed by \citet[][]{antoniucci17a}, 
RY\,Tau, RW\,Aur\,A, and HN\,Tau\,A display wide and multi-peak profiles, DG\,Tau and DL\,Tau show
multi-peak profiles, while DO\,Tau shows narrow and almost symmetric profiles. All these morphologies are
displayed by the Lupus sample studied by \citet[][]{antoniucci17a}.
Two exceptions in our sample are RY\,Tau and CQ\,Tau, where many lines are dominated by the photospheric 
absorption  and therefore the subtraction of the photospheric spectrum is needed in order to measure the residual 
emission (see Sect.~\ref{line_fluxes}).

\subsection{Flux and luminosity of lines}
\label{line_fluxes} 
The flux at Earth in permitted lines was computed by directly integrating the GIARPS flux-calibrated 
spectra using the {\em splot} package under  IRAF.
Three independent measurements per line were made, considering the lowest, highest, and the middle position of 
the local continuum, depending on the local noise level of the spectra. The flux and its error were then 
computed as the average and standard deviation of the three independent measurements, respectively.

\begin{figure}[!htbp]

\resizebox{1.05\hsize}{!}{
{\includegraphics[height=2.4cm, width=1.2cm]{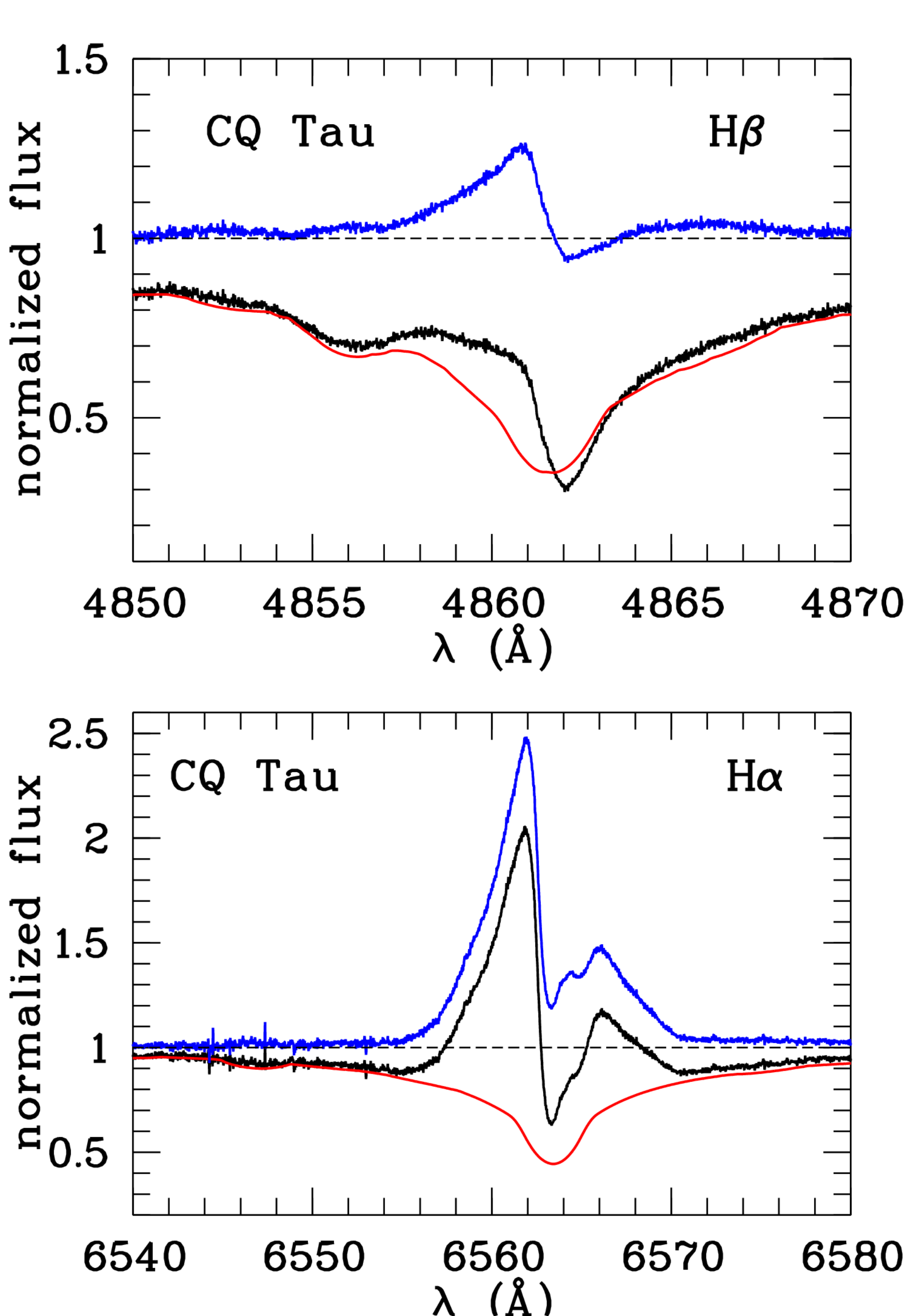}}
{\includegraphics[height=2.4cm, width=1.2cm]{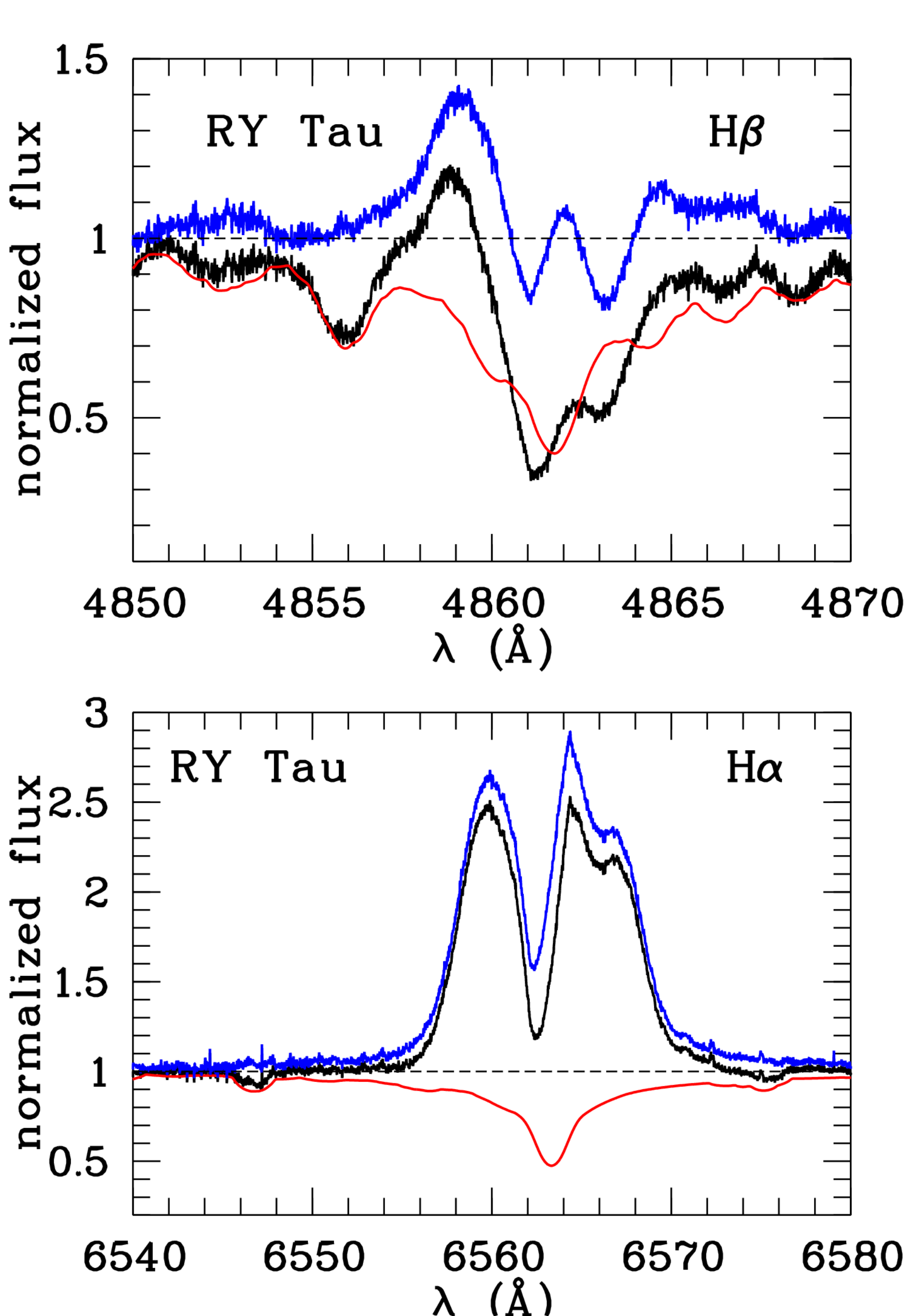}}
}
\caption{Examples of photospheric line subtraction for the \Ha\ and \Hb\ lines in CQ\,Tau
       and RY\,Tau. The black, red, and blue lines represent the observed, photospheric, 
       and photospheric-subtracted spectra, respectively. See text for details.
   \label{phot_subtract_Ha_Hb}}
\end{figure}

\begin{figure}[!ht]
\resizebox{1\hsize}{!}{{\includegraphics[height=0.6cm, width=1.1cm]{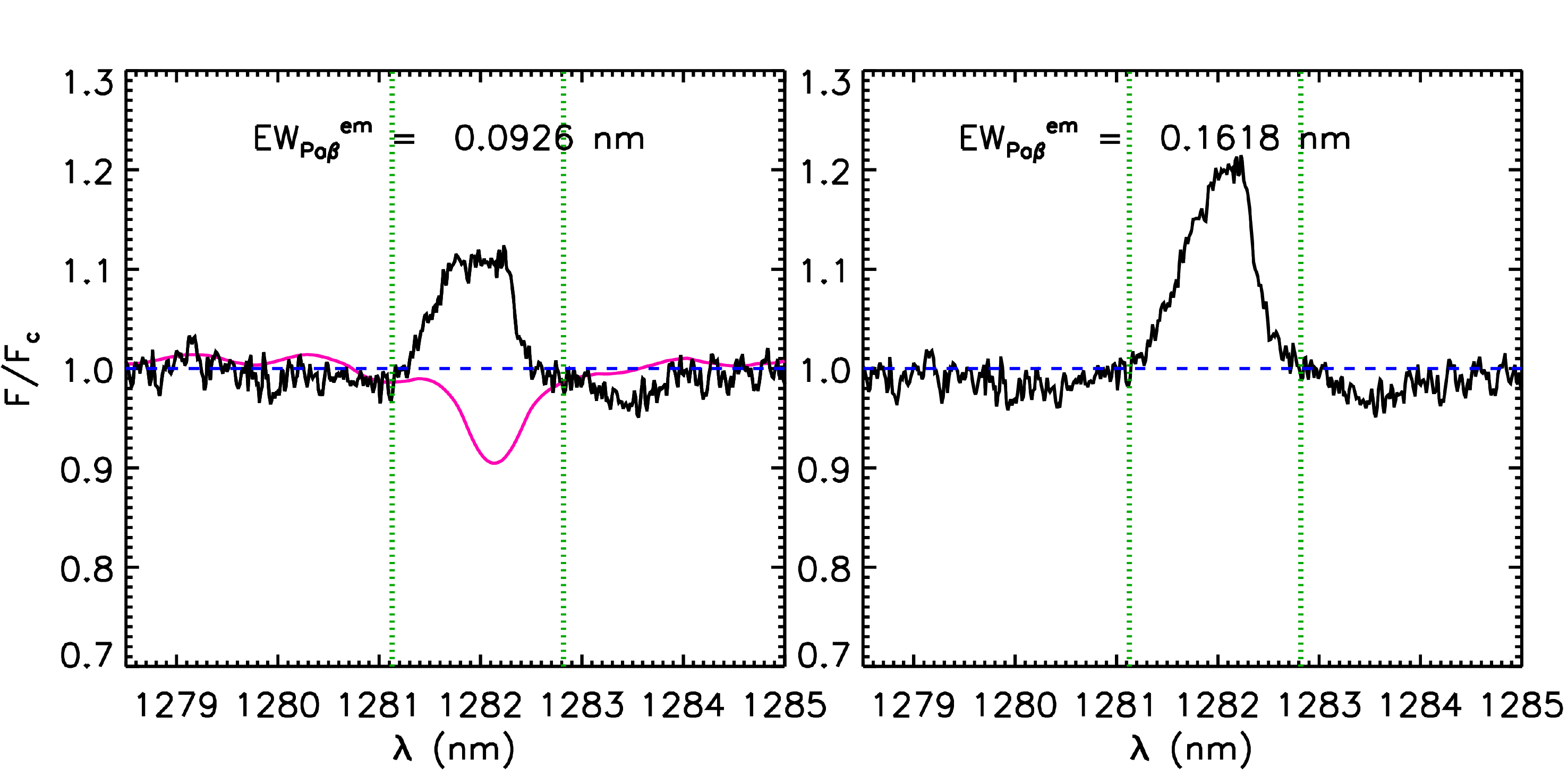}}}
\resizebox{1\hsize}{!}{{\includegraphics[height=0.6cm, width=1.1cm]{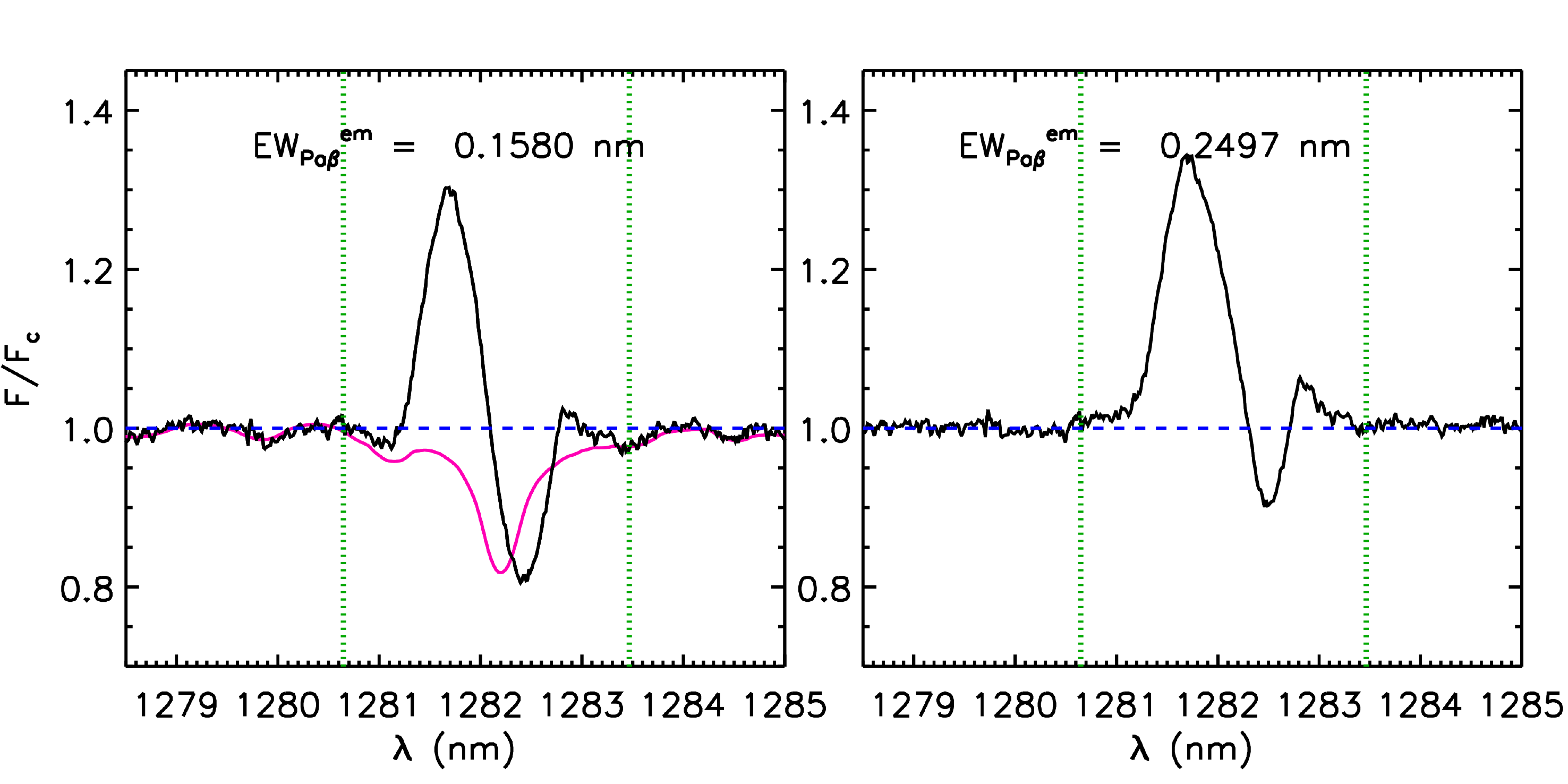}}}
\caption{Examples of photospheric line subtraction for the \pab\ line for CQ\,Tau (upper panels) 
        and RY\,Tau (lower panels). The red lines represent the photospheric templates.
        The left and right panels show, in black lines, the observed and 
        photospheric-subtracted spectra, respectively. 
        The green vertical lines mark
        the intervals for the equivalent width measurement. The equivalent widths
        of the emission line before and after the subtraction are also indicated.
   \label{phot_subtract_Pa_He}}
\end{figure}

In the case of RY\,Tau and CQ\,Tau, we performed the photospheric subtraction in the following way: 
the spectral templates used to derive the stellar parameters and NIR veiling were artificially
broadened for \vsini\ and veiled at the same values derived from ROTFIT and the NIR veiling procedure. 
The rotated and veiled templates were then subtracted from the YSO spectra. In this procedure,
both the YSO and template spectra were normalized to the continuum around the selected lines
to be analyzed. Three measurements of the equivalent widths of the corrected lines were made
and the average was computed. The error was estimated by 
the standard deviation of the observed fluxes on the difference spectra in two spectral 
regions near the line.
Finally, the flux of the corrected lines was computed as the product of
the equivalent width times the absolute continuum flux adjacent to the lines.
The procedure, illustrated  in Figure~\ref{phot_subtract_Ha_Hb} for the \Ha\ and \Hb\ lines and
in Figure~\ref{phot_subtract_Pa_He} for the \pab\ line, was applied to lines strongly affected by 
the photospheric contribution namely the Balmer, Paschen, and the \ion{He}{I} $\lambda10830$\,\AA\ lines.
We note that in some cases a residual contribution of nonphotospheric self-absorptions 
remained, in particular in the case of some Paschen and the \ion{He}{I} $\lambda10830$\,\AA\ 
lines (see example in Figure~\ref{phot_subtract_Pa_He}).  Lines with strong 
self-absorption may underestimate \Lacc\ and therefore the measurement was not considered 
when such a contribution was more than $\sim$30\% of the flux.  
The procedure adopted here to measure the integrated flux is the same adopted in 
\citet[][]{alcala17}, where the \Ll\ versus \Lacc\ calibrations were derived.


\begin{figure*}[!ht]

 \resizebox{1\hsize}{!}{
 {\includegraphics[bb=0 0 750 550]{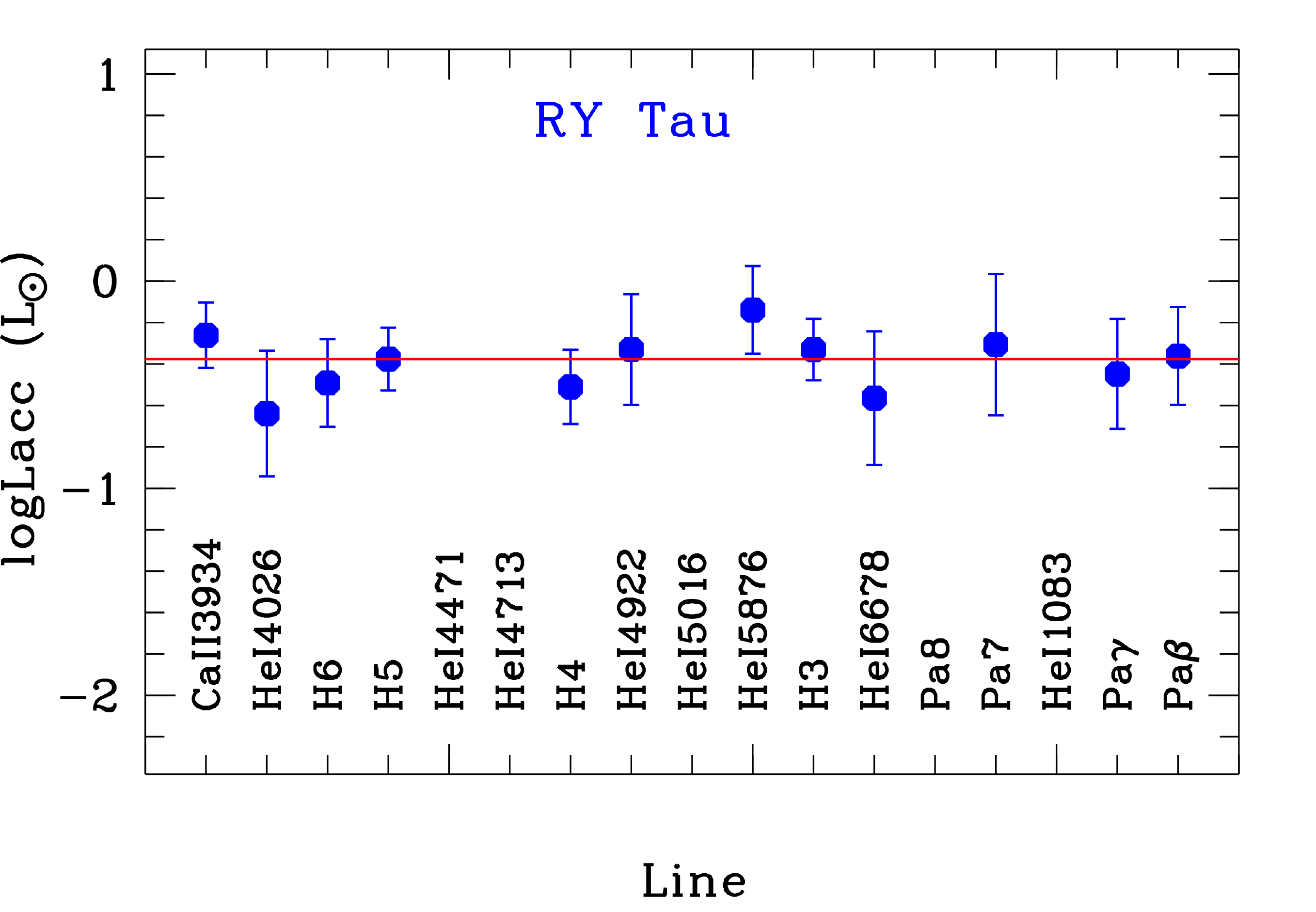}}
 {\includegraphics[bb=0 0 750 550]{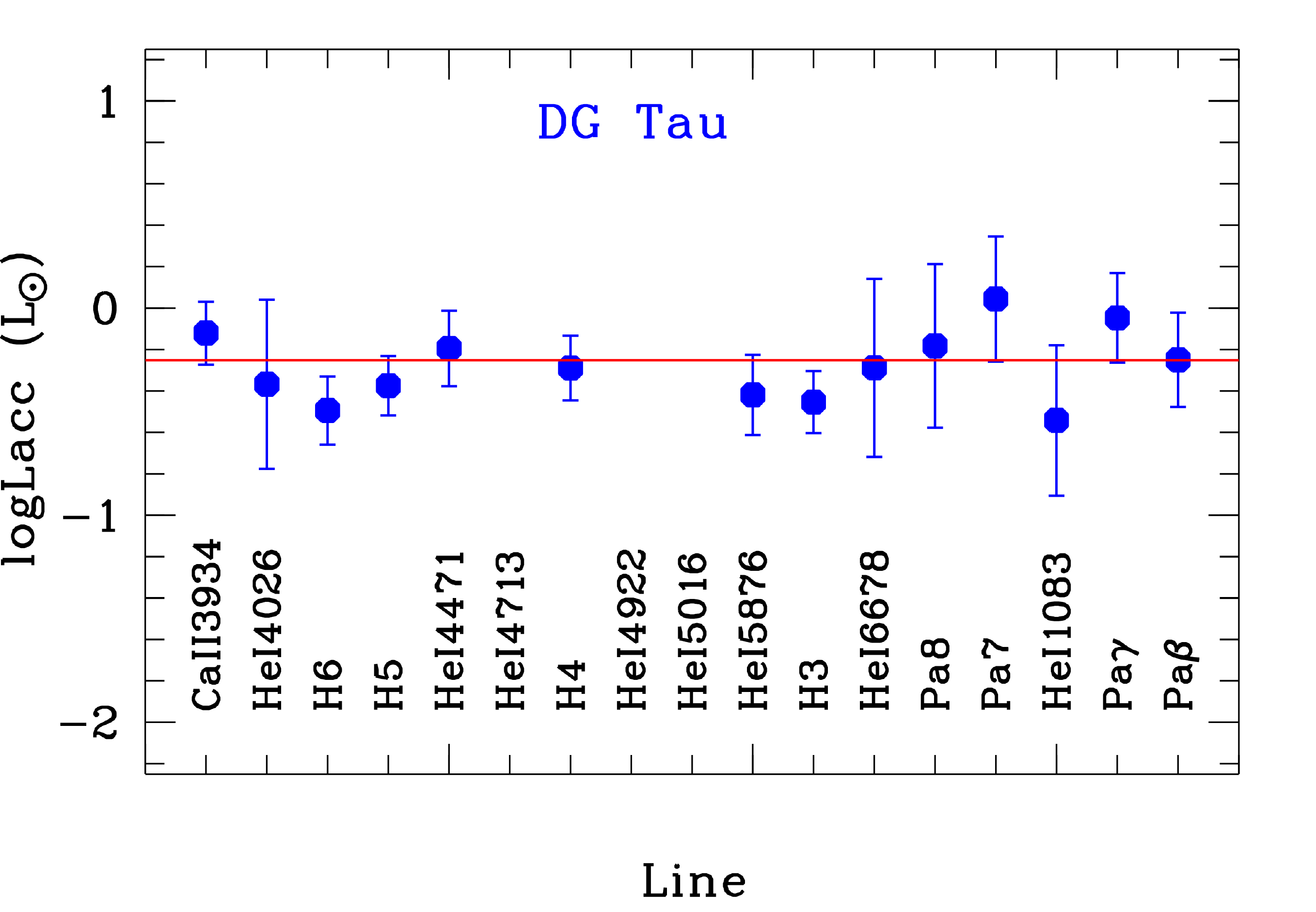}}
 {\includegraphics[bb=0 0 750 550]{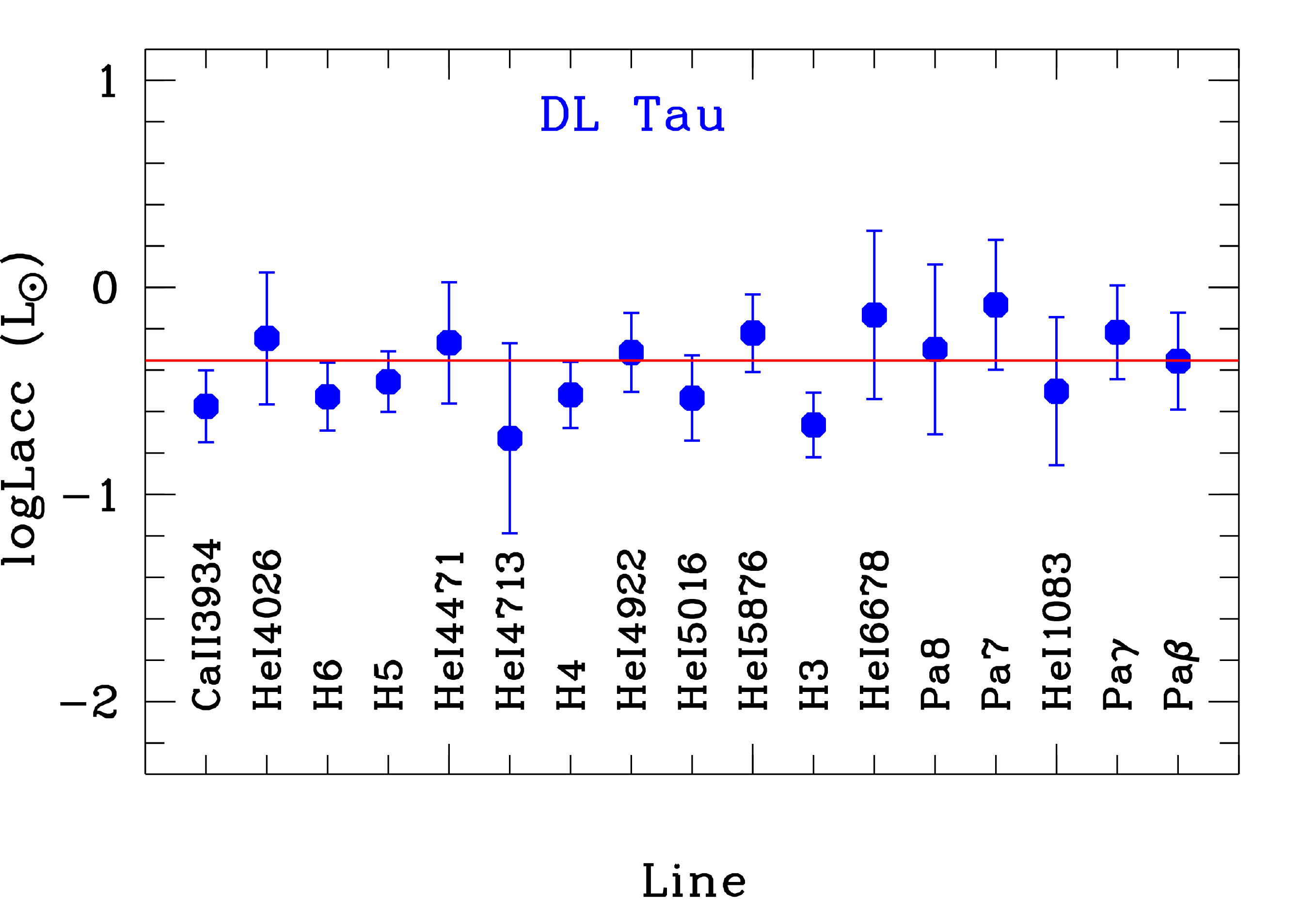}}
 }
 \resizebox{1\hsize}{!}{
 {\includegraphics[bb=0 0 750 550]{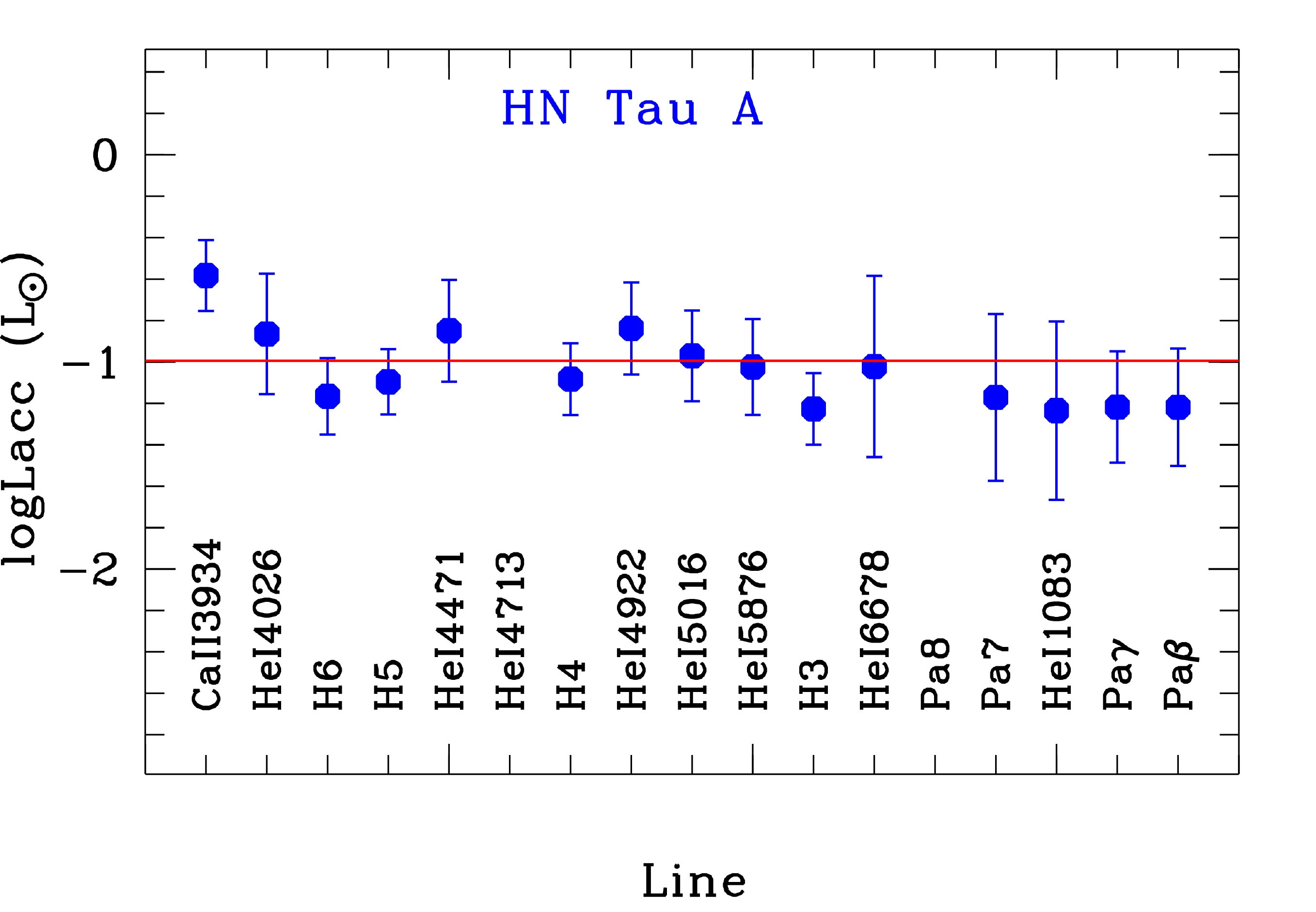}}
 {\includegraphics[bb=0 0 750 550]{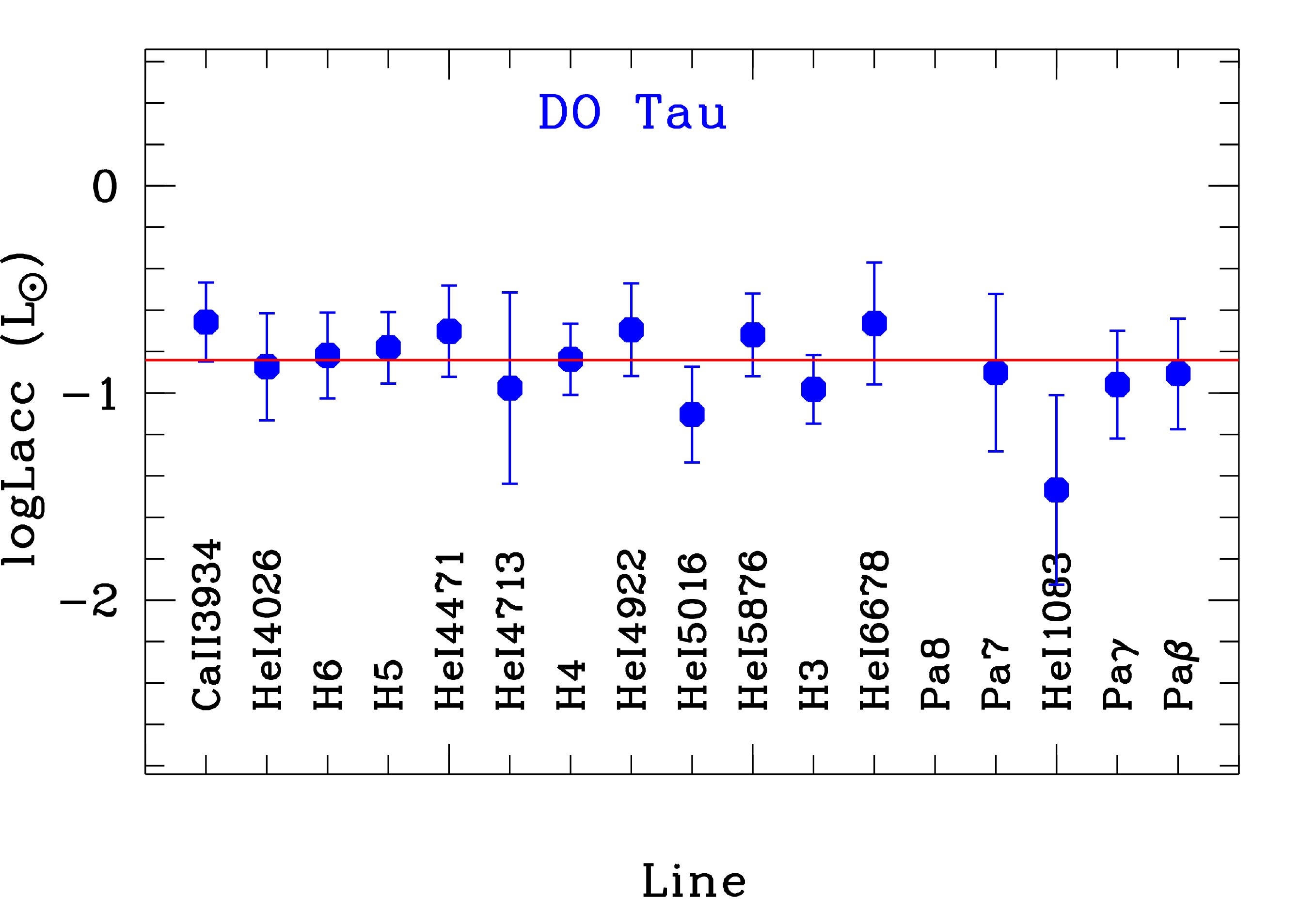}}
 {\includegraphics[bb=0 0 750 550]{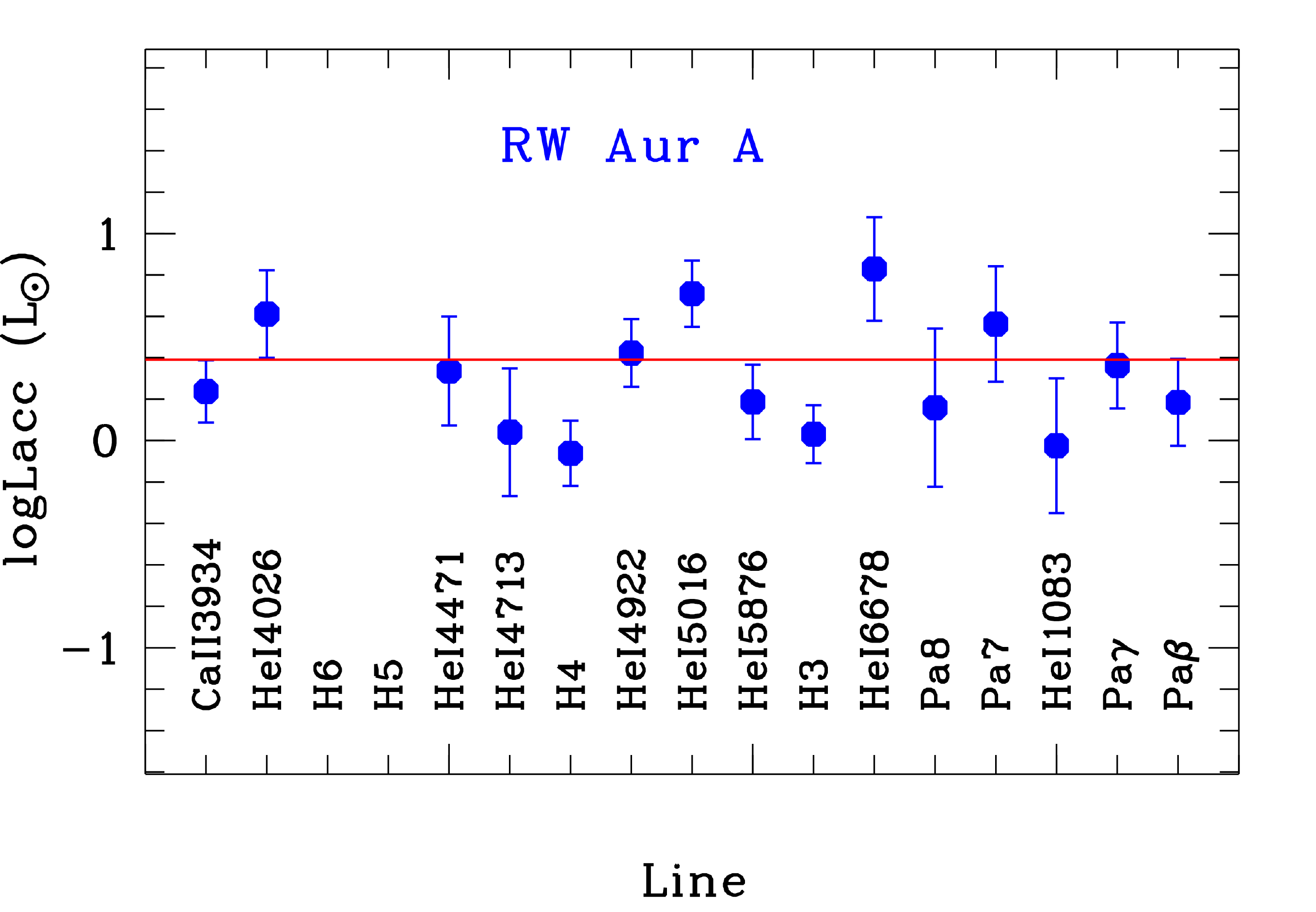}}
 }
 \resizebox{0.67\hsize}{!}{
 {\includegraphics[bb=-750 0 750 550]{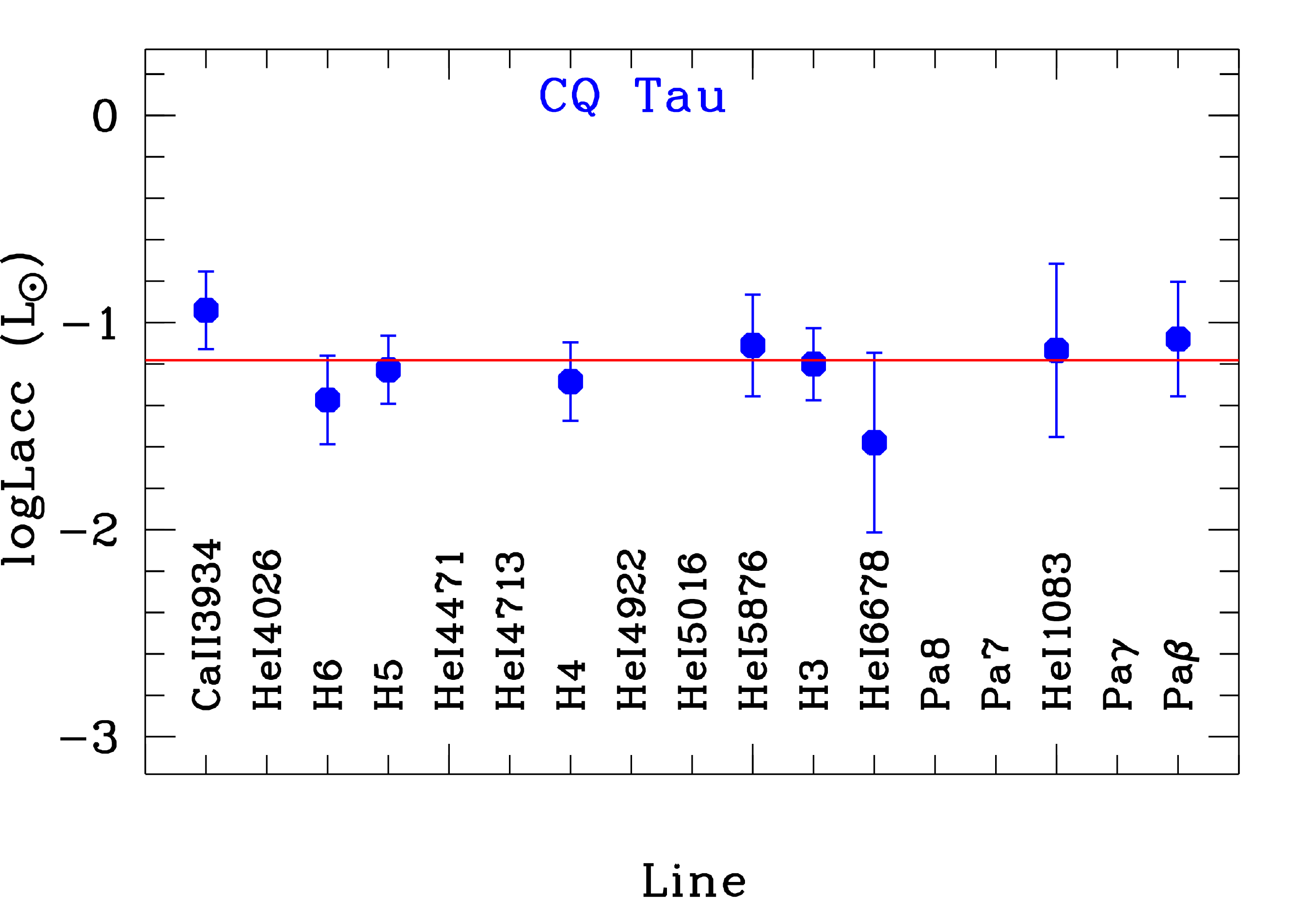}}
 }


\caption{Plots of \Lacc\ as a function of the different accretion diagnostic as labeled for the seven targets (blue dots). 
        The vertical error bars consider the error in \Ll\, as well as the errors in the \Lacc--\Ll\ relationships. 
        The horizontal red lines in each panel represent the average \Lacc. 
   \label{LaccS}}
\end{figure*}

Not all the 17 lines in the GIARPS spectral range were detected or could be measured in every CTT star.
A summary of the number of lines with flux measurements in every CTT in our sample  is provided 
in the first column of Table~\ref{line_detections}, and every panel of Figure~\ref{LaccS} 
shows the lines detected and measured. The absence of a point in these plots means that
the line was not detected or could not be extracted with the photospheric subtraction, 
mainly because of large residuals of telluric lines and/or a strong contribution of the nonphotospheric 
self-absorption component.
DL\,Tau is the only star for which the complete set of lines could be detected and measured, while 
CQ\,Tau has the lowest number of lines with measured fluxes. For most stars, fluxes have been measured for
more  than 12 lines. In the case of RW\,Aur\,A, the Balmer lines higher than H4 (\Hg, \Hd) are dominated 
by a nonphotospheric self-absorption component. Therefore, we did not attempt a measurement of the flux of 
those lines. Also, the \Ha\ and \Hb\ lines in this star may be affected by a similar self-absorption, 
but we measured their flux with the awareness that it may be underestimated. 
A contribution of absorption components in the emission lines of DG\,Tau, DL\,Tau, HN\,Tau\,A, and DO\,Tau
is not significant, and therefore we use the flux measured directly by the integration of the lines.
The observed fluxes, equivalent widths, and their errors are reported in several tables provided in Appendix~\ref{line_fluxes_EWs_Lacc} (Tables \ref{tab:fluxes_EWs_Had} to \ref{tab:fluxes_EW_CaI3934}) \footnote{The flux 
errors reported in these tables are those resulting  from the uncertainty in continuum placement. 
The estimated $\sim$20\% uncertainty of flux calibration (see Sect.~\ref{obs_datared}) should be added 
in quadrature.}.

The luminosity of the different emission lines was computed as \Ll ~$ = 4 \pi  d^2 \cdot f_{\rm line}$, 
where $d$ is the YSO distance listed in Table~\ref{ctts_prop} and $f_{\rm line}$ is the extinction-corrected 
flux of the lines.

\subsection{Accretion luminosity}
\label{accretion}

We derived \Lacc\ via empirical relationships between accretion luminosity and the luminosity 
of permitted emission lines. Such relationships  have been derived by several authors 
 \citep[e.g.,][]{HH08,rigliaco12,alcala14,alcala17}. Here we used the most recent ones by 
\citet[][]{alcala17}, where the relationships are simultaneously derived for lines from the UV 
to the NIR for a more than 90\% complete sample of Lupus YSOs.

In Figure~\ref{LaccS} the derived \Lacc\ values are plotted as a function of the line diagnostics for the seven targets.
The error bars include the errors in \Ll, as well as the errors in each \Lacc--\Ll\ calibration. 
The individual \Lacc\ values corresponding to each diagnostic are reported in the various tables in the 
Appendix~\ref{line_fluxes_EWs_Lacc} (Tables \ref{tab:fluxes_EWs_Had} to \ref{tab:fluxes_EW_CaI3934}) for every 
CTT in the sample.
An average accretion luminosity, $\langle$\Lacc$\rangle$, was then calculated for each target. These values are 
reported in Table~\ref{line_detections} together with the corresponding $\langle$\Lacc$\rangle$/\Lstar\ ratio.

The typical standard deviation of $\sim$0.25\,dex on $\log$$\langle$\Lacc$\rangle$ is within the expected error 
estimated from the fit of the UV continuum excess emission in other samples using slab models 
\citep[see][]{alcala14,alcala17,manara17a}, although the error on \Lacc\ for the individual diagnostics is 
larger than this value. This confirms that the average \Lacc, derived from several diagnostics measured 
simultaneously, has a significantly reduced error.

\subsection{Mass accretion rate}
The average accretion luminosity of each target (see Table~\ref{line_detections}) was converted into mass accretion rate, 
\Macc, using the relation

{\setlength{\mathindent}{0pt}
\begin{equation}
\label{Macc}
\dot{M}_{acc} = ( 1 - \frac{R_{\star}}{R_{\rm in}} )^{-1} ~ \frac{\langle L_{acc} \rangle R_{\star}}{G M_{\star}}
 \approx 1.25 ~ \frac{\langle L_{acc} \rangle R_{\star}}{G M_{\star}} 
,\end{equation}

\noindent
assuming $\frac{R_{\star}}{R_{\rm in}}=\frac{1}{5}$, where $R_{\star}$ and $R_{\rm in}$ are the 
stellar radius and inner-disk radius, respectively \citep[see][]{gullbring98, hart98}, and using the 
stellar parameters reported in Table~\ref{rotfit_stel_prop}. The resulting \Macc\ values are reported
in Table~\ref{line_detections}.

\setlength{\tabcolsep}{2pt}
\begin{table}
\caption[ ]{\label{line_detections} Summary of the number of lines used to derive the average \Lacc\ and accretion properties.} 
\begin{tabular}{l|c|c|c|c}
\hline \hline

Name & No. lines  &  $\log$$\langle$\Lacc$\rangle$ ($\pm\sigma$\,dex) &  $\langle$\Lacc$\rangle/$\Lstar &  $\log$\Macc  \\
     &           &  (\Lsun) &   & (\Msun\ yr$^{-1}$)  \\

\hline     
           &    &                &        &        \\
RY\,Tau              & 12 &  $-$0.38 (0.15) &  0.05  & $-$7.57 \\
DG\,Tau              & 14 &  $-$0.25 (0.18) &  1.28  & $-$7.35 \\
DL\,Tau              & 17 &  $-$0.35 (0.18) &  1.12  & $-$7.62 \\
HN\,Tau\,A$^\dagger$  & 15 &  $-$0.99 (0.23) &  0.68  & $-$8.50 \\
DO\,Tau              & 16 &  $-$0.84 (0.16) &  0.34  & $-$7.73 \\
RW\,Aur\,A           & 15 &  $+$0.39 (0.30) &  1.50  & $-$6.93 \\
CQ\,Tau              &  9 &  $-$1.18 (0.17) &  0.02  & $-$8.68 \\

\hline
\end{tabular}
\tablefoot{~\\
$\dagger$ : subluminous YSO. The values for the accretion properties may be underestimated (see Sect.~\ref{HN_Lacc}).
Corrected values are provided in Table~\ref{hntau_corr}. }
\end{table}

The sources of error in \Macc ~are the uncertainties on \Lacc, stellar mass, and radius. Propagating these, 
we estimate an average error of $\sim$0.35\,dex in \Macc\ \citep[see Appendix~A of][]{alcala17}. 
However, additional errors on these quantities come from the uncertainty in distance, as well as from 
differences in the adopted evolutionary tracks. The largest uncertainty on the YSOs distance from the 
Gaia EDR3 is estimated to be less than $\sim$10\%,  yielding relative uncertainty of $\sim$0.2\,dex in the 
mass accretion rate\footnote{We note that \Macc~$\propto$~d$^3$, as \Lacc~$\propto$~d$^2$ and \Rstar~$\propto$~d.}.
On the other hand, it has been shown \citep[see Appendix~A in][]{alcala17} that using different sets of PMS
evolutionary tracks leads to uncertainties of 0.04\,dex to 0.3\,dex in \Macc. 
We therefore estimate that the cumulative relative uncertainty in \Macc\ is about 0.4\,dex.

We note that our \Lacc$=0.08$\,\Lsun\ for CQ\,Tau  is more than an order of magnitude lower than the value reported in 
\citet[][\Lacc$=$3.8\,\Lsun; see also Tables~\ref{ctts_prop} and \ref{line_detections} for the \Macc\ estimates]{donehew11}. 
Yet, using the \brg\ luminosity reported by these authors and the \citet[][]{alcala17} relationships we derive a 
\Lacc$=$0.15\,\Lsun, which is only a factor two our value and within the limits of long-timescale variable accretion 
\citep[$<$0.4\,dex, see][]{costigan12, costigan14}.
A correction estimate due to UX\,Ori-type variability yields a  \Lacc$\sim 0.38$\,\Lsun\ for CQ\,Tau
(see Section~\ref{HN_Lacc}), which is still lower by an order of magnitude than the \citet[][]{donehew11} determination.
It is worth noting that our estimate of $\log$\Macc$=-8.68$ for this star is consistent with the upper limit derived by \citet[][$\log$\Macc$<-$8.3]{mendigutia11}, suggesting that these latter authors also observed the star in
its faint phase.

We also stress that the \Lacc\ values derived here for RY\,Tau, DL\,Tau, HN\,Tau\,A, and DO\,Tau are consistent, 
within the errors, with those by \citet[][]{ingleby09} and \citet[][]{ingleby12}, although in the case of RY\,Tau our 
value is a factor of about four lower. We think this may be due to variable accretion. 
It is also worth mentioning that our \Macc\ estimates for RY\,Tau and DG\,Tau are in good agreement with the range 
of values derived by \citet[][]{frasca18} based on the \Ha\ and He\,{\sc i}6678 emission lines.

\subsubsection{The cases of HN\,Tau\,A and CQ\,Tau}
\label{HN_Lacc}

The highly inclined disk of HN\,Tau\,A may occult, at least partially, the emission from the accretion 
flows and from the shock onto the stellar surface, and therefore  \Lacc\ for this star will be underestimated 
in a similar way to the
\Lstar\ value (see Sect.~\ref{luminosity}). Nevertheless, the \Lacc/\Lstar\ ratio for this star is at the level of the 
highest accretors, possibly showing that both  \Lacc\ and \Lstar\ of the star are obscured in the same manner.
This interpretation has proven to be correct for subluminous objects in Lupus \citep[see Sect.~7.4 in][]{alcala14}.

Following the same reasoning as in \citet[][their Appendix~C]{alcala20}, we can use the luminosity of the 
$[\ion{O}{i}]~\lambda6300$ line to estimate a correction factor on \Lacc\ and \Lstar.
 The [O\,{\sc i}] line is found to originate relatively far from the star and therefore should not be significantly affected 
 by obscuration effects from the inner disk, and the line luminosity is also correlated with \Lacc\ and \Lstar\ 
 \citep[][]{natta14, nisini18}. From the HARPS-N spectrum of HN\,Tau\,A, we measure a line flux 
F$_{[\ion{O}{i}]}=9 \times 10^{-14}$\,erg\,s$^{-1}$\,cm$^{-2}$ and derive the extinction corrected (\Av$=$1.53\,mag) 
flux F$_{[\ion{O}{i}]}^{\rm corr}=3 \times 10^{-13}$\,erg\,s$^{-1}$\,cm$^{-2}$.
From \papI\ we estimate that about one-third of the flux comes from the low-velocity component (LVC), and therefore we derive a line luminosity, in solar units, of $\log L_{[\ion{O}{i}]}^{LVC}=-4.2$. 
Using the \Lacc\--\Ll\  and \Lacc\--\Lstar\ relationships for the LVC\footnote{We use only the LVC because the 
HVC may be affected by the fact that the jet is extended. In this case, the relationships found in 
\citet[][]{nisini18} on a sample observed with the X-Shooter instrument might not give a correct value 
because of the different instrumental FOV  used. The LVC does not suffer from this problem because it forms in 
a compact region.} derived by \citet[][]{nisini18} we estimate the $\log$\Lacc\ and $\log$\Lstar\ values given 
in Table~\ref{hntau_corr}, which are a factor of  approximately 7 and 17 the subluminous values, 
respectively, meaning that both \Lacc\ and \Lstar\ are affected by a similar obscuration factor,
which is consistent with the interpretation  for subluminous objects \citep[][]{alcala14}. For comparison,
we also include the measured values in Table~\ref{hntau_corr}.
The corrected stellar luminosity implies a stellar radius a factor about 4.3 larger (i.e., 2.6\,\Rsun), 
and places the star on the HR diagram in a position consistent with the other CTTs. 

%
%
%

\setlength{\tabcolsep}{4pt}
\begin{table}[t]
\begin{center}
\caption[ ]{\label{hntau_corr} Accretion and stellar properties of HN\,Tau\,A after and before 
correction for obscuration effects and upper limits for CQ\,Tau derived as explained in 
Section~\ref{HN_Lacc}.} 
\begin{tabular}{l|c|c|c|c}
\hline \hline
 & \multicolumn{2}{c}{HN\,Tau\,A} & \multicolumn{2}{|c}{CQ\,Tau}   \\ 
\cline{2-3}
\cline{3-5}
              &  Corrected                   & Measured  & Upper lim.    & Measured     \\
Quantity      & with $L_{[\ion{O}{i}]}^{LVC}$ & values    &  with $\Delta B$   & values  \\

\hline     

$\log$\Lacc  & $-$0.12  & $-$0.99 & $<-$0.42 & $-$1.18 \\    
$\log$\Lstar & $+$0.42  & $-$0.82 & $<+$1.18 & $+$0.43 \\
$\log$\Rstar & $+$0.41  & $-$0.22 & $<+$0.44 & $+$0.07 \\
$\log$\Mstar & $+$0.20  & $-$0.10 & $<+$0.30 & $+$0.18 \\
$\log$\Macc  & $-$7.19  & $-$8.68 & $<-$7.70 & $-$8.68\\

\hline
\end{tabular}
\end{center}
\end{table}

Using the corrected \Lstar\ value and the \citet[][]{siess00} evolutionary tracks, we estimate a mass of 1.6\,\Msun\ 
for HN\,Tau\,A, which is a factor of about two higher than the value provided in Table~\ref{rotfit_stel_prop}. 
The $\log{g}=3.8$ calculated from the corrected radius and mass is consistent with the typical gravity 
for YSOs, while the subluminous values yield a much higher gravity of $\log{g}=4.7$.
The corrected \Lacc, \Rstar, and \Mstar\ values and Eq.~\ref{Macc} would imply a $\log$\Macc$\approx-7.3$, that is, 
HN\,Tau\,A would be among the strongest solar-mass accretors in Taurus.

The work by \citet[][]{nisini18} also provides \Ll\---\Mstar\ and \Ll\---\Macc\ relationships allowing us to estimate 
\Mstar\ and \Macc\ indirectly from the $\log L_{[\ion{O}{i}]}^{LVC}$ value. 
The results are also provided in Table~\ref{hntau_corr} and are in agreement with the values derived 
from the evolutionary tracks and Eq.~\ref{Macc}.

We warn the reader about the uncertainties on extinction in high-inclination objects. The above calculations 
for HN\,Tau\,A in this section are based on the assumption that the visual extinction of the LVC of the 
$[\ion{O}{i}]$ line is the same as measured for the star, i.e., \Av$=$1.53\,mag, which is not necessarily true. 
However, we note that adopting \Av$=$0\,mag yields a corrected \Lstar\ a factor of about three lower than when 
assuming \Av$=$1.53\,mag, which is still underluminous on the HR diagram. On the other hand, a much higher value of 
\Av\ would make the star unreliably luminous. We therefore conclude that a reasonable value for the visual extinction 
of the LVC of the $[\ion{O}{i}]$ line should be in the range from about 0.7\,mag to 2\,mag.
As explained above, adopting \Av$=$1.53\,mag yields a corrected \Lstar$=$2.6\,\Lsun, which leads to consistent 
results on the stellar parameters. We therefore used this value, but warn the reader that the genuine \Lstar\  value
might be in the range between $\sim$1.5\,\Lsun\ and $\sim$3.5\,\Lsun.

For CQ\,Tau, the $[\ion{O}{i}]$ line is barely seen, and therefore we cannot use the above methodologies
to correct the stellar and accretion parameters for obscuration effects. However, we can estimate upper 
limits based on the photometric variations and assuming gray extinction.
As pointed out by \citet[][]{dodin21}, CQ\,Tau did not show variations before 1940 and was approximately constant
at  $B\sim$9\,mag \citep[][]{grinin08}. Adopting this value as the unobscured magnitude of the star and based 
on our $B=$10.9\,mag measurement during the GHOsT observations, we estimate a correction factor on the 
bolometric flux of the star of $\sim$5.7. This correction provides a maximum bolometric flux, 
yielding upper limits of 15\,\Lsun, 2\,\Msun\ and 2.8\,\Rsun\ for \Lstar, \Mstar,\ and \Rstar, respectively. 
These corrected values of \Rstar\ and \Mstar\  yield a $\log{g}=3.9$, which is more consistent with the gravity
of a YSO than the $\log{g}=4.5$ derived from the observed \Rstar\ and \Mstar\ values in Table~\ref{rotfit_stel_prop}.
Assuming that \Lacc\ is affected by the same obscuration factor as \Lstar\ \citep[see Section~7.4 in][]{alcala14},  
we derive an upper limit  of $-$7.7 for $\log$\Macc. The estimated upper limits are listed in Table~\ref{hntau_corr}.

\section{Results and Discussion}
\label{results}

The results of the previous sections show that the studied CTTs in Taurus are highly accreting objects.
The novel science products and aspects of this pilot study are as follows. 

\begin{itemize}
\item  Contemporaneous low-resolution spectroscopic and photometric observations, allowing 
    an accurate flux calibration of the high-resolution spectroscopy;
  
\item  simultaneity of the high-resolution, wide-band spectroscopic observations, from
      the optical to the NIR;

\item  simultaneous use of veiling measurements, both in the optical and NIR, to determine \Av;

\item  use of more than ten line diagnostics to estimate accretion luminosity, yielding 
      a much reduced error in \Lacc\ as compared with determinations using single diagnostics.

\end{itemize}

All of the above was achieved using well-defined and assessed procedures for deriving the stellar physical and accretion 
parameters in a self-consistent and homogeneous way. Therefore, the properties derived here 
 can be considered as more robust and reliable than in previous studies.
Some limitations to the application of our procedures are related to the extremely veiled CTTs like RW\,Aur\,A, but this type of 
object is not common. We are therefore confident that the same procedures can be applied to the entire GHOsT data sample, 
which will be presented in forthcoming papers (Gangi et al., in preparation).
In the following, we discuss a few aspects of the stellar and accretion properties of 
our sample, as well as of the continuum excess emission in the NIR of these CTTs. 

\subsection{Stellar and accretion properties}

This GIARPS/TNG pilot study confirms the high levels of accretion of the selected CTTs. 
It is worth noting from Table~\ref{line_detections} that the \Lacc/\Lstar\  ratio for 
every CTT in this sample is well above the level of chromospheric noise emission in YSOs 
\citep[max (\Lacc/\Lstar)$_{\rm noise}\approx$0.01][]{manara17}. 
We point out that in three sources, namely, DG\,Tau, DL\,Tau, and RW\,Aur\,A, the total luminosity 
is accretion dominated (i.e., \Lacc/\Lstar\ $>$ 1), which is more typical of the class~I stage 
of evolution (e.g., Antoniucci et al. 2008, Fiorellino et al., in press ).

The most actively accreting object in this sample is RW\,Aur\,A. Noteworthy, the recent study by \citet[][]{takami20}
shows that the star was in its bright stage during the GIARPS observations (13 Nov. 2017). Moreover, from \papI\ we 
know that the magnitude of the star was 10.44\,mag in the $V$-band and this is consistent with the high-accretion 
activity scenario by \citet[][]{takami20}.

The apparently least active object in the sample is CQ\,Tau. This might be mostly related to the much higher 
photospheric flux in comparison with the other objects, as CQ\,Tau is an intermediate-mass YSO with \Teff $\sim$6800\,K. 
Nevertheless, its \Macc\  is comparable with that of many actively accreting stars of similar mass in other star forming regions. 
Confirming the upper limits derived in Section~\ref{HN_Lacc}  as true values would suggest CQ\,Tau has a similar level of accretion activity to RY\,Tau.
On the other hand, if the determination of \Lacc\  based on the $[\ion{O}{i}]~\lambda6300$ line for HN\,Tau\,A 
is correct (see Sect.~\ref{HN_Lacc}), this would also be one of the most actively accreting objects in Taurus.

To investigate the levels of accretion in the Taurus subsample in more detail, we use in the following the 90\% 
complete Lupus sample studied in \citet[][]{alcala17} for comparison purposes, but using the rescaled values 
reported in Appendix~A of \citet[][]{alcala19}. Given the very small number statistics of this pilot study, 
we cannot provide any result on the sample as a whole, but only at the level
of the individual objects. 

\subsubsection{Accretion luminosity versus stellar luminosity}

Figure~\ref{Lacc_Ls} shows the accretion luminosity as a function of stellar luminosity for the Lupus
sample and the seven Taurus CTTs. While most Lupus objects lie at \Lacc/\Lstar\ values between 0.01 and 
0.1, five of the Taurus CTTs have higher values, between 0.3 and 1.5 (see Table~\ref{line_detections}).
The two CTTs with values compatible with most of the Lupus YSOs are RY\,Tau and CQ\,Tau.
We note that recent ALMA data have shown that the dusty disk of the latter has a large cavity 
\citep[R$_{cavity}=$53\,au, ][]{ubeira19}, likely suggesting a transitional disk. Also, in the case of 
RY\,Tau, submillimeter data show evidence for a protoplanetary disk with structures such as rings and 
an inner cavity \citep[][]{isella10, pinilla18, long18}.
The \Lacc/\Lstar ~ratio for both CQ\,Tau and RY\,Tau is compatible with those of other 
transitional disks in Lupus.

\begin{figure}[!ht]

\resizebox{1\hsize}{!}{ {\includegraphics[bb=20 -30 750 580]{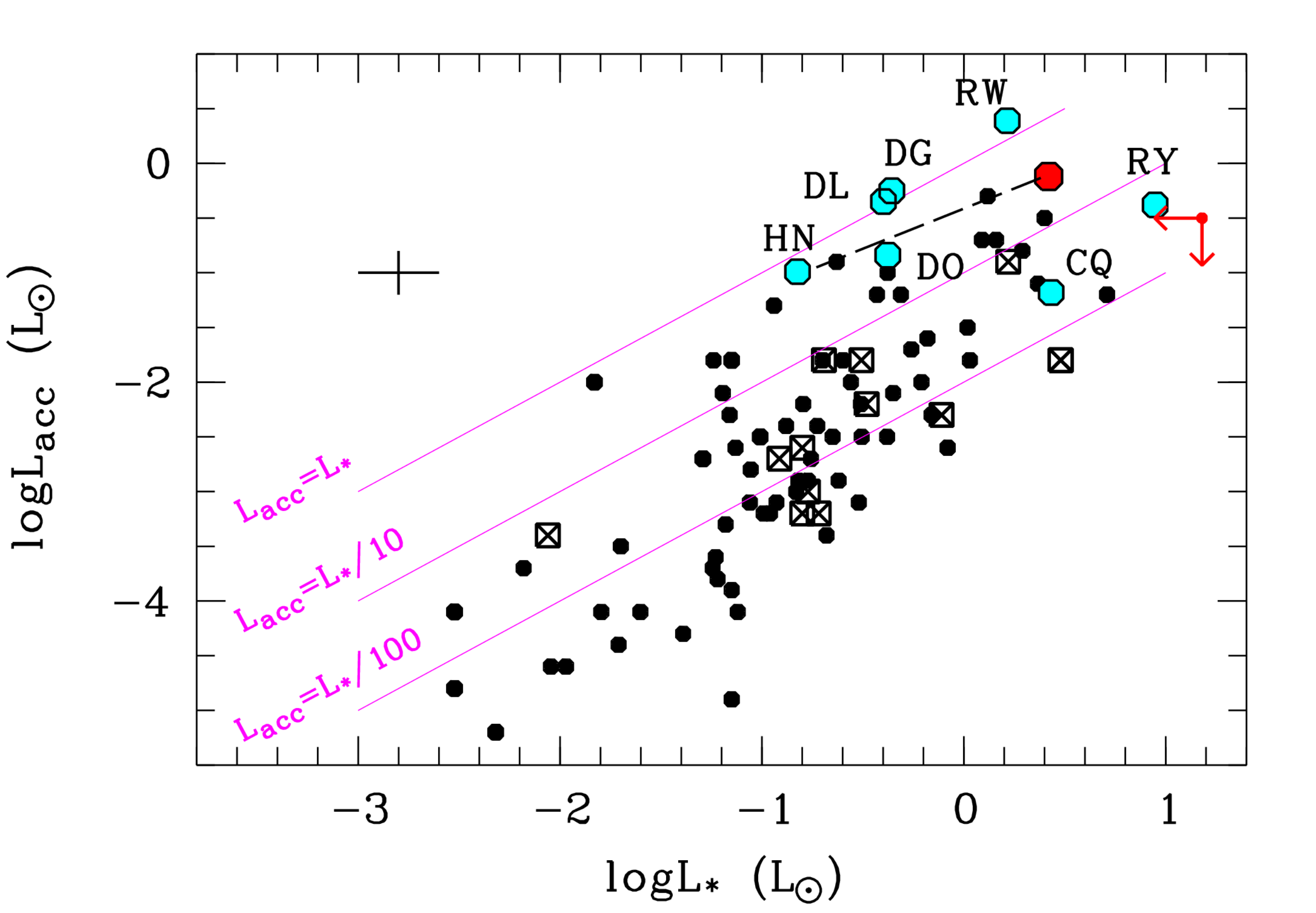}} }
 
\caption{Accretion luminosity as a function of stellar luminosity for the stars in Lupus 
 (black symbols) and the seven Taurus CTTs studied here (blue circles). The latter are labeled
 with their names. The Lupus transitional disks are shown with crossed squares. The continuous 
 lines represent the three \Lacc\ vs. \Lstar\ relations as labeled. The long-dashed line
 represents the shift of HN\,Tau\,A on the diagram when correcting its \Lacc\ and \Lstar\ 
 values for obscuration effects by the disk. The corrected values are shown with the red dot.
 The leftward and downward red arrows represent the upper limits on \Lstar\ and \Lacc\ for
 CQ\,Tau, respectively. 
 The average errors for the Taurus sample are shown in the upper left. Figure adapted 
 from \citet[][]{alcala17}.
 \label{Lacc_Ls}}
\end{figure}

\citet{tilling08} presented simplified stellar evolution calculations for stars subject to time-dependent 
accretion history, and derived evolutionary tracks on the \Lacc -- \Lstar ~diagram for a variety of 
fractional disk masses, $f_{disk}\equiv M_{disk}$/\Mstar, and YSO masses.  \citet[][]{alcala14} showed that
the relation of the Lupus data in Figure~\ref{Lacc_Ls} is steeper than the \Lacc/\Lstar~$=$constant 
lines, more or less following the slope of the  \citet{tilling08} tracks. These latter authors also concluded that 
the disks of the Lupus objects with the lowest masses should have masses lower than 0.014 $\times$ \Mstar.
The five high accretors in Taurus (DG, DL, HN, DO, and RW) fall, instead, in the region of the diagram consistent 
with the  \citet{tilling08} tracks for YSOs of  one solar mass and $f_{disk}=$0.2. These five stars have 
indeed masses on the order of 1\Msun\ hence, according to such tracks one would expect their disk mass 
to be on the order of 0.2 $\times$ \Mstar. Although a correlation between the total mass of gas$+$dust in the disk  
and the stellar mass has not  yet been confirmed, correlations between the dust mass in the disk and the stellar 
mass of YSOs in the Lupus and Chameleon star forming regions have been found \citep[][]{ansdell16, pascucci16}. 

In this scenario, the most massive disk would be RY\,Tau. The presence of the important jet in RY\,Tau was recently explained in terms of a very massive disk around this intermediate-mass T~Tauri star \citep[][]{garufi19},
which might still be fed by the interstellar matter in which the object is embedded.
For DG\,Tau, \citet[][]{podio13} estimate a total disk mass in the range 0.015--0.1\,\Msun, depending on the 
assumed dust size distribution. The upper limit would be consistent with the results from the \citet[][]{tilling08}
tracks.
We note that in the case of HN\,Tau\,A, the \Lacc\ and \Lstar\ values corrected for obscuration effects would 
still yield similar results for the estimated fractional disk mass, while confirmation of the upper limits
for CQ\,Tau would imply a disk mass similar to that of RY\,Tau.

\subsubsection{Mass accretion rate versus stellar mass}

The relationship between mass accretion rate and stellar mass is a fundamental aspect of the study
of disk evolution in YSOs. During the CTT phase,that is, after the protostar has almost entirely dispersed 
its envelope but is still actively accreting from the optically thick accretion disk, the stellar mass 
undergoes negligible changes. Therefore, the \Macc ~versus \Mstar ~relation represents a diagnostic tool for the 
evolution of \Macc ~\citep{clarke06} and for the process driving disk evolution \citep{ercolano17}.  

Figure~\ref{Macc_Ms} shows the accretion rate as a function of stellar mass for the seven CTTs in Taurus
as compared with the Lupus sample. The Taurus CTTs populate the upper right part of the diagram.
Most of the Lupus YSOs fall well below the theoretical prediction by \citet[][short-dashed black line]{vorobyov09}, 
but the latter is relatively consistent with the upper envelope of the Lupus data points distribution.
The few Lupus YSOs on the upper envelope of the \Macc--\Mstar ~relationship are the strongest Lupus 
accretors at a given mass and are also among the most luminous on the HR diagram. Interestingly, 
the five Taurus CTTs with the highest  \Lacc/\Lstar\ ratios tend to follow the upper envelope\footnote{Here 
we considered the obscuration corrected quantities for HN\,Tau\,A.}, which qualitatively is well fitted 
by the theoretical prediction. These results demonstrate that the level of accretion 
of these five CTTs is as high as that of the strongest accretors in Lupus \cite[][]{alcala17}.

On the other hand, CQ\,Tau falls in the lower envelope of the \Macc--\Mstar\  relationship, with its accretion
properties more closely resembling those of transitional disks than those of the full disks, while RY\,Tau follows
the \Macc--\Mstar\ trend for the most massive Lupus stars with full disks. Should the upper limits for CQ\,Tau
be confirmed, its accretion properties will be similar to those of RY\,Tau.

\begin{figure}[!ht]

\resizebox{1.1\hsize}{!}{{\includegraphics[bb=20 0 830 550]{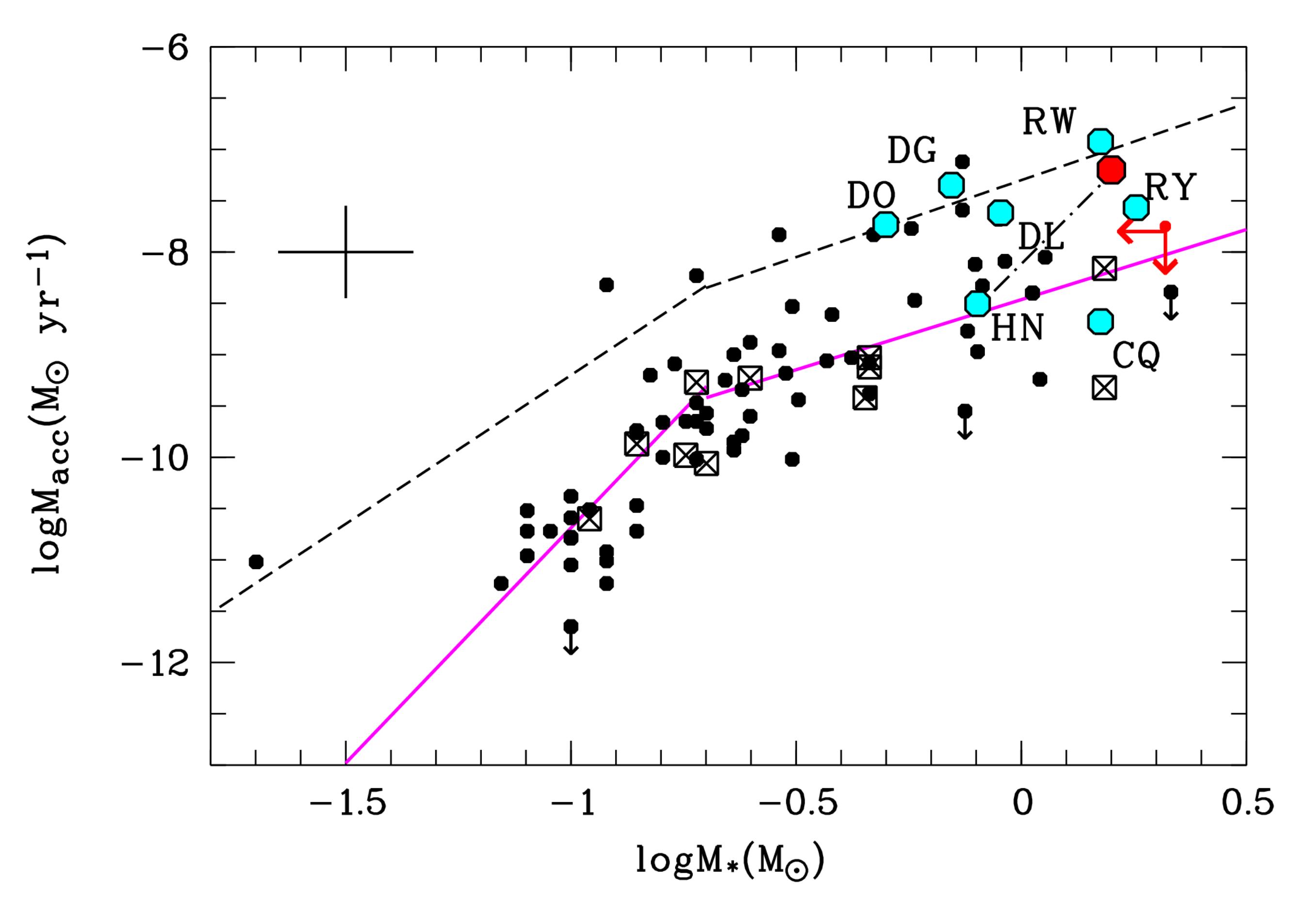}}}
 
\caption{Mass accretion rate as a function of stellar mass for the seven Taurus CTTs studied (blue circles)
 compared with stars in Lupus (black symbols). The Lupus transitional 
 disks are shown with crossed squares. Lupus objects classified as weak or negligible accretors are 
 plotted with downward arrows. The long-dashed black line represents the shift of HN\,Tau\,A on the 
 diagram when correcting its \Macc\ and \Mstar\  values for obscuration effects by the disk. 
 The corrected values are shown with the red dot.  The leftward and downward red arrows represent 
 the upper limits on \Mstar\ and \Macc\ for CQ\,Tau, respectively. 
 The black dashed line shows the double power law theoretically predicted by \citet[][]{vorobyov09}, 
 and the continuous magenta lines represent the fits to the data as in Eqs. (4) and (5) of \citet[][]{alcala17}.
 The average errors for the Taurus sample are shown in the upper left.
 Figure adapted from \citet[][]{alcala19}.
 \label{Macc_Ms}}
\end{figure}

\subsection{The continuum NIR excess emission}

Clear evidence of continuum excess emission or veiling increasing with wavelength, from the optical  
to the NIR, was found in CTTs over a decade ago\citep[][and references therein]{fischer11}.
As mentioned in Sect.~\ref{nir_veiling}, the behavior of IR veiling with wavelength can provide 
information on the physical properties of the dust in the inner edge of the disk.

Figure~\ref{veilings_nir} shows the NIR veiling as a function of wavelength for the seven CTTs in the sample.
The increase of veiling with wavelength in the NIR is evident in all the objects. Adopting the same strategies 
as in \citet[][]{antoniucci17}, a fit of r$_\lambda$ versus $\lambda$ with a black-body power law yields the temperature, 
$T_{\rm eff}^{\rm BB}$, for each CTT indicated in the corresponding panel of Figure~\ref{veilings_nir} and in
Table~\ref{veil_emiss}. 
We also include in this table the results for DR\,Tau and XZ\,Tau from \citet[][]{antoniucci17}. 
Typical errors on the $T_{\rm eff}^{\rm BB}$ 
are on the order of 100--150\,K. The fitting procedure requires another parameter, $F_{factor}^{BB}$,
namely the factor by which the black-body must be multiplied to fit the data. Such a factor is based on
the ratio of the areas of the stellar disk and the emitting region producing the NIR veiling.
The results using this factor are also given in Table~\ref{veil_emiss}. We warn the reader that the fit for 
DG\,Tau must be taken with care because only a range of values for the two reddest points could be determined 
(see Sect.~\ref{nir_veiling}).

\setlength{\tabcolsep}{4pt}
\begin{table}[t]
\begin{center}
\caption[ ]{\label{veil_emiss} Results on the veiling emission in the NIR.} 
\begin{tabular}{l|c|c|c|c}
\hline \hline
              &                          &                          &                           &   \\

Name          & $T_{\rm eff}^{\rm BB}$  & $F_{factor}^{BB}$  &  $R_{sublim}$  ($\pm$err) &  $R_{in}^{cont}$  ($\pm$err) $^{a}$ \\
              &       (K)               &                   &   (au)                   &   (au ) \\

\hline     
              &                          &                          &                       &  \\
RY\,Tau     & 1500  &  51  &  0.22 (0.05) & 0.18 (0.01) \\    
DG\,Tau     & 1700  &  39  &  0.06 (0.01) & 0.17 (0.01) \\
DL\,Tau     & 1600  &  36  &  0.06 (0.01) &   \\
HN\,Tau\,A  & 2100  &  23  &  0.07 (0.01)$^\dagger$ &   \\
DO\,Tau     & 1800  &  25  &  0.04 (0.01) &   \\
RW\,Aur\,A  & 2150  &  38  &  0.07 (0.01) & 0.10 (0.01) \\
CQ\,Tau     & 2050  &  65  &  0.07 (0.01) &   \\
   "       & "   &   "   &   $<$0.15$^\ddagger$  &    \\
           &       &      &              &    \\
DR\,Tau    &  2350 &  28  &  0.03  (0.01) & 0.12 (0.01)  \\
XZ\,Tau    &  1600 &  29  &  0.06  (0.01) & \\

\hline
\end{tabular}
\tablefoot{~\\
$\dagger$ : computed with the  \Lstar\ and \Lacc\ values in Table~\ref{hntau_corr} \\
$\ddagger$ : computed with the upper limits on \Lstar\ and \Lacc\ in Table~\ref{hntau_corr} \\
$a$: Interferometric measurements by \citet[][]{eisner10}\\
The values for DR\,Tau and XZ\,Tau are taken from \citet[][]{antoniucci17}
}
\end{center}
\end{table}

While the resulting temperatures for RY\,Tau, DG\,Tau, DL\,Tau, and within errors DO\,Tau, are consistent 
with the origin of the NIR continuum excess emission being the inner rim of the dusty disk, the
derived $T_{\rm eff}^{\rm BB}$ value for the other two CTTs is significantly higher than the dust sublimation 
temperature ($\sim$1500\,K) and is more difficult to interpret in the same way, although temperatures 
as high as 2000\,K may be expected for the sublimation of silicate dust \citep[][]{pollack94}. 

In the simple model by \citet[][]{dullemond01}, where the  inner disk is directly irradiated by the central 
star, the inner disk edge is located at a radius, $R_{sublim}$, which is given by the following equation: \\

{\setlength{\mathindent}{0pt}
\begin{equation}
\label{Rsub}
~~~~~~~~~~~R_{sublim} = \sqrt {(1 + f) \left( \frac{ L_{\star} + L_{acc} }{ 4 \pi \sigma T_{sublim}^{4} } \right) }
~~~~\\
,\end{equation}

\noindent
where $f$ is the ratio of the inner edge height to its radius, and is estimated to be 0.1 for T Tauri stars.
Assuming this model and using the \Lstar\ and \Lacc\ results of the previous sections, and adopting $T_{\rm eff}^{\rm BB}$
in Table~\ref{veil_emiss} as sublimation temperature, we computed the $R_{sublim}$ values listed in Table~\ref{veil_emiss} 
for our sample of CTTs in Taurus.  The error in $R_{sublim}$ was calculated by error propagation 
in Eq.~\ref{Rsub} and adopting errors of 0.2\,dex, 0.25\,dex, and 150\,K on $\log$\Lstar\, $\log$\Lacc,  and 
$T_{\rm eff}^{\rm BB}$, respectively. 
We can compare these values with the inner radius, $R_{in}^{cont}$, derived from interferometric 
observations of the continuum NIR emission by \citet[][]{eisner10} for a few of the CTTs in common.
These values are listed in the last column of Table~\ref{veil_emiss}.
We note that except for RY\,Tau, where $R_{sublim}$ and $R_{in}^{cont}$ are in very good agreement, 
the $R_{sublim}$ values for the other four stars with interferometric observations are lower than 
the $R_{in}^{cont}$ measurements. Interestingly, the upper limits on \Lstar\ and \Lacc\ for CQ\,Tau
would yield a $R_{sublim}$ value more similar to those typically found using interferometric observations 
(see Table~\ref{veil_emiss}).

Previous works \citep[i.e.,][]{eisner07,anthonioz15} have shown that the irradiated 
disk model for CTTs predicts radii for the inner rim of the dusty disk that are underestimated with respect to 
interferometric measurements. We therefore expect the $R_{sublim}$ values in Table~\ref{veil_emiss} 
for DL\,Tau, HN\,Tau\,A, DO\,Tau, CQ\,Tau, and XZ\,Tau to be underestimated by a factor of between about two and three.
However, one possibility is that the $T_{\rm eff}^{\rm BB}$ we measure is not the sublimation 
temperature, but being instead related to the excess emission in the NIR, may be the 
temperature of hot gas inside the sublimation radius.

\begin{figure*}[!ht]

 \resizebox{1\hsize}{!}{
 {\includegraphics[bb=100 100 500 520]{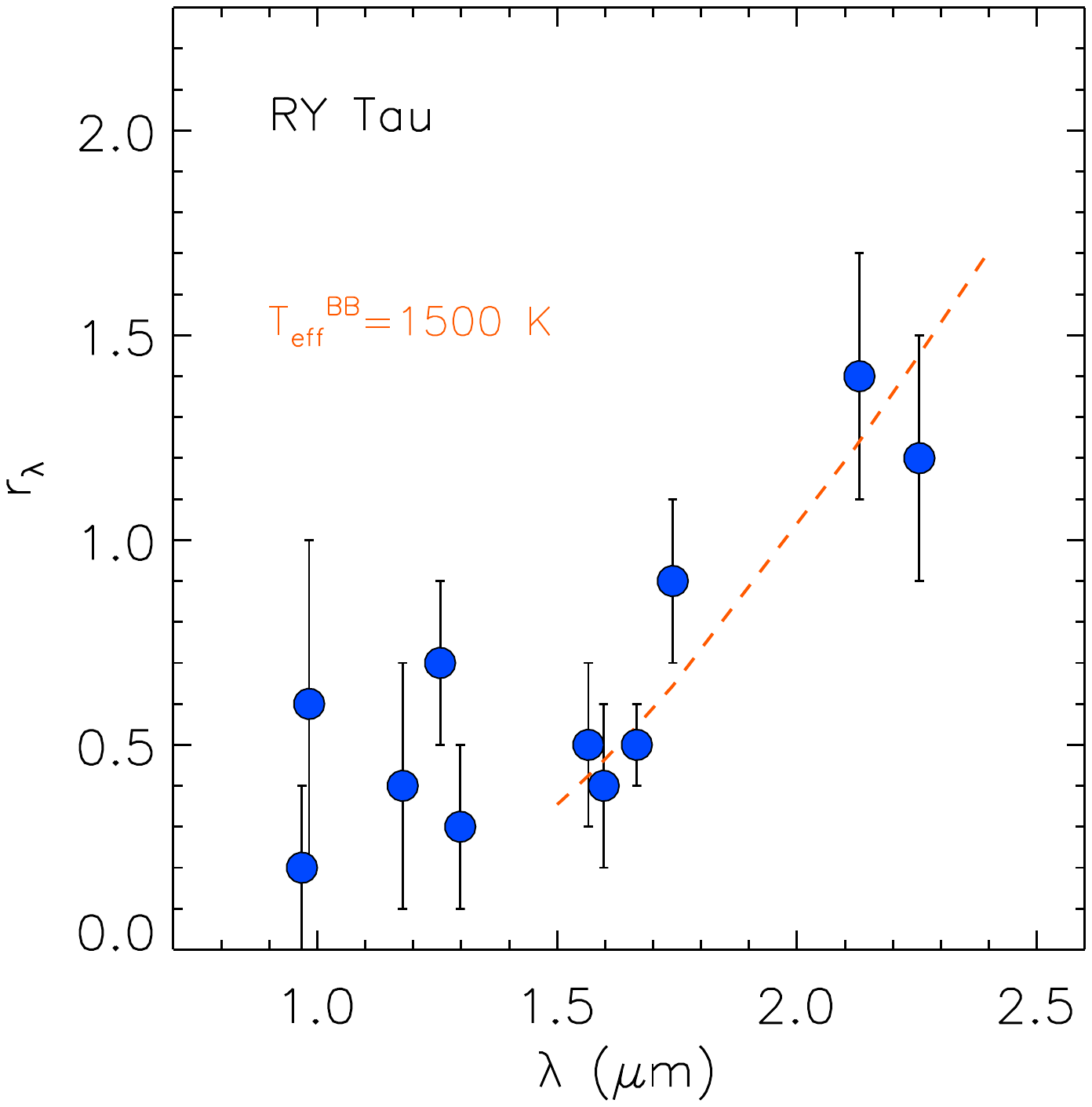}}
 {\includegraphics[bb=100 100 500 520]{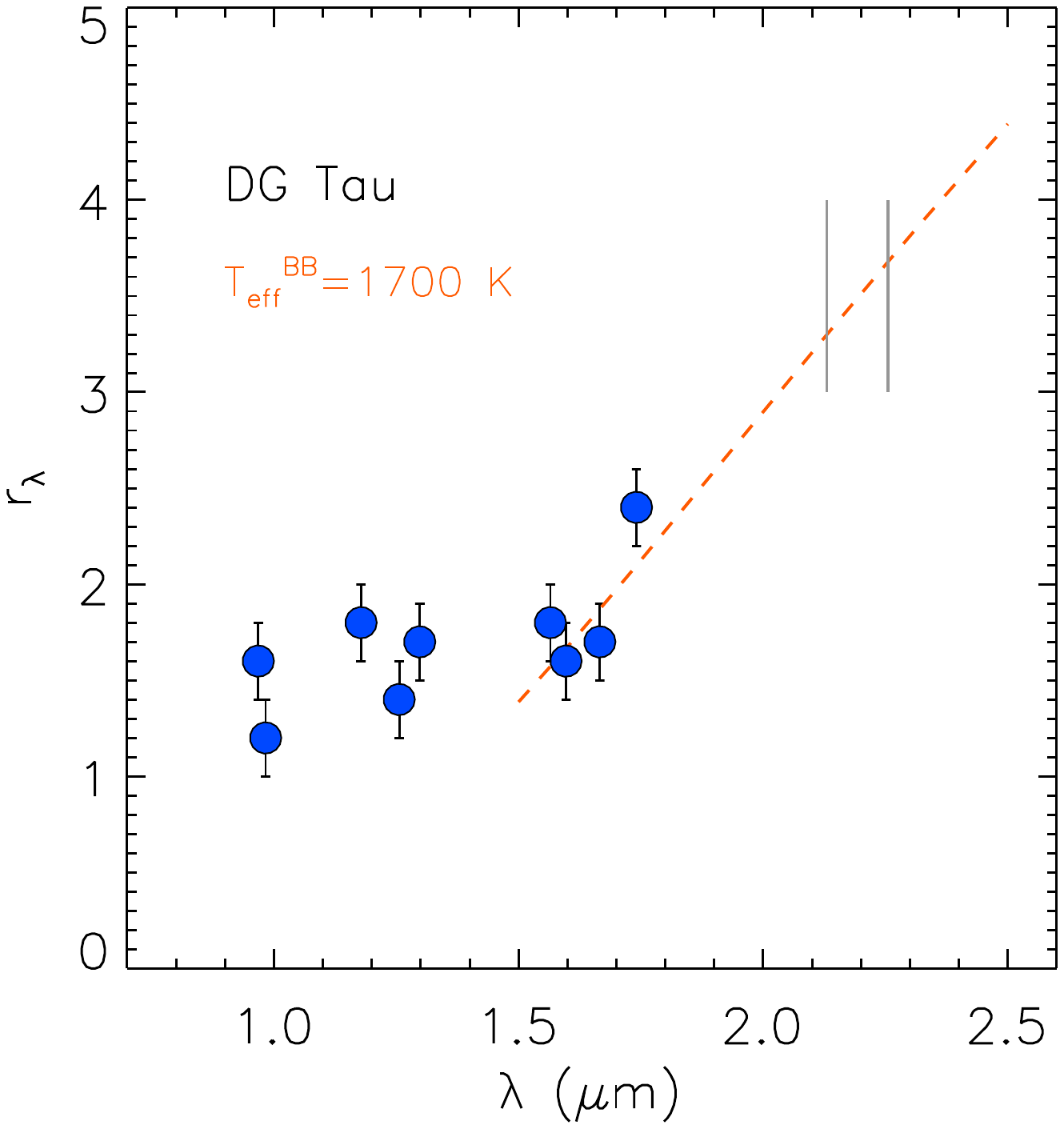}}
 {\includegraphics[bb=100 100 500 520]{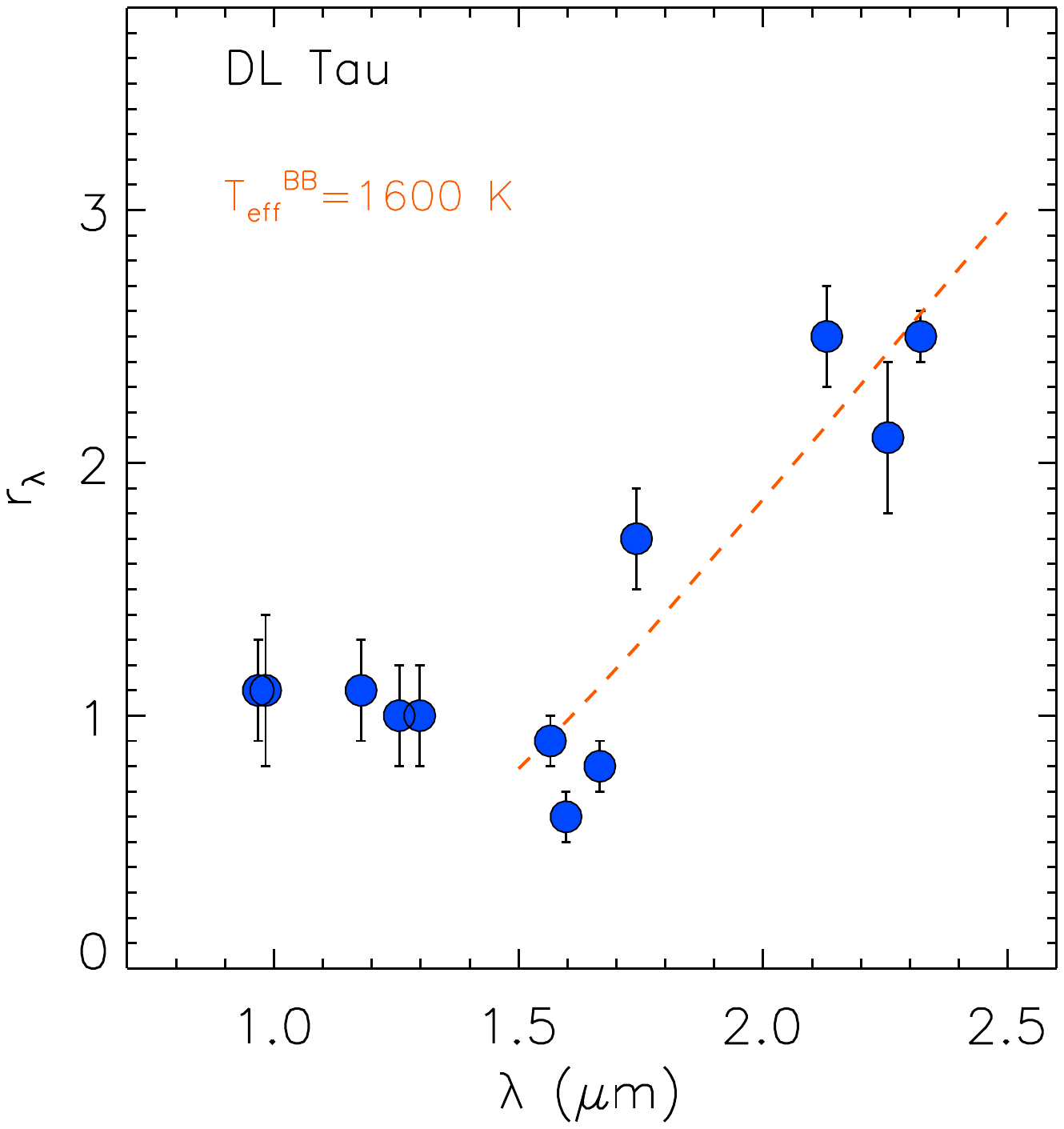}}
 }
 \resizebox{1\hsize}{!}{
 {\includegraphics[bb=100 100 500 520]{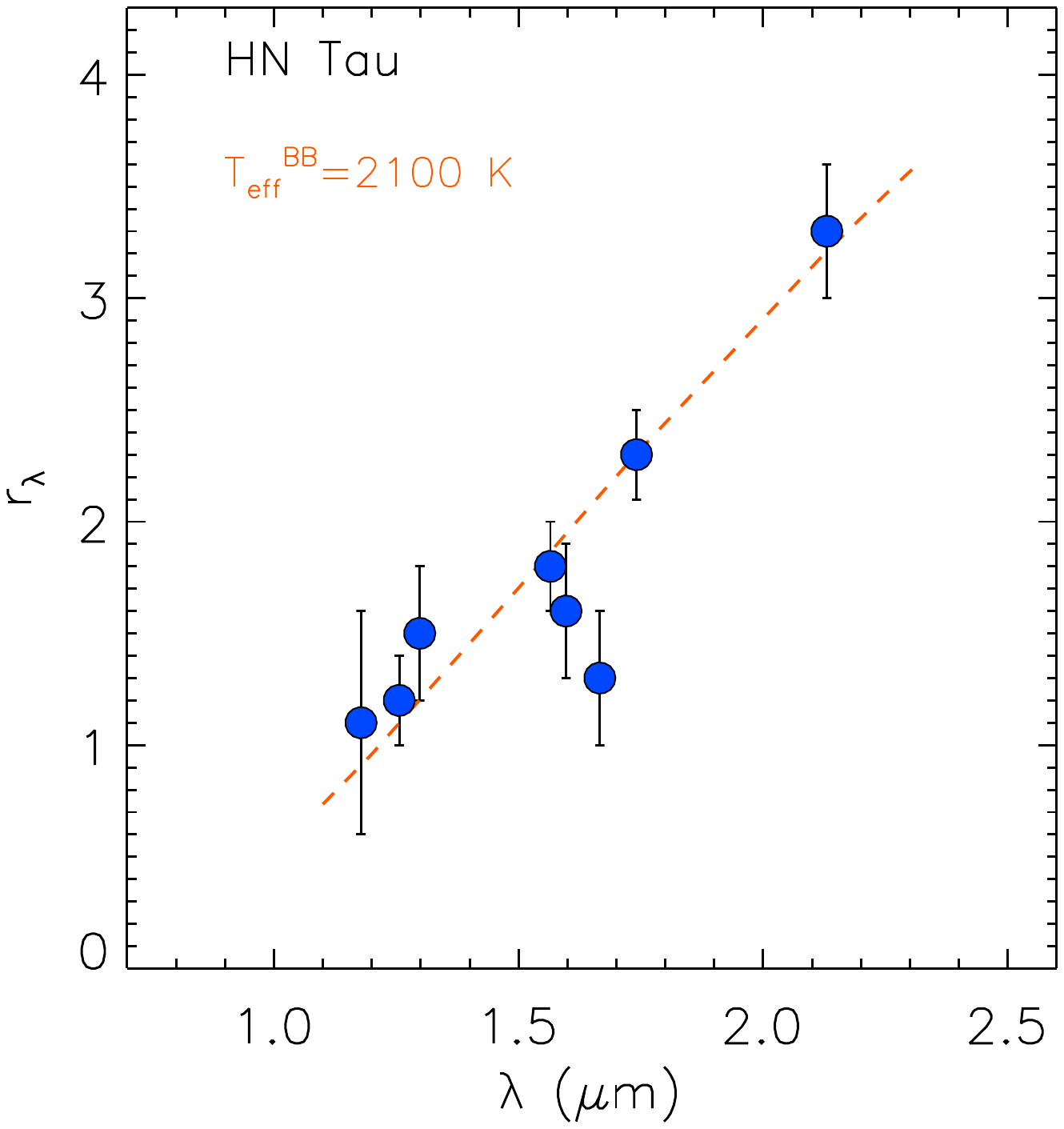}}
 {\includegraphics[bb=100 100 500 520]{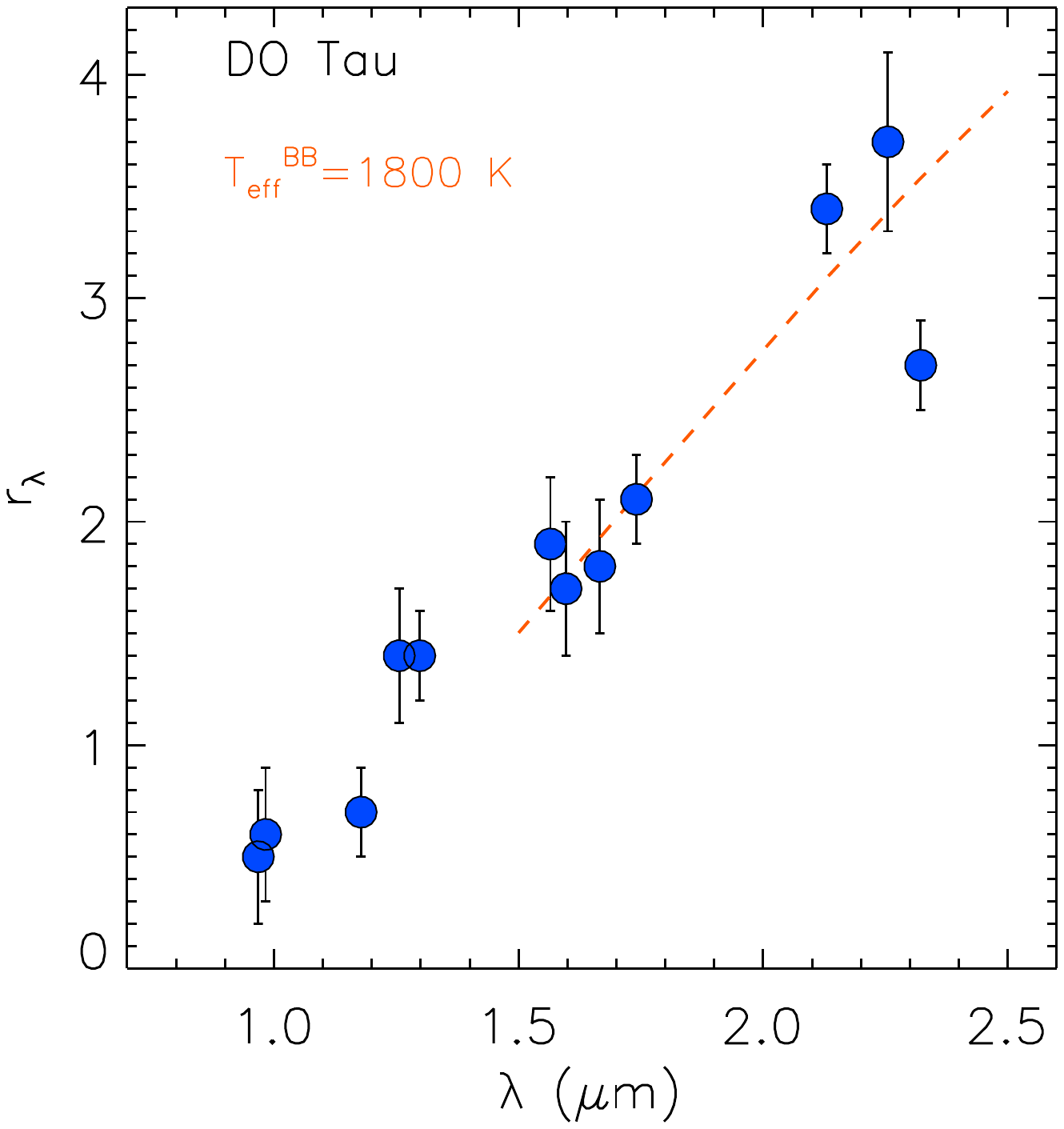}}
 {\includegraphics[bb=100 100 500 520]{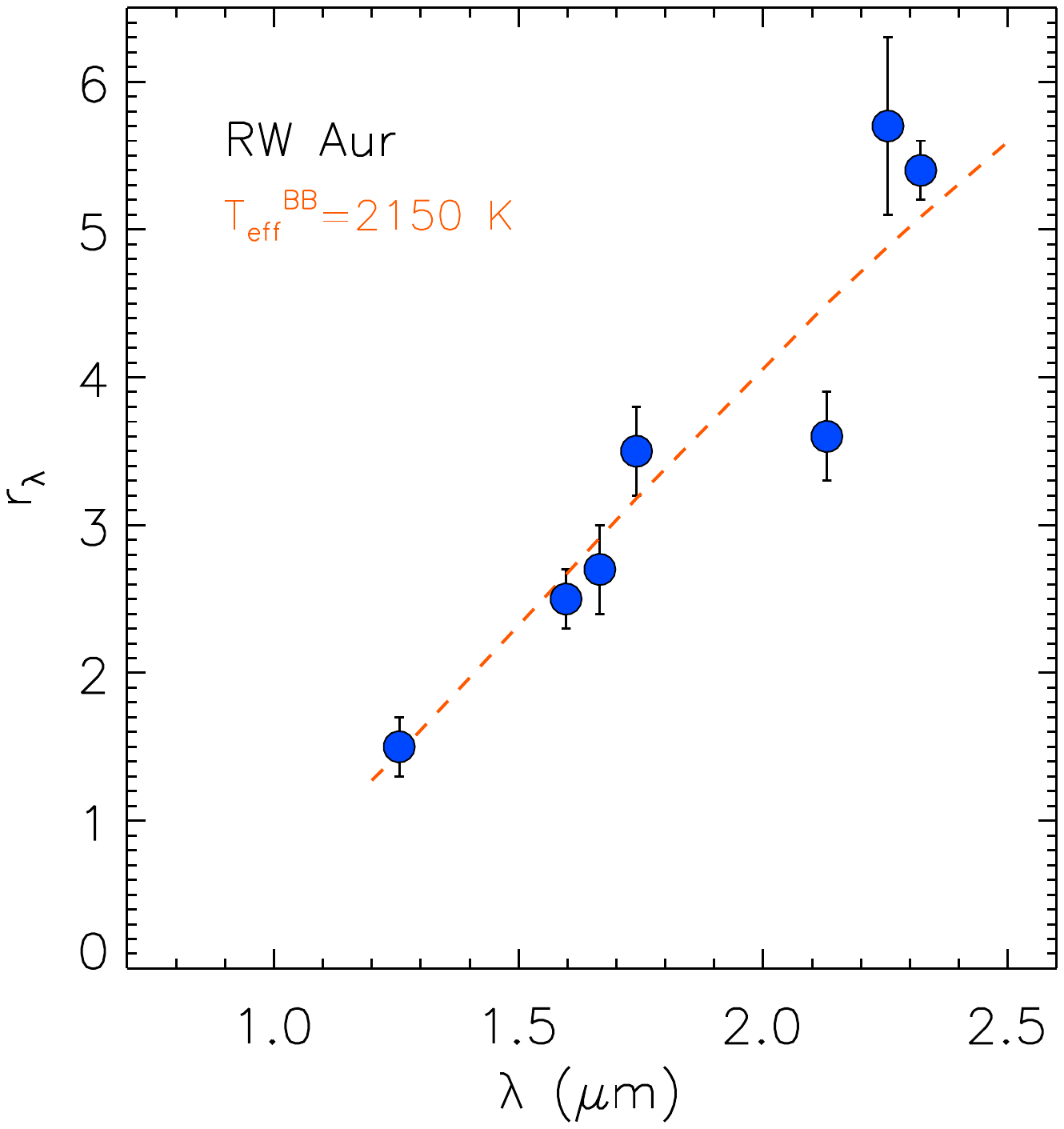}}
}
 \resizebox{0.665\hsize}{!}{
 {\includegraphics[bb=-310 100 500 520]{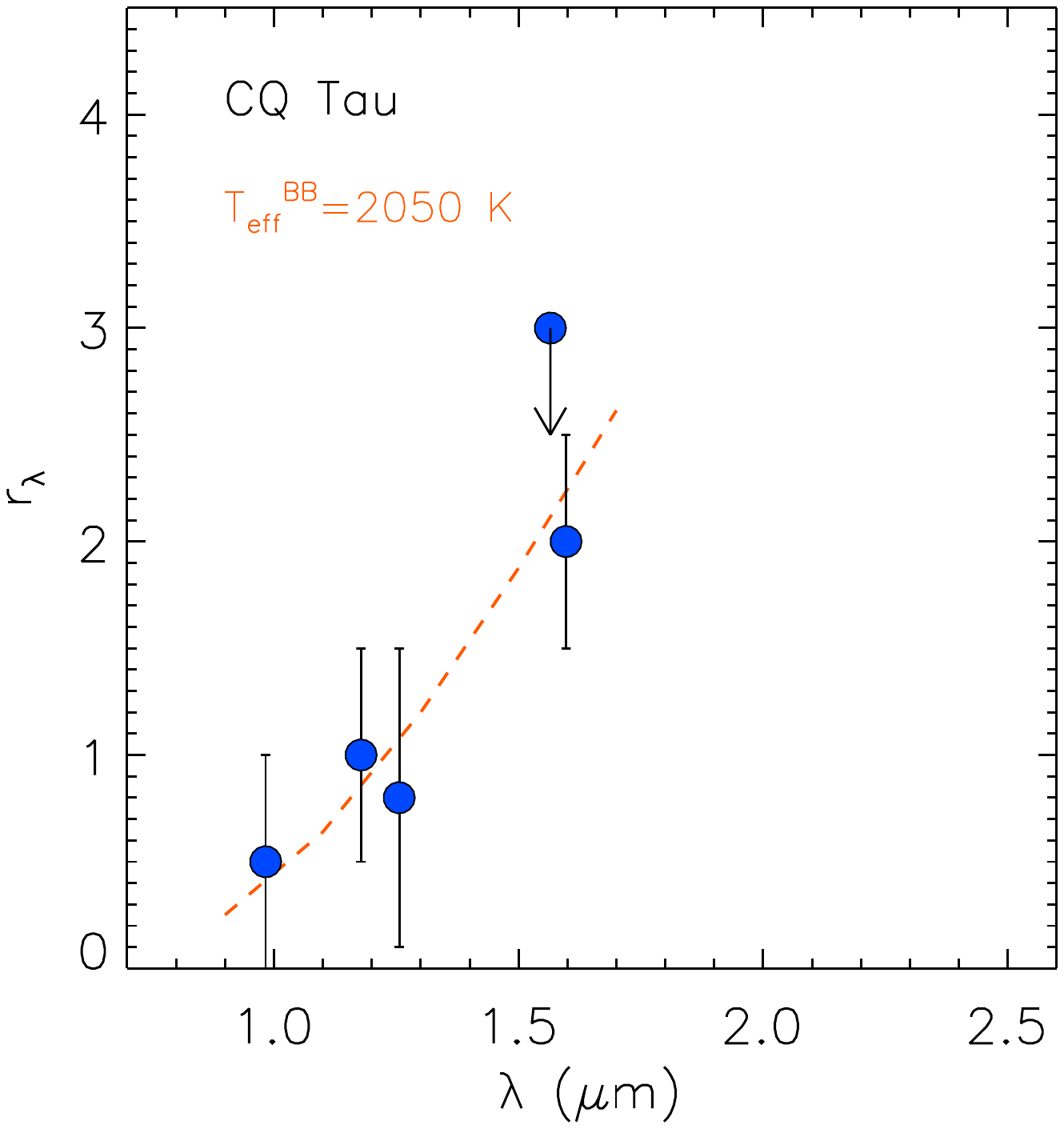}}
}
 
\caption{Plots of NIR veiling as a function of wavelength for the stars in our sample (blue dots). The red dashed lines
       show the black body fits to the data with T$_{\rm eff}^{\rm BB}$ as labeled. See text for details. 
   \label{veilings_nir}}
\end{figure*}

Several authors \citep[e.g.,][]{fischer11} discussed possible scenarios for the NIR continuum emission. 
Among the suggested contributions that may enhance the NIR continuum excess at the shorter $IYJ$ wavelengths 
there are warm annuli around accretion hot spots, hot gas inside the dust sublimation radius, 
and hot gas in the accretion flows and/or winds. \citet[][]{fischer11} modeled and discussed the presence 
of multiple temperature components in the shocked photosphere. Also, detailed multiple-component
modeling of the accretion emission has been performed by \citet[][]{ingleby13}.

The results in \papI\ showed that DG\,Tau and HN\,Tau\,A share similar physical conditions in their jets. 
The temperature and ionization gradients of the jets in these two objects would favor a magneto-hydrodynamical 
shock heating in which the warm and ionized streamlines originate in the internal and mainly gaseous disk, 
while the low-velocity and almost neutral stream lines come from the dusty regions of the outer 
disk. Therefore, the hypothesis of a gaseous hot disk inside the dust sublimation radius in HN\,Tau\,A and the 
similarity of the physical conditions of its jet with those in the DG\,Tau jet would also favor a much 
higher accretion rate in HN\,Tau\,A than what is measured without obscuration correction.
The possibility for the formation of the jets in DO\,Tau and RW\,Aur\,A in a gaseous inner disk was also 
explored in \papI. In this context, jets signify the presence of dense hot regions inside the gaseous 
inner disk. 

In conclusion, it may be possible that some of the NIR continuum excess emission of the CTTs, 
in particular in the case of HN\,Tau\,A, RW\,Aur\,A,  and DR\,Tau, originates in a thick gaseous 
disk inside the dust sublimation radius, as already suggested by \citet[][]{fischer11} and 
\citet[][]{antoniucci17} for other similar cases. 
In fact, the size of an emitting region estimated as $R_{emiss}^{cont} \sim$ \Rstar $\cdot \sqrt{F_{factor}^{BB}}$
is always smaller than the typical inner rim radius of $\sim$0.1\,au in CTTs measured from interferometric
observations in the NIR. This is consistent with the results by \citet[][]{koutoulaki19} who detected ro-vibrational 
emission of CO in the NIR X-Shooter spectrum of RW\,Aur\,A. The modeling presented by these latter authors of five band-heads shows that the CO 
emission comes from a region at a distance of $\sim$0.06--0.08\,au  from the star.

\section{Conclusions}
\label{conclusions}

In this pilot study, we report GIARPS@TNG high-resolution observations of seven CTTs in the Taurus-Auriga star 
forming region, namely RY\,Tau, DG\,Tau, DL\,Tau, HN\,Tau\,A, DO\,Tau, RW\,Aur\,A, and CQ\,Tau. The spectra simultaneously 
cover a wide spectral range from the optical to the NIR. Contemporaneous spectrophotometric 
and photometric observations were also performed, allowing us to flux-calibrate the high-resolution 
spectra with an estimated accuracy of $<$20\%. 
We show that the GIARPS@TNG data, together with the ancillary observations, allows  
the stellar and accretion parameters of YSOs to be derived in a self-consistent and homogeneous way. 

High-resolution spectroscopy was used to derive the veiling throughout the wide spectral range.
The impact of veiling on the estimates of extinction was accounted for. Deriving extinction on the 
basis of well-flux-calibrated spectral templates alone may lead to underestimation of \Av\ by up to $\sim$30\%, 
if veiling is neglected. Best matching the spectral type of templates and YSOs reduces errors in the 
\Av\ estimates.
Simultaneously deriving veiling, stellar parameters, and \vsini\ avoids degeneracy, which in turn allows 
more accurate stellar parameters to be obtained.

A large number of emission line diagnostics were used to calculate  the accretion luminosity in the seven CTTs via accretion
luminosity versus line luminosity 
relationships. We confirmed that the 
average \Lacc\  derived from several diagnostics measured simultaneously has a significantly reduced error. 

We therefore conclude that the GHOsT data sets and the procedures adopted here yield more robust results 
on the stellar and accretion parameters than those in previous studies of Taurus CTTs. However, in the 
case of extremely veiled objects, our procedures may fail or provide uncertain results, although this 
type of object is expected to be rare.

Assuming magnetospheric accretion, we calculated the mass accretion rate of the CTTs in the sample, 
confirming the high-levels of accretion in these objects. 
Being in its bright stage during the GIARPS observations, RW\,Aur\,A is the most actively accreting object 
in this sample, which is consistent with the high-accretion activity scenario by \citet[][]{takami20}.
The apparently least active objects in the sample are CQ\,Tau and HN\,Tau. We identify these two
objects as subluminous on the HR diagram, the former because of its UX\,Ori-type variability
and the latter because of the high inclination of its disk. Correction for disk obscuration 
makes HN\,Tau one of the most actively accreting objects in the sample, at a level close to
RW\,Aur\,A.

A comparison of the accretion properties of the Taurus CTTs with those of Lupus YSOs yields the
following results: the Taurus CTTs have values of \Lacc/\Lstar\ of between 0.3 and 1.5, which are normally higher 
than those of the Lupus YSOs. The two Taurus objects classified as transitional disks, namely RY\,Tau and
CQ\,Tau, have values similar to those of the Lupus transitional disks.
The five CTTs with the highest \Lacc/\Lstar\ ratios, namely DG\,Tau, DL\,Tau, HN\,Tau\,A, DO\,Tau, and RW Aur\,A,
tend to follow the upper envelope of the \Macc\ -- \Mstar\ relationship for the Lupus population, 
and have accretion rates comparable to those of the strongest accretors in Lupus.
 
The NIR veiling increases with wavelength in all the studied CTTs. The analysis of this behavior shows 
that these CTTs display a significant continuum excess emission in the NIR. In some cases, such excess
can be ascribed to the thermal emission from the inner rim of the dusty disk, while in others 
may be more compatible with emission from a thick gaseous disk inside the dust sublimation radius. 
The origin of the jets studied in \papI\ is compatible with the latter possibility.

\begin{acknowledgements}
We very much thank the anonymous referee for her/his comments and suggestions.
This work has been supported by PRIN-INAF-MAIN-STREAM 2017 "Protoplanetary disks seen 
through the eyes of new-generation instruments" and by PRIN-INAF 2019 "Spectroscopically tracing the
disk dispersal evolution (STRADE)". We warmly thank the GAPS team for sharing the solar spectrum
acquired with GIARPS, and the TNG personnel for their help during the observations.
We dedicate this work to Mr.~G.Atusino, whose premature loss deeply saddened all of us; 
for his immense presence, unconditional support and long-life great friendship. 
This research made use of the SIMBAD database, operated at the CDS (Strasbourg, France). 
This work has made use of data from the European Space Agency (ESA) mission
{\it Gaia} (https://www.cosmos.esa.int/gaia), processed by the {\it Gaia}
Data Processing and Analysis Consortium (DPAC, https://www.cosmos.esa.int/web/gaia/dpac/consortium). 
Funding for the DPAC has been provided by national institutions, in particular the institutions
participating in the {\it Gaia} Multilateral Agreement.
\end{acknowledgements}



\begin{appendix}


\section{Measurements of continuum fluxes for the CTTs and templates}
\label{Ext_Av_veil}

In Tables~\ref{veil_anal_ry} to \ref{veil_anal_cq} we report the observed continuum fluxes for the CTTs and 
the spectral templates adopted  for the determination of the extinction according to the 
methods  described in \citet[][]{fischer11}. We define the quantity: 

\begin{equation}
\Gamma_\lambda \equiv 2.5 \cdot \log {\rm [(1 + r_\lambda) \cdot \frac{F^T_\lambda}{F^O_\lambda}] }
,\end{equation}

\noindent
where ${\rm r_\lambda}$, ${\rm F^T_\lambda}$, and ${\rm F^O_\lambda}$ are the veiling,  the observed continuum flux of 
the spectral template, and the observed continuum flux of the object, as a function of wavelength, respectively.
As mentioned in Sect.~\ref{extinction}, the method is based on the fact that

\vspace{0.5cm}
\begin{equation}
\Gamma_\lambda = {\rm (A_v^O - A_v^T) \cdot  \frac{\rm A_\lambda}{\rm A_v}  - 2.5 \cdot \log{C} }
,\end{equation}

\noindent
so that $\Gamma_\lambda$ is a lineaar function of $\frac{\rm A_\lambda}{\rm A_v}$,  where ${\rm A_v^O}$ and ${\rm A_v^T}$ 
are the visual extinction of the the object and template, respectively, and $C$ is a constant. 
The tables also provide  $\Gamma_\lambda$ as a function of wavelength and the  r$_\lambda$ values are also included 
for convenience.

\setlength{\tabcolsep}{2pt}
\begin{table}[ht]
\caption[]{\label{veil_anal_ry} $\Gamma_\lambda$  versus A$_\lambda$/A$_{\rm v}$ for RY\,Tau.}
\begin{tabular}{r|c|c|c|c|c}
\hline \hline
  Wavelength  &  A$_\lambda$/\Av  & $ F^T_\lambda $  & $ F^O_\lambda $  & r$_\lambda$ & $\Gamma_\lambda$  \\
 (nm)   &      &  erg/s/cm$^2$/nm & erg/s/cm$^2$/nm  &   &  \\

 \hline
   &      &   &  &   &   \\
  450.0 & 1.169 &  1.70e-11 & 2.00e-12 & 0.0 & 2.324 \\
  500.0 & 1.080 &  2.10e-11 & 2.20e-12 & 0.0 & 2.449 \\
  550.0 & 0.990 &  2.10e-11 & 2.60e-12 & 0.0 & 2.268 \\
  600.0 & 0.900 &  1.90e-11 & 3.00e-12 & 0.0 & 2.004 \\
  650.0 & 0.820 &  1.80e-11 & 3.20e-12 & 0.0 & 1.875 \\
  968.0 & 0.450 &  8.60e-12 & 5.00e-12 & 0.2 & 0.787 \\
  983.0 & 0.430 &  8.40e-12 & 5.10e-12 & 0.6 & 1.052 \\
 1178.0 & 0.300 &  5.60e-12 & 5.30e-12 & 0.4 & 0.425 \\
 1256.0 & 0.264 &  4.90e-12 & 5.10e-12 & 0.7 & 0.533 \\
 1298.0 & 0.250 &  4.60e-12 & 4.90e-12 & 0.3 & 0.216 \\
 1565.0 & 0.178 &  2.90e-12 & 3.50e-12 & 0.5 & 0.236 \\
 1597.0 & 0.173 &  2.80e-12 & 3.30e-12 & 0.4 & 0.187 \\
 1666.0 & 0.163 &  2.40e-12 & 2.90e-12 & 0.5 & 0.235 \\
 1741.0 & 0.151 &  2.00e-12 & 2.40e-12 & 0.9 & 0.499 \\
 2130.0 & 0.114 &  1.00e-12 & 1.60e-12 & 1.4 & 0.440 \\
 2255.0 & 0.110 &  8.30e-13 & 2.20e-12 & 1.2 & 0.202 \\
 2322.0 & 0.100 &  7.50e-13 & 2.80e-12 & ... & ...   \\
 & & & & &  \\
\hline
\end{tabular}
\end{table}

\setlength{\tabcolsep}{2pt}
\begin{table}
\caption[]{\label{veil_anal_dg}  $\Gamma_\lambda$  versus A$_\lambda$/A$_{\rm v}$ for DG\,Tau.}
\begin{tabular}{r|c|c|c|c|c}
\hline \hline
  Wavelength  &  A$_\lambda$/\Av & $ F^T_\lambda $  & $ F^O_\lambda $  & r$_\lambda$ & $\Gamma_\lambda$  \\
 (nm)   &      &  erg/s/cm$^2$/nm & erg/s/cm$^2$/nm  &   &  \\

 \hline
   &      &   &   &  & \\
  450.0 & 1.169 & 4.20e-14 & 2.20e-13 & ... &   ...  \\
  500.0 & 1.080 & 4.70e-14 & 2.40e-13 & ... &   ...  \\
  550.0 & 0.990 & 6.90e-14 & 3.20e-13 & 2.0 & -0.473 \\
  600.0 & 0.900 & 7.70e-14 & 3.90e-13 & 1.5 & -0.767 \\
  650.0 & 0.820 & 8.60e-14 & 4.60e-13 & 1.0 & -1.068 \\
  968.0 & 0.450 & 1.00e-13 & 9.60e-13 & 1.6 & -1.418 \\
  983.0 & 0.430 & 9.00e-14 & 9.80e-13 & 1.2 & -1.736 \\
 1178.0 & 0.300 & 7.00e-14 & 1.10e-12 & 1.8 & -1.873 \\
 1256.0 & 0.264 & 6.70e-14 & 1.10e-12 & 1.4 & -2.088 \\
 1298.0 & 0.250 & 6.50e-14 & 1.10e-12 & 1.7 & -1.993 \\
 1565.0 & 0.178 & 5.50e-14 & 1.00e-12 & 1.8 & -2.031 \\
 1597.0 & 0.173 & 5.40e-14 & 9.90e-13 & 1.6 & -2.121 \\
 1666.0 & 0.163 & 5.10e-14 & 9.60e-13 & 1.7 & -2.108 \\
 1741.0 & 0.151 & 4.50e-14 & 9.30e-13 & 2.4 & -1.959 \\
 2130.0 & 0.114 & 2.70e-14 & 7.90e-13 & 3.0 & -2.161 \\
 2255.0 & 0.110 & 2.20e-14 & 7.70e-13 & 4.0 & -2.113 \\
 2322.0 & 0.100 & 2.00e-14 & 7.70e-13 & ... &    ... \\
 & & & & & \\
\hline
\end{tabular}
\end{table}

\setlength{\tabcolsep}{2pt}
\begin{table}
\caption[]{\label{veil_anal_dl}  $\Gamma_\lambda$  versus A$_\lambda$/A$_{\rm v}$ for DL\,Tau.}
\begin{tabular}{r|c|c|c|c|c}
\hline \hline
  Wavelength  &  A$_\lambda$/\Av & $ F^T_\lambda $  & $ F^O_\lambda $  & r$_\lambda$ & $\Gamma_\lambda$  \\
 (nm)   &      &  erg/s/cm$^2$/nm & erg/s/cm$^2$/nm  &   &  \\

 \hline
   &      &   &   &  & \\
  450.0 &  1.169  & 6.10e-13  & 1.40e-13  &  3.0 & 3.103  \\
  500.0 &  1.080  & 6.20e-13  & 1.50e-13  &  2.5 & 2.901  \\
  550.0 &  0.990  & 8.90e-13  & 2.00e-13  &  2.0 & 2.814  \\
  600.0 &  0.900  & 9.00e-13  & 2.50e-13  &  1.5 & 2.386  \\
  650.0 &  0.820  & 9.60e-13  & 2.90e-13  &  1.5 & 2.295  \\
  968.0 &  0.450  & 9.50e-13  & 4.90e-13  &  1.1 & 1.524  \\
  983.0 &  0.430  & 1.00e-12  & 4.90e-13  &  1.1 & 1.580  \\
 1178.0 &  0.300  & 1.10e-12  & 4.70e-13  &  1.1 & 1.729  \\
 1256.0 &  0.264  & 1.00e-12  & 4.70e-13  &  1.0 & 1.572  \\
 1298.0 &  0.250  & 9.70e-13  & 4.60e-13  &  1.0 & 1.563  \\
 1565.0 &  0.178  & 8.60e-13  & 4.20e-13  &  0.9 & 1.475  \\
 1597.0 &  0.173  & 8.00e-13  & 4.10e-13  &  0.6 & 1.236  \\
 1666.0 &  0.163  & 7.80e-13  & 4.00e-13  &  0.8 & 1.363  \\
 1741.0 &  0.151  & 6.90e-13  & 3.90e-13  &  1.7 & 1.698  \\
 2130.0 &  0.114  & 3.70e-13  & 2.90e-13  &  2.5 & 1.625  \\
 2255.0 &  0.110  & 3.10e-13  & 2.60e-13  &  2.1 & 1.419  \\
 2322.0 &  0.100  & 2.70e-13  & 2.40e-13  &  2.5 & 1.488  \\
 & & & & & \\
\hline
\end{tabular}
\end{table}

\setlength{\tabcolsep}{2pt}
\begin{table}
\caption[]{\label{veil_anal_hn}  $\Gamma_\lambda$  versus A$_\lambda$/A$_{\rm v}$ for HN\,Tau\,A.}
\begin{tabular}{r|c|c|c|c|c}
\hline \hline
  Wavelength  &  A$_\lambda$/\Av & $ F^T_\lambda $  & $ F^O_\lambda $  & r$_\lambda$ & $\Gamma_\lambda$  \\
 (nm)   &      &  erg/s/cm$^2$/nm & erg/s/cm$^2$/nm  &   &  \\

 \hline
   &      &   &   &  & \\

  450.0  & 1.169  & 7.50e-13  & 6.40e-14  &  ...  &   ...    \\
  500.0  & 1.080  & 7.60e-13  & 7.40e-14  &  0.8  &   3.167  \\
  550.0  & 0.990  & 8.60e-13  & 8.80e-14  &  0.8  &   3.113  \\
  600.0  & 0.900  & 9.00e-13  & 1.00e-13  &  0.8  &   3.024  \\
  650.0  & 0.820  & 8.70e-13  & 1.10e-13  &  0.5  &   2.686  \\
  968.0  & 0.450  & 5.90e-13  & 1.50e-13  &  ...  &     ...  \\
  983.0  & 0.430  & 5.80e-13  & 1.50e-13  &  ...  &     ...  \\
 1178.0  & 0.300  & 4.60e-13  & 1.50e-13  &  1.1  &   2.022  \\
 1256.0  & 0.264  & 4.00e-13  & 1.40e-13  &  1.2  &   1.996  \\
 1298.0  & 0.250  & 3.80e-13  & 1.40e-13  &  1.5  &   2.079  \\
 1565.0  & 0.178  & 3.00e-13  & 1.40e-13  &  1.8  &   1.945  \\
 1597.0  & 0.173  & 2.90e-13  & 1.40e-13  &  1.6  &   1.828  \\
 1666.0  & 0.163  & 2.70e-13  & 1.30e-13  &  1.3  &   1.698  \\
 1741.0  & 0.151  & 2.20e-13  & 1.30e-13  &  2.3  &   1.867  \\
 2130.0  & 0.114  & 1.20e-13  & 1.30e-13  &  3.3  &   1.497  \\
 2255.0  & 0.110  & 1.00e-13  & 1.20e-13  &  5.0  &   1.747  \\
 2322.0  & 0.100  & 8.30e-14  & 1.20e-13  &  ...  &     ...  \\
 & & & & & \\
\hline
\end{tabular}
\end{table}

\setlength{\tabcolsep}{2pt}
\begin{table}[ht]
\caption[]{\label{veil_anal_do}  $\Gamma_\lambda$ versus A$_\lambda$/A$_{\rm v}$ for DO\,Tau.}
\begin{tabular}{r|c|c|c|c|c}
\hline \hline
  Wavelength  &  A$_\lambda$/\Av & $ F^T_\lambda $  & $ F^O_\lambda $  & r$_\lambda$ & $\Gamma_\lambda$  \\
 (nm)   &      &  erg/s/cm$^2$/nm & erg/s/cm$^2$/nm  &   &  \\

 \hline
   &      &   &   &  & \\

  450.0  & 1.169  & 4.60e-14  & 1.50e-13  &  1.8  &  -0.165  \\
  500.0  & 1.080  & 5.30e-14  & 1.60e-13  &  1.5  &  -0.205  \\
  550.0  & 0.990  & 7.10e-14  & 1.80e-13  &  1.5  &  -0.015  \\
  600.0  & 0.900  & 7.80e-14  & 2.10e-13  &  1.0  &  -0.323  \\
  650.0  & 0.820  & 8.70e-14  & 2.30e-13  &  1.5  &  -0.061  \\
  968.0  & 0.450  & 9.90e-14  & 5.00e-13  &  0.5  &  -1.318  \\
  983.0  & 0.430  & 8.90e-14  & 5.10e-13  &  0.6  &  -1.385  \\
 1178.0  & 0.300  & 7.10e-14  & 5.80e-13  &  0.7  &  -1.704  \\
 1256.0  & 0.264  & 6.60e-14  & 6.00e-13  &  1.4  &  -1.446  \\
 1298.0  & 0.250  & 6.40e-14  & 6.00e-13  &  1.4  &  -1.479  \\
 1565.0  & 0.178  & 5.50e-14  & 6.10e-13  &  1.9  &  -1.456  \\
 1597.0  & 0.173  & 5.50e-14  & 6.10e-13  &  1.7  &  -1.534  \\
 1666.0  & 0.163  & 5.20e-14  & 6.00e-13  &  1.8  &  -1.537  \\
 1741.0  & 0.151  & 4.70e-14  & 5.90e-13  &  2.1  &  -1.518  \\
 2130.0  & 0.114  & 2.70e-14  & 4.90e-13  &  3.4  &  -1.538  \\
 2255.0  & 0.110  & 2.10e-14  & 4.50e-13  &  3.7  &  -1.647  \\
 2322.0  & 0.100  & 1.90e-14  & 4.30e-13  &  2.7  &  -1.966  \\
  & & & & & \\
\hline
\end{tabular}
\end{table}

\setlength{\tabcolsep}{2pt}
\begin{table}
\caption[]{\label{veil_anal_rw}  $\Gamma_\lambda$  versus A$_\lambda$/A$_{\rm v}$ for RW\,Aur\,A.}
\begin{tabular}{r|c|c|c|c|c}
\hline \hline
  Wavelength  &  A$_\lambda$/\Av & $ F^T_\lambda $  & $ F^O_\lambda $  & r$_\lambda$ & $\Gamma_\lambda$  \\
 (nm)   &      &  erg/s/cm$^2$/nm & erg/s/cm$^2$/nm  &   &  \\

 \hline
   &      &   &   &  & \\

  450.0  & 1.169  & 7.70e-13  & 2.20e-12  & ...  &   ...  \\
  500.0  & 1.080  & 7.50e-13  & 2.30e-12  & ...  &   ...  \\
  550.0  & 0.990  & 7.60e-13  & 2.20e-12  &  1.2 & -0.298  \\
  600.0  & 0.900  & 7.10e-13  & 2.20e-12  & ...  &   ...  \\
  650.0  & 0.820  & 6.70e-13  & 2.30e-12  & ...  &   ...  \\
  968.0  & 0.450  & 3.80e-13  & 1.80e-12  & ...  &   ...  \\
  983.0  & 0.430  & 3.70e-13  & 1.80e-12  & ...  &   ...  \\
 1178.0  & 0.300  & 2.60e-13  & 1.40e-12  & ...  &   ...  \\
 1256.0  & 0.264  & 2.30e-13  & 1.30e-12  &  1.5 & -0.886  \\
 1298.0  & 0.250  & 2.20e-13  & 1.30e-12  & ...  &   ...  \\
 1565.0  & 0.178  & 1.60e-13  & 1.00e-12  & ...  &   ...  \\
 1597.0  & 0.173  & 1.50e-13  & 1.00e-12  &  2.5 & -0.700  \\
 1666.0  & 0.163  & 1.30e-13  & 9.50e-13  &  2.7 & -0.739  \\
 1741.0  & 0.151  & 1.10e-13  & 9.10e-13  &  3.5 & -0.661  \\
 2130.0  & 0.114  & 5.50e-14  & 6.60e-13  &  3.6 & -1.041  \\
 2255.0  & 0.110  & 4.40e-14  & 5.70e-13  &  5.7 & -0.716  \\
 2322.0  & 0.100  & 3.70e-14  & 5.10e-13  &  5.4 & -0.833  \\
  & & & & & \\
\hline
\end{tabular}
\end{table}

\setlength{\tabcolsep}{2pt}
\begin{table}
\caption[]{\label{veil_anal_cq}  $\Gamma_\lambda$ versus A$_\lambda$/A$_{\rm v}$ for CQ\,Tau.}
\begin{tabular}{r|c|c|c|c|c}
\hline \hline
  Wavelength  &  A$_\lambda$/\Av & $ F^T_\lambda $  & $ F^O_\lambda $  & r$_\lambda$ & $\Gamma_\lambda$  \\
 (nm)   &      &  erg/s/cm$^2$/nm & erg/s/cm$^2$/nm  &   &  \\

 \hline
   &      &   &   &  & \\
  450.0  & 1.169  & 7.90e-12  & 4.90e-12  &  0.0   & 0.519   \\
  500.0  & 1.080  & 7.00e-12  & 4.60e-12  &  0.0   & 0.456   \\
  550.0  & 0.990  & 6.00e-12  & 4.30e-12  &  0.0   & 0.362   \\
  600.0  & 0.900  & 5.30e-12  & 3.90e-12  &  0.0   & 0.333   \\
  650.0  & 0.820  & 4.50e-12  & 3.40e-12  &  0.0   & 0.304   \\
  968.0  & 0.450  & 1.90e-12  & 2.60e-12  &  ...   &  ...    \\
  983.0  & 0.430  & 1.80e-12  & 2.50e-12  &  0.5   & 0.084   \\
 1178.0  & 0.300  & 1.10e-12  & 2.00e-12  &  1.0   & 0.103   \\
 1256.0  & 0.264  & 9.50e-13  & 1.90e-12  &  0.8   & -0.114  \\
 1298.0  & 0.250  & 8.60e-13  & 1.80e-12  &  ...   &   .....  \\
 1565.0  & 0.178  & 5.00e-13  & 1.50e-12  & $<$3.0 & $<$0.312 \\
 1597.0  & 0.173  & 4.80e-13  & 1.50e-12  &  2.0   & -0.044 \\
 1666.0  & 0.163  & 4.10e-13  & 1.50e-12  &  ...   &  ...    \\
 1741.0  & 0.151  & 3.50e-13  & 1.40e-12  &  ...   &  ...    \\
 2130.0  & 0.114  & 1.70e-13  & 1.20e-12  &  ...   &  ...    \\
 2255.0  & 0.110  & 1.30e-13  & 1.10e-12  &  ...   &  ...    \\
 2322.0  & 0.100  & 1.20e-13  & 1.10e-12  &  ...   &  ...    \\
  & & & & & \\
\hline
\end{tabular}
\end{table}

 

\section{Individual fluxes, equivalent widths and \Lacc\ estimates}
\label{line_fluxes_EWs_Lacc}

 Tables~\ref{tab:fluxes_EWs_Had} to \ref{tab:fluxes_EW_CaI3934} report the observed fluxes, equivalent widths,
 for every CTTs in the sample, as well as the corresponding \Lacc\ values derived from the individual accretion 
 diagnostics and using the \Lacc\---\Ll\ relationships by \citet[][]{alcala17}. 

\onecolumn

\setlength{\tabcolsep}{5pt}

\begin{landscape}
\begin{longtable}{l|c|c|c|c|c|c|c|c}
\caption[ ]{\label{tab:fluxes_EWs_Had} Measured fluxes and equivalent widths of Balmer lines and accretion luminosity for the CTTs sample: H$\alpha$ (H3) to H$\delta$ (H6).}\\
\hline\hline
Object & $f_{\rm H\alpha}$ &  $EW_{\rm H\alpha}$&$f_{\rm H\beta}$& $EW_{\rm H\beta}$&$f_{\rm H\gamma}$& $EW_{\rm H\gamma}$&$f_{\rm H\delta}$& $EW_{\rm H\delta}$\\
    & (erg\,s$^{-1}$\,cm$^{-2}$)&  (\AA)             &(erg\,s$^{-1}$\,cm$^{-2}$)& (\AA)           &(erg\,s$^{-1}$\,cm$^{-2}$)& (\AA)         &(erg\,s$^{-1}$\,cm$^{-2}$)& (\AA)       \\
\hline
&&&&&&&& \\
RY\,Tau     &  4.78($\pm$0.22)e$-$12  &  $-$14.49$\pm$0.54  & 3.30($\pm$0.56)e$-$13  &  $-$1.72$\pm$0.13   & 2.44($\pm$0.30)e$-$13   &  $-$1.36$\pm$0.20   & 1.54($\pm$0.46)e$-$13   &  $-$0.95$\pm$0.14   \\ 
DG\,Tau     &  6.30($\pm$0.09)e$-$12  & $-$108.43$\pm$7.28  & 1.01($\pm$0.06)e$-$12  & $-$43.72$\pm$8.80   & 5.02($\pm$0.35)e$-$13   & $-$33.92$\pm$8.01   & 3.20($\pm$0.27)e$-$13   & $-$24.81$\pm$6.66   \\ 
DL\,Tau     &  3.19($\pm$0.06)e$-$12  &  $-$91.93$\pm$6.09  & 5.41($\pm$0.30)e$-$13  & $-$39.20$\pm$6.85   & 3.73($\pm$0.33)e$-$13   & $-$33.30$\pm$9.29   & 2.65($\pm$0.21)e$-$13   & $-$30.15$\pm$7.32   \\ 
HN\,Tau\,A  &  1.42($\pm$0.02)e$-$12  & $-$113.60$\pm$7.57  & 2.39($\pm$0.10)e$-$13  & $-$32.39$\pm$4.53   & 1.36($\pm$0.13)e$-$13   & $-$25.86$\pm$6.27   & 9.25($\pm$0.96)e$-$14   & $-$19.64$\pm$5.24    \\ 
DO\,Tau     &  1.92($\pm$0.07)e$-$17  &  $-$64.87$\pm$6.99  & 3.10($\pm$0.29)e$-$13  & $-$19.87$\pm$4.51   & 2.03($\pm$0.39)e$-$13   & $-$12.69$\pm$8.19   & 1.51($\pm$0.39)e$-$13   & $-$10.13$\pm$8.38   \\ 
RW\,Aur\,A  &  1.74($\pm$0.02)e$-$11  &  $-$69.88$\pm$2.47  & 2.28($\pm$0.20)e$-$12  & $-$10.18$\pm$1.29   &         ...             &        ...         &         ...            &        ...         \\ 
CQ\,Tau     &  2.29($\pm$0.18)e$-$12  &  $-$7.53$\pm$0.42   & 3.09($\pm$0.45)e$-$13  &  $-$0.83$\pm$0.17   & 2.18($\pm$0.29)e$-$13   &  $-$0.56$\pm$0.32   & 1.29($\pm$0.31)e$-$13   &  $-$0.29$\pm$0.10  \\ 
&&&&&&&& \\

\hline  
\hline  
&&&&&&&& \\

Object & $\log$(\Lacc/\Lsun)  &  $\pm\sigma$  & $\log$(\Lacc/\Lsun)& $\pm\sigma$ &$\log$(\Lacc/\Lsun) & $\pm\sigma$ &$\log$(\Lacc/\Lsun) & $\pm\sigma$ \\
       & ${\rm H\alpha}$  & (dex)            & ${\rm H\beta}$ &    (dex)     & ${\rm H\gamma}$ &   (dex)    & ${\rm H\delta}$ &   (dex)   \\
\hline
&&&&&&&& \\ 
RY\,Tau     &  $-$0.33  &  0.15  &  $-$0.51  &  0.18  & $-$0.38  & 0.15 & $-$0.49  &  0.21   \\
DG\,Tau     &  $-$0.45  &  0.15  &  $-$0.29  &  0.16  & $-$0.38  & 0.14 & $-$0.50  &  0.16   \\
DL\,Tau     &  $-$0.67  &  0.16  &  $-$0.52  &  0.16  & $-$0.46  & 0.15 & $-$0.53  &  0.16   \\
HN\,Tau\,A  &  $-$1.23   & 0.17  &  $-$1.08   & 0.17  & $-$1.10  & 0.16 & $-$1.17  &  0.18   \\
DO\,Tau     &  $-$0.98  &  0.17  &  $-$0.84  &  0.17  & $-$0.78  & 0.17 & $-$0.82  &  0.21   \\
RW\,Aur\,A  &  $-$0.29  &  0.14  &  $-$0.38  &  0.16  &   ...    &  ... &   ...   &    ...  \\
CQ\,Tau     &  $-$1.20  &  0.17  &  $-$1.29   & 0.17  & $-$1.23  & 0.17 & $-$1.37  &  0.21   \\

\hline  
                                                  
\end{longtable}
\end{landscape}

\setlength{\tabcolsep}{5pt}

\begin{landscape}
\begin{longtable}{l|c|c|c|c|c|c|c|c}
\caption[ ]{\label{tab:fluxes_EWs_Pabe} Measured fluxes and equivalent widths of Paschen lines and accretion luminosity for the CTTs sample: \pab\ (Pa5) to Pa$\epsilon$ (Pa8).}\\
\hline\hline
Object & $f_{\rm Pa\beta}$ &  $EW_{\rm Pa\beta}$&$f_{\rm Pa\gamma}$& $EW_{\rm Pa\gamma}$&$f_{\rm Pa\delta}$& $EW_{\rm Pa\delta}$&$f_{\rm Pa\epsilon}$& $EW_{\rm Pa\epsilon}$ \\
    & (erg\,s$^{-1}$\,cm$^{-2}$)&  (\AA)             &(erg\,s$^{-1}$\,cm$^{-2}$)& (\AA)           &(erg\,s$^{-1}$\,cm$^{-2}$)& (\AA)         &(erg\,s$^{-1}$\,cm$^{-2}$)& (\AA)       \\
\hline
&&&&&&&& \\           
RY\,Tau     &  1.20($\pm$0.08)e$-$12  &  $-$2.66$\pm$0.32  & 4.92($\pm$1.16)e$-$13  &  $-$1.05$\pm$0.35   & 3.71($\pm$0.71)e$-$13   &  $-$0.80$\pm$0.22   &       ...              &       ...        \\ 
DG\,Tau     &  1.98($\pm$0.07)e$-$12  & $-$18.50$\pm$0.98  & 1.40($\pm$0.02)e$-$12  & $-$13.33$\pm$0.83   & 1.00($\pm$0.06)e$-$12   & $-$10.15$\pm$0.89   & 8.01($\pm$1.01)e$-$13  & $-$7.93$\pm$1.67   \\ 
DL\,Tau     &  1.03($\pm$0.07)e$-$12  & $-$22.01$\pm$2.15  & 6.86($\pm$0.53)e$-$13  & $-$14.29$\pm$1.67   & 5.37($\pm$0.06)e$-$13   & $-$10.76$\pm$1.72   & 4.35($\pm$0.81)e$-$13  & $-$8.19$\pm$2.43   \\ 
HN\,Tau\,A  &  2.28($\pm$0.12)e$-$13  & $-$16.06$\pm$1.26  & 1.52($\pm$0.17)e$-$13  & $-$10.41$\pm$1.56   & 9.80($\pm$2.40)e$-$14   &  $-$6.67$\pm$2.13   &      ...               &        ...        \\ 
DO\,Tau     &  3.97($\pm$0.33)e$-$13  &  $-$6.64$\pm$0.75  & 2.16($\pm$0.27)e$-$13  &  $-$3.85$\pm$0.65   & 1.42($\pm$0.03)e$-$13   &  $-$2.80$\pm$0.76   &       ...              &       ...         \\ 
RW\,Aur\,A  &  2.80($\pm$0.10)e$-$12  & $-$21.37$\pm$1.24  & 1.91($\pm$0.11)e$-$12  & $-$12.99$\pm$1.12   & 1.80($\pm$0.15)e$-$12   & $-$10.71$\pm$1.26    & 1.13($\pm$0.25)e$-$12  &  $-$6.46$\pm$1.88  \\ 
CQ\,Tau     &  3.02($\pm$0.40)e$-$13  &  $-$1.66$\pm$0.38  &         ...           &        ...         &         ...             &        ...         &       ...             &       ...          \\ 
&&&&&&&& \\

\hline  
\hline  

&&&&&&&& \\

Object & $\log$(\Lacc/\Lsun)  &  $\pm\sigma$  & $\log$(\Lacc/\Lsun)& $\pm\sigma$ &$\log$(\Lacc/\Lsun)& $\pm\sigma$ &$\log$(\Lacc/\Lsun)& $\pm\sigma$  \\
       & ${\rm Pa\beta}$  & (dex)            & ${\rm Pa\gamma}$ &    (dex)     & ${\rm Pa\delta}$ &   (dex)           & ${\rm Pa\epsilon}$ &        (dex)   \\
\hline
&&&&&&&& \\
RY\,Tau     &  $-$0.36  &  0.24  &  $-$0.45  & 0.27  & $-$0.31  & 0.34  &   ...    & ...  \\
DG\,Tau     &  $-$0.25  &  0.23  &  $-$0.05  & 0.22  & $+$0.04  & 0.30  & $-$0.18  & 0.40 \\
DL\,Tau     &  $-$0.36  &  0.23  &  $-$0.22  & 0.23  & $-$0.08  & 0.31  & $-$0.30  & 0.41 \\
HN\,Tau\,A  &  $-$1.22   & 0.28  &  $-$1.22  & 0.27  & $-$1.17  & 0.40  &   ...    & ...  \\
DO\,Tau     &  $-$0.91  &  0.27  &  $-$1.00  & 0.26  & $-$0.90  & 0.38  &   ...    & ...  \\
RW\,Aur\,A  &  $+$0.19  &  0.21  &  $+$0.36  & 0.21  & $+$0.56  & 0.28  & $+$0.16  & 0.38 \\
CQ\,Tau     &  $-$1.08  &  0.28  &    ...   &  ...  &    ...   & ...   &   ...    & ...   \\

\hline                                                    
\end{longtable}
\end{landscape}

\setlength{\tabcolsep}{5pt}

 \begin{landscape}
\begin{longtable}{l|c|c|c|c|c|c|c|c}
\caption[ ]{\label{tab:fluxes_EWs_HeI1} Measured fluxes and equivalent widths of He\,{\sc i} lines and accretion luminosity for the CTTs sample: He\,{\sc i}4026 to He\,{\sc i}4922.}\\
\hline\hline
Object & $f_{\rm \ion{He}{i}~\lambda4026}$ &  $EW_{\rm \ion{He}{i}~\lambda4026}$&$f_{\rm \ion{He}{i}~\lambda4471}$& $EW_{\rm \ion{He}{i}~\lambda4471}$&$f_{\rm \ion{He}{i}~\lambda4713}$& $EW_{\rm \ion{He}{i}~\lambda4713}$&$f_{\rm \ion{He}{i}~\lambda4922}$& $EW_{\rm \ion{He}{i}~\lambda4922}$ \\
    & (erg\,s$^{-1}$\,cm$^{-2}$)&  (\AA)             &(erg\,s$^{-1}$\,cm$^{-2}$)& (\AA)           &(erg\,s$^{-1}$\,cm$^{-2}$)& (\AA)         &(erg\,s$^{-1}$\,cm$^{-2}$)& (\AA)       \\
\hline
&&&&&&&& \\           
RY\,Tau     &  1.02($\pm$0.51)e$-$14  &  $-$0.07$\pm$0.06  &       ...             &       ...        &         ...           &          ...     & 5.28($\pm$2.56)e$-$14  & $-$0.25$\pm$0.05  \\ 
DG\,Tau     &  3.89($\pm$3.08)e$-$14  &  $-$3.98$\pm$1.56  & 9.36($\pm$0.90)e$-$14  & $-$5.23$\pm$0.75  &         ...          &          ...     &      ...               &          ...     \\ 
DL\,Tau     &  4.54($\pm$2.60)e$-$14  & $-$12.04$\pm$1.15  & 6.99($\pm$3.52)e$-$14  & $-$6.59$\pm$3.30  & 1.35($\pm$1.07)e$-$14 & $-$1.18$\pm$1.00  & 9.19($\pm$1.86)e$-$14  & $-$6.81$\pm$2.28   \\ 
HN\,Tau\,A  &  1.61($\pm$0.74)e$-$14  &  $-$4.51$\pm$3.50  & 2.73($\pm$0.89)e$-$14  & $-$5.35$\pm$2.47  &         ...           &           ...    & 3.70($\pm$1.04)e$-$14  & $-$5.55$\pm$2.26  \\ 
DO\,Tau     &  1.22($\pm$0.43)e$-$14  &  $-$1.27$\pm$0.82  & 2.93($\pm$0.69)e$-$14  & $-$1.83$\pm$0.80  & 7.47($\pm$5.83)e$-$15 & $-$0.56$\pm$0.50  & 4.08($\pm$1.23)e$-$14  & $-$3.00$\pm$1.43  \\ 
RW\,Aur\,A  &  5.41($\pm$1.63)e$-$13  &  $-$2.79$\pm$0.99  & 4.48($\pm$1.99)e$-$13  &  $-$2.18$\pm$1.13 & 1.71($\pm$0.35)e$-$13  & $-$0.85$\pm$0.18 & 8.33($\pm$1.20)e$-$13  &  $-$3.97$\pm$0.81 \\ 
CQ\,Tau     &       ...              &       ...         &         ...           &        ...        &         ...          &        ...       &       ...             &       ...        \\ 
&&&&&&&& \\

\hline  
\hline  

&&&&&&&& \\

Object & $\log$(\Lacc/\Lsun)  &  $\pm\sigma$  & $\log$(\Lacc/\Lsun)& $\pm\sigma$ &$\log$(\Lacc/\Lsun)& $\pm\sigma$ &$\log$(\Lacc/\Lsun)& $\pm\sigma$  \\
       & ${\ion{He}{i}~\lambda4026}$  & (dex)        & ${\ion{He}{i}~\lambda4471}$ &    (dex)     & ${\ion{He}{i}~\lambda4713}$ &   (dex) & ${\ion{He}{i}~\lambda4922}$ & (dex)   \\
\hline
&&&&&&&& \\
RY\,Tau     &  $-$0.64  & 0.30  &    ...    & ...   &    ...   & ...  &  $-$0.33  & 0.27  \\
DG\,Tau     &  $-$0.37  & 0.41  &  $-$0.20  & 0.18   &    ...  & ...  &    ...  &   ... \\
DL\,Tau     &  $-$0.25  & 0.32  &  $-$0.27  & 0.29  & $-$0.73  & 0.46  &  $-$0.32  & 0.19 \\
HN\,Tau\,A  &  $-$0.87  & 0.29  &  $-$0.85  & 0.25  &    ...   &  ...  &  $-$0.84  & 0.22  \\
DO\,Tau     &  $-$0.87  & 0.26  &  $-$0.70  & 0.22  & $-$0.98  & 0.46  &  $-$0.70  & 0.22  \\
RW\,Aur\,A  &  $+$0.61  & 0.21  &  $+$0.34  & 0.26  & $+$0.04  & 0.31  &  $+$0.42  & 0.16 \\
CQ\,Tau     &  ...     &   ...  &    ...   &  ...  &    ...   & ...   &   ...    & ...   \\

\hline                                                    
\end{longtable}
\end{landscape}

\setlength{\tabcolsep}{5pt}

\begin{landscape}
\begin{longtable}{l|c|c|c|c|c|c|c|c}
\caption[ ]{\label{tab:fluxes_EWs_HeI2} Measured fluxes and equivalent widths of He\,{\sc i} lines and accretion luminosity for the CTTs sample: He\,{\sc i}5016 to He\,{\sc i}10830.}\\
\hline\hline
Object & $f_{\rm \ion{He}{i}~\lambda5016}$ &  $EW_{\rm \ion{He}{i}~\lambda5016}$&$f_{\rm \ion{He}{i}~\lambda5876}$& $EW_{\rm \ion{He}{i}~\lambda5876}$&$f_{\rm \ion{He}{i}~\lambda6678}$& $EW_{\rm \ion{He}{i}~\lambda6678}$&$f_{\rm \ion{He}{i}~\lambda10830}$& $EW_{\rm \ion{He}{i}~\lambda10830}$ \\
    & (erg\,s$^{-1}$\,cm$^{-2}$)&  (\AA)             &(erg\,s$^{-1}$\,cm$^{-2}$)& (\AA)           &(erg\,s$^{-1}$\,cm$^{-2}$)& (\AA)         &(erg\,s$^{-1}$\,cm$^{-2}$)& (\AA)       \\
\hline
&&&&&&&& \\           
RY\,Tau     &       ...              &      ...          &  1.24($\pm$0.29)e$-$13  & $-$0.45$\pm$0.10  &  2.09($\pm$0.59)e$-14$  &  $-$0.07$\pm$0.05 &  1.02($\pm$0.10)e$-$12  &  $-$0.25$\pm$0.02  \\ 
DG\,Tau     &       ...              &      ...          &  1.27($\pm$0.16)e$-$13  & $-$3.57$\pm$0.61  &  5.84($\pm$3.57)e$-14$  &  $-$1.17$\pm$0.78 &  1.65($\pm$0.07)e$-$12  &  $-$1.70$\pm$0.15  \\ 
DL\,Tau     &  2.55($\pm$0.45)e$-$14  &  $-$1.84$\pm$0.56  &  1.52($\pm$0.18)e$-$13  & $-$6.55$\pm$1.12  &  5.98($\pm$3.40)e$-14$  &  $-$2.06$\pm$1.34 &  1.21($\pm$0.06)e$-$12  &  $-$2.49$\pm$0.22  \\ 
HN\,Tau\,A  &  1.29($\pm$0.23)e$-$14  &  $-$1.88$\pm$0.45  &  4.23($\pm$0.99)e$-$14  & $-$4.76$\pm$1.44  &  1.62($\pm$0.94)e$-14$  &  $-$1.53$\pm$0.98 &  3.78($\pm$0.21)e$-$13  &  $-$2.61$\pm$0.25  \\ 
DO\,Tau     &  7.44($\pm$1.67)e$-$15  &  $-$0.55$\pm$0.18  &  6.31($\pm$0.72)e$-$14  & $-$3.02$\pm$0.57  &  2.59($\pm$0.33)e$-14$  &  $-$1.00$\pm$0.18 &  2.04($\pm$0.16)e$-$13  &  $-$0.59$\pm$0.08  \\ 
RW\,Aur\,A  &  7.22($\pm$0.72)e$-$13  &  $-$3.20$\pm$0.40  &  5.00($\pm$0.56)e$-$13  & $-$2.33$\pm$0.31  &  4.72($\pm$0.83)e$-13$  &  $-$2.21$\pm$0.45 &  3.00($\pm$0.24)e$-$12  &  $-$1.91$\pm$0.21  \\ 
CQ\,Tau     &       ...              &       ...         &  5.97($\pm$1.70)e$-$14  & $-$0.15$\pm$0.05  &  8.63($\pm$4.58)e$-15$  &  $-$0.03$\pm$0.02 &  4.92($\pm$0.67)e$-$13  &  $-$0.22$\pm$0.04   \\ 
&&&&&&&& \\

\hline  
\hline  

&&&&&&&& \\

Object & $\log$(\Lacc/\Lsun)  &  $\pm\sigma$  & $\log$(\Lacc/\Lsun)& $\pm\sigma$ &$\log$(\Lacc/\Lsun) & $\pm\sigma$ &$\log$(\Lacc/\Lsun) & $\pm\sigma$  \\
       &  ${\ion{He}{i}~\lambda5016}$ & (dex)        & ${\ion{He}{i}~\lambda5876}$ &    (dex)     & ${\ion{He}{i}~\lambda6678}$ &   (dex) & ${\ion{He}{i}~\lambda10830}$ &  (dex)   \\
\hline
&&&&&&&& \\
RY\,Tau     &   ...     & ...    &  $-$0.14  & 0.21  & $-$0.57  & 0.32  & $-$0.63  &  0.38 \\
DG\,Tau     &   ...     & ...    &  $-$0.42  & 0.19  & $-$0.29  & 0.43  & $-$0.54  &  0.36 \\
DL\,Tau     &  $-$0.54  & 0.21   &  $-$0.22  &  0.19  & $-$0.13  & 0.41 & $-$0.50   & 0.36 \\
HN\,Tau\,A  &  $-$0.97  & 0.22   &  $-$1.03  &  0.23  & $-$1.02  & 0.44  & $-$1.24  & 0.43  \\
DO\,Tau     &  $-$1.10  & 0.23   &  $-$0.72  &  0.20  & $-$0.66  & 0.30 & $-$1.47   & 0.46  \\
RW\,Aur\,A  &  $+$0.71  & 0.16   &  $+$0.19  &  0.18  & $+$0.83  & 0.25  & $-$0.02  & 0.33  \\
CQ\,Tau     &  ...     &   ...   &  $-$1.11  & 0.25  &  $-$1.58 & 0.43  & $-$1.13  &  0.42 \\

\hline                                                    
\end{longtable}
 \end{landscape}

\setlength{\tabcolsep}{5pt}

 \begin{landscape}
\begin{longtable}{l|c|c|c|c}
\caption[ ]{\label{tab:fluxes_EW_CaI3934} Measured fluxes and equivalent widths of the $\ion{Ca}{i}~\lambda3934$ line and accretion luminosity for the CTTs sample.}\\
\hline\hline   
Object & $f_{\rm \ion{Ca}{i}~\lambda3934}$ & $EW_{\rm \ion{Ca}{i}~\lambda3934}$   &  $\log$(\Lacc/\Lsun)  &  $\pm\sigma$  \\
    & (erg\,s$^{-1}$\,cm$^{-2}$)    & (\AA)    &    ${\rm \ion{Ca}{i}~\lambda3934}$      &  (dex)   \\
\hline
 &  & & & \\           
RY\,Tau     &  2.57($\pm$0.29)e$-$13 &   -6.61$\pm$2.97  & $-$0.26 & 0.16 \\ 
DG\,Tau     &  7.45($\pm$0.72)e$-$13 &  -40.05$\pm$16.00 & $-$0.12 & 0.15 \\ 
DL\,Tau     &  2.43($\pm$0.39)e$-$13 &  -56.38$\pm$15.00 & $-$0.57 & 0.17 \\ 
HN\,Tau\,A  &  3.29($\pm$0.49)e$-$13 &  -94.81$\pm$25.50 & $-$0.58 & 0.17 \\ 
DO\,Tau     &  2.14($\pm$0.49)e$-$13 &  -25.94$\pm$10.40 & $-$0.66 & 0.19 \\ 
RW\,Aur\,A  &  2.72($\pm$0.33)e$-$12 &  -13.27$\pm$2.86  & $+$0.24 & 0.15 \\ 
CQ\,Tau     &  3.30($\pm$0.64)e$-$13 &   -2.71$\pm$0.70  & $-$0.95 & 0.19 \\ 

\hline                                                    
\end{longtable}
\end{landscape}

\end{appendix}

\end{document}